\pdfoutput=1
\documentclass[usenatbib]{mn2e}
\usepackage{apjfonts}
\usepackage{amssymb}
\usepackage{amsmath}
\usepackage{ctable}
\usepackage{lipsum,cuted}
\usepackage{multicol}
\usepackage{fixltx2e} 
\usepackage{verbatim}

\newcommand{\be}{\begin{equation}}
\newcommand{\ee}{\end{equation}}

\newcommand{\msun}{M_{\sun}}

\newcommand{\paperone}{Paper {\small I}}
\newcommand{\papertwo}{Paper {\small II}}

\newcommand{\partialAB}[2]{\frac{\partial{#1}}{\partial{#2}}}
\newcommand{\ispecial}{i_{0}}
\newcommand{\vfastest}{v_{f,\,0}}

\newcommand{\driftvel}{{\bf w}_{s}}
\newcommand{\driftvelmag}{w_{s}}
\newcommand{\driftvelhat}{\hat{{\bf w}}_{s}}
\newcommand{\driftvelX}{{\bf w}_{0}}
\newcommand{\driftvelXmag}{w_{0}}
\newcommand{\driftvelXhat}{\hat{{\bf w}}_{0}}
\newcommand{\BV}{Brunt-V\"ais\"al\"a}

\newcommand{\coeffTSrho}{\zeta_{s}} 
\newcommand{\coeffTSrhoV}{\zeta_{s/w}}
\newcommand{\coeffTLrho}{\zeta_{q}} 
\newcommand{\tildecoeffTLrho}{\tilde{\zeta}_{q}}
\newcommand{\coeffTLrhoV}{\zeta_{q/w}}
\newcommand{\coeffTSv}{\zeta_{w}}
\newcommand{\tildeCoeffTSv}{\tilde{\zeta}_{w}}
\newcommand{\coeffTSqCoulomb}{\zeta_{C}}

\newcommand{\coulombcoeff}{a_{C}} 
\newcommand{\iimag}{i} 
\newcommand{\omegaZ}{\varomega} 

\newcommand{\apjl}{ApJL}
\newcommand{\apjs}{ApJS}

\newcommand{\aap}{A\&A}

\newcommand{\acknowledgments}{\begin{small}\section*{Acknowledgments}\end{small}}
\newcommand\altaffilmark[1]{$^{#1}$}
\newcommand\altaffiltext[1]{$^{#1}$}
\title[The MHD RDIs]{Ubiquitous Instabilities of Dust Moving in Magnetized Gas\vspace{-0.5cm}}

\vspace{-0.2cm}
\author[Hopkins \&\ Squire]{
\parbox[t]{\textwidth}{ 
Philip F.~Hopkins\altaffilmark{1}, \&\ 
Jonathan Squire\altaffilmark{1}
} 
\vspace*{6pt} \\
\altaffiltext{1}{TAPIR, Mailcode 350-17, California Institute of Technology, Pasadena, CA 91125, USA} 
\vspace{-0.5cm}
}

\date{Submitted to MNRAS, January 2018\vspace{-0.6cm}}
\begin{document}
\maketitle
\label{firstpage}

\vspace{-0.2cm}
\begin{abstract}
\vspace{-0.2cm}

Squire \&\ Hopkins (2017) showed that coupled dust-gas mixtures are generically subject to ``resonant drag instabilities'' (RDIs), which drive violently-growing fluctuations in both. But the role of magnetic fields and charged dust has not yet been studied. We therefore explore the RDI in gas which obeys ideal MHD and is coupled to dust via both Lorentz forces and drag, with an external acceleration (e.g., gravity, radiation) driving dust drift through gas. We show this is always unstable, at {\em all} wavelengths and non-zero values of dust-to-gas ratio, drift velocity, dust charge, ``stopping time'' or drag coefficient (for any drag law), or field strength; moreover growth rates depend only weakly (sub-linearly) on these parameters. Dust charge and magnetic fields do not suppress instabilities, but give rise to a large number of new instability ``families,'' each with distinct behavior. The ``MHD-wave'' (magnetosonic or Alfv\'en) RDIs exhibit maximal growth along ``resonant'' angles where the modes have a phase velocity matching the corresponding MHD wave, and growth rates increase {\em without limit} with wavenumber. The ``gyro'' RDIs are driven by resonances between drift and Larmor frequencies, giving growth rates sharply peaked at specific wavelengths. Other instabilities include ``acoustic'' and ``pressure-free'' modes (previously studied), and a family akin to cosmic ray instabilities which appear when Lorentz forces are strong and dust streams super-Alfv\'enically along field lines. We discuss astrophysical applications in the warm ISM, CGM/IGM, HII regions, SNe ejecta/remnants, Solar corona, cool-star winds, GMCs, and AGN. 

\end{abstract}

\begin{keywords}
instabilities --- turbulence --- ISM: kinematics and dynamics --- star formation: general --- 
galaxies: formation --- cosmology: theory --- planets and satellites: formation --- accretion, accretion disks\vspace{-0.5cm}
\end{keywords}

\vspace{-1.1cm}
\section{Introduction}
\label{sec:intro}

Almost all astrophysical fluids are laden with dust, and that dust is critical for a wide range of phenomena including planet and star formation, extinction and reddening, stellar evolution (in cool stars), astro-chemistry, feedback and launching of winds from star-forming regions and active galactic nuclei (AGN), the origins and evolution of heavy elements, inter-stellar gas cooling or heating, and many more. It is therefore of paramount importance to understand how dust and gas interact dynamically. 

\citet{squire.hopkins:RDI} (henceforth \paperone) showed that dust-gas mixtures are {\em generically} unstable to a broad class of previously-unrecognized instabilities. These ``resonant drag instabilities'' (RDIs) appear whenever a gas system that supports any wave (or linear perturbation/mode) with frequency $\omega_{g}$ also contains dust streaming with a finite drift velocity $\driftvel$ (relative to the gas). Although a very broad range of wavenumbers are typically unstable, the ``resonance,'' which produces the fastest-growing modes, arises when $\driftvel \cdot {\bf k} = \omega_{g}$ 
(
or equivalently, the dust drift velocity in the direction of wave propagation is equal to the natural phase velocity of the wave in the gas). 

Given the fact that dust is ubiquitous, and that essentially any type of gas system can meet these conditions, we expect the RDI to arise across a wide range of astrophysical contexts in the ISM, stars, galaxies, AGN, and more. Dust is almost always expected to have some non-vanishing drift velocity owing to combinations of radiative forces on grains (e.g., absorption of light, photo-desorption, photo-electric, and Poynting-Robertson effects) or gas (e.g., line-driving the gas, pushing gas instead of dust), or gravity (which causes dust to ``settle'' when the gas is pressure supported), or any hydrodynamic/pressure forces on the gas (accelerating or decelerating gas, but not [directly] dust).

\paperone\ briefly noted several representative examples of the RDI, where the resonance could be between gas and acoustic modes (sound waves), magnetosonic waves, \BV\ oscillations, or epicyclic oscillations (which turns out to be the well-studied \citealt{youdin.goodman:2005.streaming.instability.derivation} ``streaming instability''). Each of these modes is associated with a corresponding RDI. Many of these -- particularly the epicyclic RDI and new variations with faster growth rates -- are explored in more detail in the specific context of proto-planetary and proto-stellar disks, in \citet{squire:rdi.ppd}. In \citet{hopkins:2017.acoustic.RDI} (hereafter \papertwo), we explored the ``acoustic RDI'' in detail. This is perhaps the simplest example of the RDI -- ideal, inviscid, neutral hydrodynamics, where the only wave (absent dust) is a sound wave. However, there  still exists  an entire family of instabilities, with a range of non-trivial scalings for their growth rates depending on wavenumber and drift velocity. We showed that the growth rates were particularly interesting for cool-star winds, and regions of the cold, dense ISM (e.g., star-forming GMCs and ``dusty torii'' around AGN). However, because there is only one wavespeed in the problem, $c_{s}$, the acoustic RDI in particular requires $|\driftvel| \ge c_{s}$ for the ``resonance condition'' to be met; otherwise the system is still unstable but growth rates are significantly lower. 

Of course, a huge range of astrophysical systems are ionized (at least partially), and magnetic fields cannot be neglected. This introduces two important changes to the RDI. First, even in ideal MHD with neutral dust grains, there are now three wave families: fast and slow magnetosonic waves, and Alfv\'en waves. Each of these has a corresponding associated family of RDIs. Second, if dust grains are charged (as they are expected to be), then at low densities and/or in sufficiently warm/hot gas, the Lorentz forces on grains can be significantly stronger than the drag forces (and, if the grains are moving sub-sonically, electrostatic Coulomb drag may dominate over collisional Epstein or Stokes drag; see e.g., \citealt{elmegreen:1979.charged.grain.diffusion.gmcs}). This again introduces new families of RDIs.

Our purpose in this paper is therefore to study the linear instability of the RDI in magnetized gas, allowing for arbitrarily charged dust. We will show that all of these changes introduce new associated instabilities and behaviors of the RDI -- a wide variety of previously-unrecognized families of instabilities appear, each with associated resonances and different mode structure. Critically, none of these changes uniformly suppresses the RDI. In fact, we will show that the presence of slow and Alfv\'en waves, which have phase velocities that can become arbitrarily slow (at the appropriate propagation angles), means that the resonance condition can {\em always} be satisfied, regardless of the dust drift velocity. We will show that for {\em any} non-zero dust drift velocity $\driftvel$, dust-to-gas mass ratio $\mu$, magnetic field strength $\beta$, grain charge, or dust drag law, the instabilities persist and {\em all} wavelengths are unstable, with growth rates that formally become infinitely fast at small wavelengths (absent dissipative effects such as viscosity).

In \S~\ref{sec:overview.modes}, we provide a brief, high-level overview of the most important new instability families described here. \S~\ref{sec:deriv}-\ref{sec:scales.validity} are largely technical: in \S~\ref{sec:deriv}, we present the relevant derivation, equilibrium (background) solutions (\S~\ref{sec:equilibrium}), linearized equations-of-motion (\S~\ref{sec:linearized.equations}), detailed scalings for different drag laws and Lorentz forces (\S~\ref{sec:stopping.time.scalings}), and the resulting dispersion relation (\S~\ref{sec:dispersion.relation}; more detail in appendices). \S~\ref{sec:parallel.modes}-\ref{sec:gyro} are devoted to detailed discussion of the origins, instability conditions, resonance conditions (wavevector angles or wavelengths where growth rates are fastest), and mode structure of the different instabilities. \S~\ref{sec:parallel.modes} focuses on the families of ``parallel'' pressure-free and cosmic-ray-like modes; \S~\ref{sec:mhdwave.rdi} on the MHD wave (fast and slow magnetosonic and Alfv\'en) RDIs; and \S~\ref{sec:gyro} on the gyro RDIs. \S~\ref{sec:other.modes} briefly notes additional modes and \S~\ref{sec:scales.validity} discusses the range of scales where our derivations are valid. In \S~\ref{sec:application} we discuss astrophysical applications. We first present some relevant scalings (\S~\ref{sec:application:scalings}) then provide simple estimates of the growth rates and modes of greatest interest in different contexts, including the warm ionized and warm neutral medium (\S~\ref{sec:application.env:wim.wnm}), the circum- and inter-galactic medium (\S~\ref{sec:application.env:cgm.igm}), HII regions (\S~\ref{subsub: HII regions}), SNe ejecta and remnants (\S~\ref{sec:application.env:sne}), the Solar and stellar coronae (\S~\ref{sec:application.env:coronae}), cool-star winds (\S~\ref{sec:application.env:coolstars}), the cold ISM in GMCs and around AGN (\S~\ref{sec:application.env:cold.ism}), and protoplanetary disks and planetary atmospheres (\S~\ref{sec:application.env:ppd}). We conclude in \S~\ref{sec:discussion}. 

Readers primarily interested in astrophysical applications may wish to simply read the overview of the instabilities in \S~\ref{sec:overview.modes}, then skip directly to \S~\ref{sec:application}. Table~\ref{tbl:variables} defines a number of variables to which we will refer throughout the text. 

\begin{figure}
\begin{center}
\includegraphics[width=1.0\columnwidth]{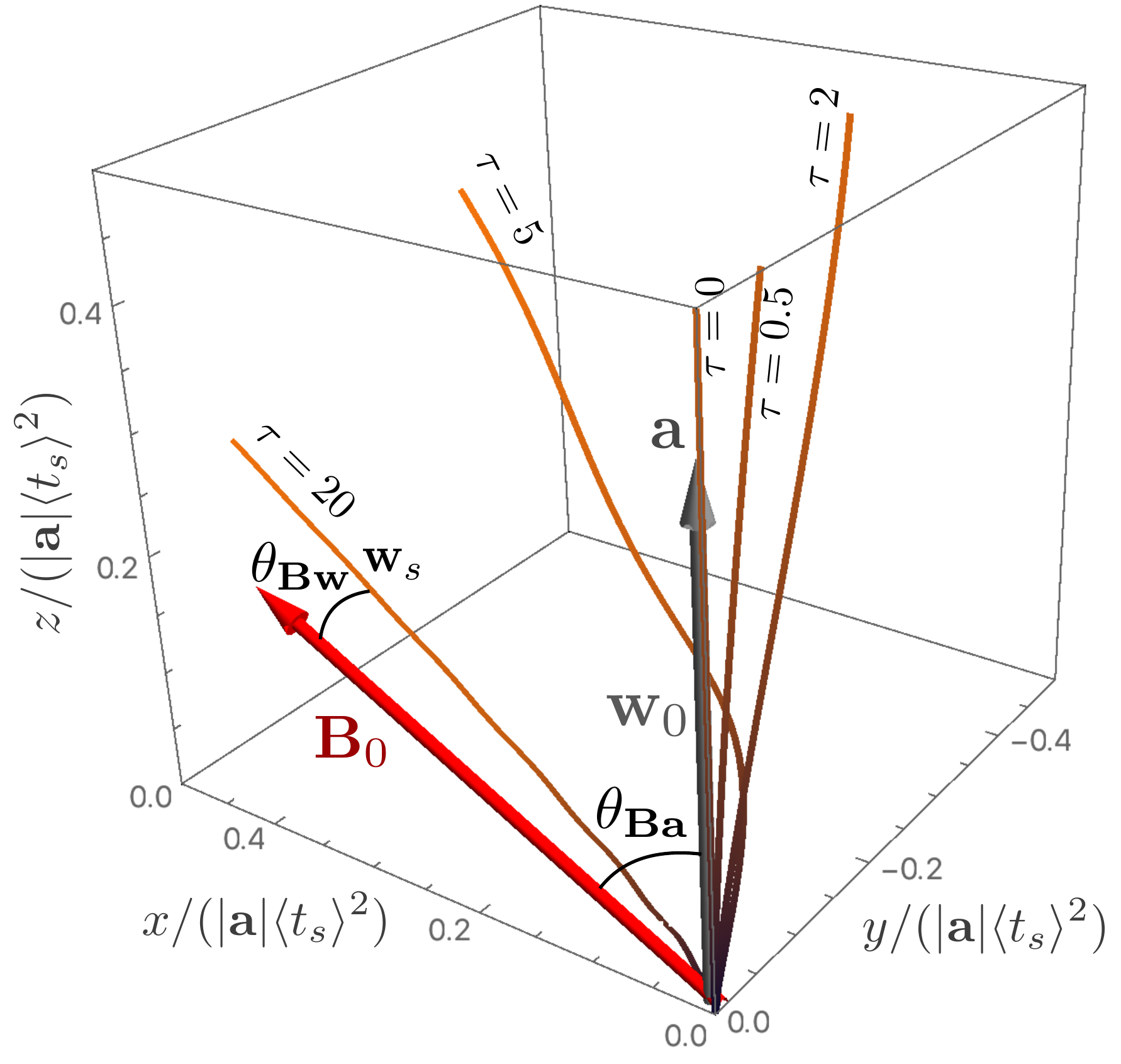}
\caption{Illustration of the geometry of charged dust grains accelerated by a force (e.g., gravity, radiation pressure, etc., producing ${\bf a}$) at some angle to the magnetic field (${\bf B}_{0}$). Lines show the trajectories of dust at different $\tau = \langle t_{s}\rangle/\langle t_{L}\rangle$ (ratio of dust drag stopping time $t_{s}$ to gyro/Larmor time $t_{L}$), accelerated from rest at the origin (we assume constant $t_{s}$ for simplicity, rather than, for example, Epstein drag). The line color indicates the time along the trajectory from $t=0$ (dark brown) to $t=2.5\langle t_{s}\rangle$ (orange). The equilibrium drift velocity is $\driftvel$, while $\driftvelX$ is the drift velocity in the absence of magnetic (Lorentz) forces on grains ($\tau=0$). We also indicate, with the gray and orange arrows respectively, the acceleration direction ${\bf a}$ (with $\driftvelXhat = \hat{{\bf a}}$) and the magnetic field direction. Some relevant angles defined in the text, $\theta_{{\bf B a}}$ and  $\theta_{{\bf B w}}$ (for the $\tau=20 $ trajectory), are also marked. We see that the dust quickly (by $t\sim \langle t_{s}\rangle$, or even faster for $\tau\gg 1$) reaches the  equilibrium trajectory given in Eq.~\eqref{eqn:W0}, with drift velocity $\driftvel$ which is preferentially aligned with the magnetic field at large-$\tau$.}
\label{fig:geometry}\vspace{-0.25cm}
\end{center}
\end{figure}

\begin{footnotesize}
\ctable[caption={{\normalsize Variables used throughout the text (defined here)}\label{tbl:variables}},center,star]{lcl}{\tnote[ ]{
}}{
\hline\hline 
Variable & Equation/Definition & Notes (explanation)\\
\hline\hline 
$X_{0}$ & -- & value of quantity ``$X$'' in the equilibrium homogeneous medium \\ 
$c_{s}$, $v_{A}$, $\beta$ & $c_{s}^{2} \equiv \frac{\partial P}{\partial \rho}$, $v_{A}^{2} \equiv \frac{|{\bf B}_{0}|^{2}}{4\pi\,\rho}$, $\beta \equiv \frac{c_{s}^{2}}{v_{A}^{2}}$ & gas thermal sound speed ($c_{s}$), Alfv\'en speed ($v_{A}$), and plasma $\beta$ (field strength) \\
$\vfastest$ & $v_{f,\,0}^{2} \equiv c_{s}^{2} + v_{A}^{2}$ & fastest gas wavespeed \\
$v_{\pm}$ & $v_{\pm}^{2} = \frac{1}{2}\left[ \vfastest^{2} \pm \left({\vfastest^{4} - 4\,c_{s}^{2}\,v_{A}^{2}\,\cos^{2}{\theta_{\bf Bk}}}\right)^{1/2}\right]$ & fast (``$+$'' branch) and slow (``$-$'') magnetosonic wave speeds \\
$\mu$, $\hat{\mu}$ & $\mu\equiv\frac{\rho_{d,\,0}}{\rho_{0}}$, $\hat{\mu} \equiv \frac{\rho_{d,\,0}}{\rho_{0}+\rho_{d,\,0}}=\frac{\mu}{1+\mu}$ & mean dust-to-gas ($\mu$) or dust-to-total ($\hat{\mu}$) mass ratio \\
$\langle t_{s} \rangle$, $\langle t_{L} \rangle$, $\tau$ & (see \S~\ref{sec:stopping.time.scalings}), $\tau \equiv \frac{\langle t_{s} \rangle}{\langle t_{L} \rangle}$ & equilibrium ``stopping time'' or drag coefficient $\langle t_{s} \rangle$, gyro/Larmor time $\langle t_{L} \rangle$, and ratio $\tau$  \\
$\coeffTSrho$, $\coeffTSv$, $\coeffTLrho$ & (see \S~\ref{sec:stopping.time.scalings}); $\tilde{\zeta}_{x}\equiv 1+\zeta_{x}$; $\zeta_{x/w}\equiv\zeta_{x}/\tilde{\zeta}_{w}$ & 
dimensionless scaling of perturbations to $t_{s}$ from $\rho$ ($\coeffTSrho$) or $\driftvel$ ($\coeffTSv$), or to $t_{L}$ from $\rho$ ($\coeffTLrho$) \\
$\driftvel$ & Eq.~\ref{eqn:mean.v.offset} & relative dust-gas drift velocity ($\driftvel = \langle {\bf v}_{0,\,{\rm dust}} - {\bf u}_{0,\,{\rm gas}} \rangle$; $\driftvelmag \equiv |\driftvel|$) \\ 
$\driftvelX$ & $\driftvelX \equiv {{\bf a}\,\langle t_{s} \rangle}/({1+\mu})$ (Eq.~\ref{eqn:W0}) & value of $\driftvel$ in the absence of Lorentz forces ($\driftvelXmag \equiv |\driftvelX|$) \\ 
${\bf a}$ &  ${\bf a} \equiv {\bf a}_{\rm dust} - {\bf a}_{\rm gas}$ (Eq.~\ref{eqn:general}) & difference in external acceleration (e.g., gravity, radiation, pressure) between dust \&\ gas \\ 
$\cos{\theta_{\bf XY}}$ & $\hat{\bf X}\cdot \hat{\bf Y}$ & cosine of angle between the specified vectors ${\bf X}$ and ${\bf Y}$ (e.g., $\cos{\theta}_{\bf Ba} \equiv \hat{\bf B} \cdot \hat{\bf a}$; see Fig.~\ref{fig:geometry}) \\ 
$\omega$, ${\bf k}$, $k$ & $\delta X = \delta X_{0}\,\exp{[\iimag\,({\bf k}\cdot{\bf x} - \omega\,t )]}$, $k\equiv |{\bf k}|$ & frequency $\omega$ and wavenumber ${\bf k}=(k_{x},\,k_{y},\,k_{z})$ of a mode. $\Im{(\omega)} > 0$ is the growth rate. \\ 
$\hat{x}$, $\hat{y}$, $\hat{z}$ & $\hat{x} \propto (\driftvel\times{\bf B}_{0})\times\driftvel$, $\hat{y}\propto \driftvel\times\hat{\bf B}_{0}$, $\hat{z}=\driftvelhat$ & unit vectors parallel to $(\driftvel\times {\bf B}_{0})\times\driftvel$, $\driftvel\times{\bf B}_{0}$, and $\driftvel$, respectively \\ 
\hline\\
}
\end{footnotesize}

\vspace{-0.5cm}
\section{Overview of Different Instabilities}
\label{sec:overview.modes}
 
%
%
%
%

We will show that, with magnetic fields present, the dust-gas mixture gives rise to several different families of instabilities, each of which can behave differently. To guide the reader, we summarize here the modes which will be explored in detail in this paper. Variable names used here and throughout the manuscript are defined in Table~\ref{tbl:variables}.

\vspace{-0.5cm}
\subsection{Parallel/Aligned Modes}

First we introduce mode families which are primarily ``parallel,'' in that the fastest-growing mode has wavevector along the direction of the dust drift ($\hat{\bf k} \approx \driftvelhat$). This is primarily of interest either (a) at very long wavelengths (low-$k$) at any level of magnetization, or (b) at intermediate/short wavelengths when the ratio of drag stopping time to Larmor time ($\tau \equiv \langle t_{s} \rangle / \langle t_{L} \rangle$) is very large, so that the dust motion (drift and perturbed velocities) becomes increasingly confined along field lines (so $\hat{\bf k} \approx \driftvelhat \approx \hat{\bf B}_{0}$; for explanation, see Eq.~\ref{eqn:mean.v.offset} and \S~\ref{sec:equilibrium}). 

\begin{itemize}

\item{\bf ``Pressure-Free'' or Long-Wavelength Mode}: At sufficiently long wavelength (low-$k$), the system is always unstable\footnote{Technically, some modes, such as the long-wavelength mode, can be stabilized if and only if some (normally order-unity) complex pre-factor $\mathcal{C}$ (which is usually a complicated function of the various parameters here) satisfies $\| \mathcal{C} \| = 0$ exactly, which is possible only for specific, singular values of certain parameters (typically the drift speed, equation of state parameter $\gamma$, $\tau$, and magnetic field angles must all have exactly one certain -- and often un-physical -- value). Even then, we show in \papertwo\ this often does not eliminate the instability, but only the leading-order term in the series expansions used to estimate the growth rates here. We will therefore simply refer to these modes as ``always unstable.''} with a mode that is fastest growing in the parallel direction, with growth rate $\Im(\omega)\sim (\hat{\mu}\,\driftvelXmag^{2}\,k^{2}/\langle t_{s}\rangle)^{1/3}$. This is a strongly compressible, longitudinal mode ($\delta {\bf v}$ and $\delta {\bf u}$ are also aligned with $\hat{\bf k}$), which only occurs on wavelengths much larger than the dust free-streaming length and gyro radius. When the wavelength is sufficiently large, the aerodynamic force from dust on gas (``back-reaction'') becomes larger than the gas pressure/magnetic forces, so this is fundamentally an instability of two frictionally-coupled co-spatial pressure-free fluids, and is essentially identical in non-magnetized gas (\papertwo). 

\item{\bf ``Quasi-Sound'' and ``Quasi-Drift'' Modes}: Even at very large $\tau$, two longitudinal, compressible modes exist when $\hat{\bf k} \approx \hat{\bf B}_{0}$, which are only weakly modified by magnetic fields. These are field-aligned acoustic modes, and have identical scalings and mode structure as shown in \papertwo. The ``quasi-sound'' mode is a modified sound wave, propagating at the sound speed, and is unstable when $\driftvelmag$ is trans or super-sonic, with growth rate $\sim \mu\,\driftvelmag/(2\,c_{s}\,\langle t_{s} \rangle)$. The ``quasi-drift'' mode is modified dust advection, propagating at $\driftvelmag$, and is unstable either for (a) super-sonic drift or (b) sub-sonic drift if Coulomb drag dominates over Epstein drag, with growth rate $\Im(\omega)\sim (\mu/\langle t_{s} \rangle)\,{\rm min}(\driftvelmag^{2}/c_{s}^{2} \, , \, 1)$. These are the ``out of resonance'' modes that merge and become the ``fast magnetosonic'' RDI at the appropriate $\hat{\bf k}$, however over some conditions they can be faster-growing than the resonant mode either because no resonant angle exists, or because non-aligned modes are suppressed by $\tau$ (and they can be faster-growing than the cosmic ray-like modes below at low-$k$ or intermediate-$\tau$).

\item{\bf ``Cosmic Ray Streaming'' Mode}: At large $\tau$, other parallel modes appear ($\hat{\bf k} \approx \driftvelhat \approx \hat{\bf B}_{0}$). These fall into two broad categories, which are manifestations of well-known 
resonant and non-resonant cosmic-ray instabilities \citep{kulsrud.1969:streaming.instability,bell.2004.cosmic.rays}. The resonant variety is the high-$\tau$ limit of the gyro-resonant RDI (see \S\ref{sub:gyro.modes.overview} below). The non-resonant variety is a transverse, weakly-compressible mode, featuring large transverse perturbations to the magnetic field with corresponding gyro motion of the dust and gas, as the dust drifts super-Alfv\'enically along the field (it is unstable for $\driftvelmag > v_{A}\,\hat{\mu}^{-1/2}$ along $\hat{\bf B}_{0}$). It has a growth rate $\Im(\omega)\sim k\,\driftvelmag\,\hat{\mu}^{1/2}$ at large scales, a growth rate $\Im(\omega)\sim  ( \mu\,\driftvelmag\langle t_L \rangle)^{1/2} k_z^{1/2}$ for an intermediate range of scales, $\mu\,v_A/\driftvelmag \lesssim k\,v_A\langle t_L \rangle \lesssim \mu\,\driftvelmag/v_A$, and is stabilized at very short wavelengths.
\end{itemize}

\vspace{-0.5cm}
\subsection{The MHD-Wave RDI Modes}

These are the simple RDI modes described in \S~\ref{sec:intro}, which have growth rates that depend on the direction of $\hat{\bf k}$ and usually peak at the ``resonant angle'' when $\omega_{d} = \driftvel \cdot {\bf k} = \omega_{g} = {v}_{p}(\hat{\bf k})\, { k}$, where $v_{p}(\hat{\bf k})$ is the phase velocity of either the fast magnetosonic, slow magnetosonic, or Alfv\'en wave in the gas in the direction $\hat{\bf k}$ (without dust). Because ${v}_{p}$ depends on angle, there are a range of angles that satisfy the resonant criterion, so each wave sources a different sub-family of resonant instabilities.\footnote{The forward and backward-traveling wave groups behave essentially identically for the normal MHD-Wave RDI modes, so we do not distinguish them here.} These are always unstable {\em at all wavelengths} if the resonant criterion can be satisfied. At intermediate wavelengths,\footnote{As shown in \paperone\ and \papertwo, technically the intermediate-wavelength (``mid-$k$'') and short-wavelength (``high-$k$'') RDI modes (for a given MHD wave family) are different branches of the dispersion relation, which produce faster growth rates at the same angle at mid-$k$ and high-$k$, respectively. Since their behavior is similar and the resonance condition is identical (and for some parameter choices they become degenerate at both mid-$k$ and high-$k$), we will refer to them for simplicity as a single mode, with different behavior in different limiting regimes.}
 the growth rates scale as $\sim (\hat{\mu}\,k\,v_{f,\,0} / 2\,\langle t_{s} \rangle)^{1/2}$, and at sufficiently high $k$ as $\sim (\mu\,k\,v_{f,\,0}/2\,\langle t_{s} \rangle^{2})^{1/3}$. Out of-resonance the modes are still present but with (usually) lower growth rates. 

\begin{itemize} 

\item{\bf Fast Magnetosonic RDI}: Here the resonance is with the fast wave. Since this has a minimum phase speed $|v_{p}| = {\rm max}(v_{A},\,c_{s})$, the resonance condition can only be satisfied if $\driftvelmag \ge {\rm max}(v_{A},\,c_{s}) \sim v_{f,\,0}$. This is the simple MHD extension of the acoustic RDI from \papertwo\ (it reduces to that for $\beta\rightarrow \infty$), and the mode structure, growth rates (above), and resonant angles ($\cos{\theta}_{\bf wk} \sim \pm \driftvelmag / v_{f,\,0}$) are very similar to the acoustic RDI after the replacement $c_{s} \rightarrow v_{f,\,0}$. These are strongly-compressible, fast-wave-like\footnote{By ``fast-wave-like,'' we mean that the mode structure/eigenfunction resembles a fast wave in both the dust and the gas to leading order, with approximately the usual ratios and phases of the density ($\delta\rho$), longitudinal (acoustic/pressure) velocity ($\delta{\bf u} \cdot \hat{\bf k}$), and transverse (electromagnetic/tension) velocity ($\delta{\bf u} \cdot \hat{\bf B}_{0}$ and $\delta {\bf B}$ in the direction perpendicular $\hat{\bf k}$) perturbations. The presence of drift and Lorentz forces adds a perturbation to these, and to terms which normally vanish (e.g., $\delta{\bf u}$ and $\delta{\bf B}$ in the mutually-perpendicular direction ${\bf k} \times {\bf B}$), and (more importantly) introduces a phase offset between the dust and gas perturbations, which drives the growth of the instability (see \S~\ref{sec:mhd.wave.mode.structure}).} modes that source proportionally very large fluctuations in the dust-to-gas ratio at high-$k$ (see \papertwo). 

\item{\bf Slow Magnetosonic RDI}: Since the slow wave has a phase speed which vanishes as $\hat{\bf k} \cdot \hat{\bf B}_{0} \rightarrow 0$, there always exists a range of angles that satisfy the resonant condition with the slow mode (for any $\driftvel$). However, when $\driftvelmag \ll {\rm min}(c_{s},\,v_{A})$, the resonant condition can only be satisfied at angles $\hat{\bf k}$ nearly perpendicular to $\hat{\bf B}_{0}$, where the phase speed is low, so the growth rates are suppressed by a factor $\sim \driftvelmag/v_{f,\,0}$ (at intermediate-$k$) or $\sim (\driftvelmag/v_{f,\,0})^{2/3}$ (at high-$k$). The perpendicular nature also means the mid-$k$ (but not high-$k$) mode can be suppressed at large $\tau$ when $\driftvelmag$ is small. This is a compressible, slow-wave-like mode that also sources proportionally large fluctuations in the dust-to-gas ratio at high-$k$.

\item{\bf Alfv\'en RDI}: Like the slow mode, the phase speed of Alfv\'en waves also vanishes as $\hat{\bf k} \cdot \hat{\bf B}_{0} \rightarrow 0$, so there is again always a range of angles which satisfy the Alfv\'en RDI condition. These are Alfv\'en-wave-like modes, so are primarily transverse and weakly-compressible in the gas (although the dust has a large $\delta \rho_{d}$ component). The transverse nature of Alfv\'en modes means that coupling to the dust occurs through the Lorentz force (coupling transverse field perturbations to compressible dust fluctuations), and the instability vanishes in the mid-$k$ regime if $\tau\rightarrow 0$. Interestingly, the growth rates are multiplied by $\sim \driftvelmag/v_{f,\,0}$ (mid-$k$) or $\sim (\driftvelmag/v_{f,\,0})^{1/3}$ (high-$k$), in {\em both} super-and-sub-sonic limits, so when drift is super-sonic the growth rates can be even faster than the fast RDI (while with sub-sonic drift they are similar to the slow RDI). Also, like the slow mode, the mid-$k$ growth rates can be suppressed at large $\tau$ when $\driftvelmag \ll {\rm min}(c_{s},\,v_{A})$ because the resonance requires $\hat{\bf k}$ nearly-perpendicular to $\hat{\bf B}_{0}$. However, the high-$k$ mode growth rate is {\em enhanced} by a factor $\sim \tau^{1/3}$ when $\tau \gg 1$ (but this may require very large $k$ to be realized). 

\end{itemize}

\vspace{-0.5cm}
\subsection{The Gyro-Resonant RDI Modes}\label{sub:gyro.modes.overview}

With non-zero Lorentz forces on grains, there is another dust eigenmode featuring coupled advection and gyro motion. The dispersion relation of this mode at high-$k$ or large-$\tau$ is given by $\omega_{d} \sim \driftvel \cdot {\bf k} \pm \langle t_{L} \rangle^{-1}$. There is thus another set of resonances, the ``gyro-resonant'' RDI modes, which occur at the resonant condition $\omega_{d} \sim \driftvel \cdot {\bf k} \pm \langle t_{L} \rangle^{-1} = \pm v_{p}(\hat{\bf k})\,k$, where again $v_{p}$ is the phase velocity of any of the natural gas modes (Alfv\'en, fast, or slow). So there are three gyro-resonant wave families, each of which can satisfy resonance for any of the four differently-signed versions of the resonance equation (each of which, in turn, has a range of angles that satisfy the equation). However, when $k$ is sufficiently large, the $\tau$ term is negligible and the resonance condition reduces to $\driftvel \cdot {\bf k} \approx \pm v_{p}(\hat{\bf k})\, { k}$, i.e., these become degenerate with the MHD RDIs above. When $| \driftvel \cdot {\bf k} \pm \langle t_{L} \rangle^{-1} |$ is small, terms we neglected in this dust eigenmode (discussed in detail in \S~\ref{sec:gyro:overview}) cannot be ignored and these stabilize the mode. So the interesting behavior primarily comes from the gyroresonances when $\langle t_{L} \rangle^{-1} \sim \pm v_{p}\,k$, and instability at these resonances requires $\tau \gtrsim 1$. Thus, fundamentally unlike the MHD RDIs, for a given angle $\hat{\bf k}$ the resonance occurs at a specific wavenumber $k$. 

\begin{itemize} 

\item{\bf Fast Gyroresonance}: Because the fast-wave phase velocity $v_{p} \sim v_{f,\,0}$, depends only weakly on angle, this resonance is sharply peaked in $k$ around $k^{-1}\approx \langle t_{L} \rangle\,v_{f,\,0}$ (barring special cases where the resonance overlaps with the MHD-wave RDI angles). At the resonant wavenumber, the growth rates are $\sim \hat{\mu}^{1/2} / \langle t_{L} \rangle$, although the growth rates are suppressed for the wave angles far from either alignment with the drift or MHD-wave RDI angles.

\item{\bf Slow Gyroresonance}: Here the phase speed varies smoothly from $v_{p}={\rm min}(c_{s},\,v_{A})$ for field-parallel modes to $v_{p}=0$ for field-perpendicular modes, so at every angle $\hat{\bf k}$ the resonance occurs at a different $k^{-1} \sim \langle t_{L} \rangle\,v_{p}(\hat{\bf k})$ ($\sim  \langle t_{L} \rangle\,{\rm min}(c_{s},\,v_{A})\,\hat{\bf k}\cdot\hat{\bf B}_{0}$ for small $|\hat{\bf k}\cdot\hat{\bf B}_{0}|$). For sufficiently perpendicular angles, $|\hat{\bf k}\cdot\hat{\bf B}_{0}| \ll \driftvelmag / {\rm min}(c_{s},\,v_{A})$, this implies sufficiently large $k$ such that the mode becomes degenerate with the high-$k$ RDI. The growth rates around resonance scale similarly to the fast gyro-resonance at sufficiently high-$k$, but the suppression factor for modes perpendicular to the dust-drift means that gyro-modes that simultaneously satisfy the slow-wave RDI condition are less interesting. At sufficiently sub-Alfv\'enic drift velocities $\driftvelmag \ll v_{A}$, the slow gyro-RDI becomes stable for most angles.   

\item{\bf Alfv\'en Gyroresonance}: The Alfv\'en-wave phase velocity is $v_{p} = v_{A}\,\hat{\bf k}\cdot\hat{\bf B}_{0}$,  so again, the resonant wavenumber, $k^{-1} \sim \langle t_{L} \rangle\,v_{A}\,\hat{\bf k}\cdot\hat{\bf B}_{0}$, depends relatively strongly on mode angle, $\hat{\bf k}$. For angles $|\hat{\bf k}\cdot\hat{\bf B}_{0}| \ll \driftvelmag / v_{A} $ this goes to high-$k$ and becomes degenerate with the high-$k$ RDI. The scaling of the growth rate is similar to the slow-gyro RDI above.

\end{itemize}

\vspace{-0.5cm}
\section{Basic Equations \&\ Linear Perturbations}
\label{sec:deriv}

\vspace{-0.05cm}
\subsection{Equations Solved \&\ Equilibrium Solution}
\label{sec:equilibrium}

As in \papertwo, consider a mixture of gas and a second component which can be approximated as a pressure-free fluid, which we will refer to as ``dust'' henceforth  (see, e.g.,  \citealt{youdin.goodman:2005.streaming.instability.derivation} and App.~A of \citealt{Jacquet:2011cy}, as well as \S~\ref{sec:scales.validity} below). We consider a magnetized gas which obeys the ideal MHD equations,\footnote{Non-ideal and kinetic effects, as well as the effects of current carried by grains in the induction equation, are briefly discussed in \S~\ref{sec:scales.validity}, where we show they are negligible for many astrophysical cases discussed in \S~\ref{sec:application}.} and dust which feels both a generalized arbitrary drag force (e.g., neutral/aerodynamic and/or Coulomb drag) and Lorentz forces from the magnetic fields. The system is described by the conservation equations:
\begin{align}
	\nonumber \partialAB{\rho}{t} + \nabla\cdot ({\bf u}\,\rho) &= 0\, ,\\
\nonumber \left(\partialAB{}{t} + {\bf u}\cdot\nabla \right){\bf u} &=  -\frac{\nabla P}{\rho} - \frac{1}{4\pi\,\rho}\,{\bf B} \times\left( \nabla \times {\bf B} \right) + {\bf g} \\ 
\nonumber & + \frac{\rho_{d}}{\rho}\,\left[ \frac{({\bf v}-{\bf u})}{t_{s}} + \frac{({\bf v}-{\bf u})\times\hat{\bf B}}{t_{L}} \right]\, ,\\
\nonumber \partialAB{\bf B}{t} &= \nabla\times\left( {\bf u} \times {\bf B} \right) \, ,\\ 
\nonumber  \nabla \cdot {\bf B} &= 0\, , \\
\nonumber \partialAB{\rho_{d}}{t} + \nabla\cdot ({\bf v}\,\rho_{d}) &= 0\, ,\\
\label{eqn:general} \left(\partialAB{}{t} + {\bf v}\cdot\nabla \right){\bf v} &= {\bf g} + {\bf a} - \left[ \frac{({\bf v}-{\bf u})}{t_{s}} + \frac{({\bf v}-{\bf u})\times\hat{\bf B}}{t_{L}} \right] \, .
\end{align}
Here ($\rho,\,{\bf u}$) and ($\rho_{d},\,{\bf v}$) are the (density,\,velocity) of the gas and dust, respectively, and $P$, ${\bf B}$ are the gas (thermal) pressure and magnetic field. Throughout this manuscript $\hat{\bf x} \equiv {\bf x}/|{\bf x}|$ denotes the unit vector in direction ${\bf x}$. The external acceleration of the gas is ${\bf g}$, while ${\bf g}+{\bf a}$ is the external acceleration of dust (i.e., ${\bf a}$ is any difference in the dust vs.\ gas acceleration). The dust experiences a drag acceleration ${\bf a}_{\rm drag} = -({\bf v}-{\bf u})/t_{s}$ where $t_{s}$ is the arbitrary drag coefficient or ``stopping time'' (which is generally a function of other properties; see below). Similarly the dust, if it charged, feels a Lorentz force with acceleration ${\bf a}_{\rm Lorentz} = -({\bf v}-{\bf u})\times \hat{\bf B} / t_{L}$, where we define the Larmor/gyro time $t_{L} = m_{\rm grain}\,c / |q_{\rm grain}\,{\bf B}|$ in terms of the individual dust grain's charge ($q_{\rm grain}$) and mass ($m_{\rm grain}$), and the speed-of-light $c$ (the sign convention here is arbitrary but convenient, so $t_{L} > 0$).\footnote{In Eq.~\ref{eqn:general}, taking $q_{\rm grain}\rightarrow -q_{\rm grain}$ is mathematically identical to taking $\hat{\bf B}\rightarrow -\hat{\bf B}$. Since we will consider all possible signs/directions of ${\bf B}$, we can define $t_{L}$ in terms of $|q_{\rm grain}\,{\bf B}|$ to be positive definite, without loss of generality.}

Equation \eqref{eqn:general} has the spatially homogeneous, steady-state solution: 
\begin{align}
\label{eqn:mean.v.offset} \rho^{h} &= \langle \rho \rangle = \rho_{0}, \\ 
\nonumber \rho_{d}^{h} &= \langle \rho_{d} \rangle = \rho_{d,\,0} \equiv \mu\,\rho_{0}, \\ 
\nonumber t_{s}^{h} &= \langle t_{s} \rangle = t_{s}^{h}(\rho^{h},\,\driftvel,\,..), \\ 
\nonumber t_{L}^{h} &= \langle t_{L} \rangle = t_{L}^{h}(\rho^{h},\,\driftvel,\,..) \equiv \tau^{-1}\,\langle t_{s} \rangle, \\ 
\nonumber {\bf B}^{h} &= \langle {\bf B} \rangle = {\bf B}_{0}, \\
\nonumber {\bf u}^{h} &= \langle {\bf u} \rangle = {\bf u}_{0} + \left[{\bf g} + {\bf a}\,\left(\frac{\mu}{1+\mu}\right) \right]\,t,\\
\nonumber {\bf v}^{h} &= \langle {\bf v} \rangle = \langle {\bf u} \rangle + \driftvel,\\
\nonumber \driftvel &\equiv \frac{|{\bf a}|\,\langle t_{s} \rangle}{1+\mu}\,\left[ \frac{\hat{\bf a} - \tau\,(\hat{\bf a}\times\hat{\bf B}_{0}) + \tau^{2}\,( \hat{\bf a}\cdot \hat{\bf B}_{0} )\,\hat{\bf B}_{0}}{1+\tau^{2}} \right] .
\end{align}
In equilibrium, the system features the dust and gas both moving with constant acceleration $\tilde{\bf a} \equiv {\bf g} + {\bf a}\,\mu/(1+\mu)$, and a constant relative drift velocity $\driftvel$. The total mass-ratio between dust and gas is defined as $\mu\equiv \langle \rho_{d} \rangle / \langle \rho \rangle$, and $\langle t_{s} \rangle$, $\langle t_{L} \rangle$ are the values of $t_{s}$ and $t_{L}$ in the homogeneous solution (note these can, in principle, depend on $\driftvelmag\equiv|\driftvel|$, as discussed below, which makes Eq.~\ref{eqn:mean.v.offset} a non-linear equation for $\driftvel$). The parameter $\tau \equiv \langle t_{s} \rangle / \langle t_{L} \rangle$ is introduced here for convenience. 

Some examples of these equilibrium solutions, illustrating the geometry of the problem and dependence on $\tau$, are shown in Fig.~\ref{fig:geometry}. The scaling of $\driftvel$ with $\tau$ is intuitive: for small $\tau$, i.e., $t_{s} \ll t_{L}$, drag dominates, and $\driftvel \approx {\bf a}\,\langle t_{s} \rangle / (1+\mu)$, which is just the solution from \papertwo\ for the ``terminal velocity'' in a system with aerodynamic drag only (neglecting Lorentz forces). For large $\tau$, i.e., $t_{s} \gg t_{L}$, Lorentz forces suppress motion perpendicular to the field, giving $\driftvel \approx ({\bf a}\cdot \hat{\bf B}_{0})\,\hat{\bf B}_{0}\,\langle t_{s} \rangle / (1+\mu)$, i.e., the drift is given only by the projection of the differential acceleration ${\bf a}$ onto the field direction $\hat{\bf B}_{0}$. 
It is useful below to define a parameter $\driftvelX$ as the drift velocity in the {\em absence} of Lorentz forces ($\tau\rightarrow0$), giving: 
\begin{align}
\label{eqn:W0} \driftvelX &\equiv \frac{{\bf a}\,\langle t_{s} \rangle}{1+\mu} \ \ \ \ , \ \ \ \ \driftvel = \frac{\driftvelX - \tau\,(\driftvelX \times \hat{\bf B}_{0}) + \tau^{2}\,(\driftvelX\cdot\hat{\bf B}_{0})\,\hat{\bf B}_{0}}{1+\tau^{2}} \, .
\end{align}
Note $\driftvelXhat = \hat{\bf a}$, so when $\tau \ll 1$ we have $\driftvel \approx \driftvelX$ and $\driftvelhat \approx \hat{\bf a}$ (so e.g., $\cos{\theta}_{\bf Bw} \equiv \hat{\bf B}\cdot\hat\driftvel \approx \cos{\theta}_{\bf Ba} = \hat{\bf B}\cdot \hat{\bf a}$, etc.). Accounting for finite $\tau$, we have $\cos{\theta}_{\bf Bw} = \cos{\theta}_{\bf Ba}\,[(1+\tau^{2})/(1+\tau^{2}\,\cos^{2}\theta_{\bf Ba})]^{1/2}$ and $\sin{\theta}_{\bf Bw} = \sin{\theta}_{\bf Ba}\,[{1+\tau^{2}\,\cos^{2}\theta_{\bf Ba}}]^{-1/2}$, which for $\tau \gg 1$ become $\cos{\theta}_{\bf Bw} \rightarrow 1$, $\sin{\theta}_{\bf Bw} \rightarrow (1/\tau)\,\tan{\theta}_{\bf Ba}$. These relations will prove useful below.

\vspace{-0.5cm} 
\subsection{Linearized Perturbation Equations}
\label{sec:linearized.equations}

Now consider small perturbations $\delta$: $\rho = \rho^{h} + \delta\rho$, etc., and boost to a free-falling frame moving with the homogeneous gas solution $\langle {\bf u} \rangle$. Linearizing Eq.~\eqref{eqn:general}, we obtain,
\begin{align}
\label{eqn:linearized}  \partialAB{\delta\rho}{t} =& -\rho_{0}\,\nabla\cdot \delta{\bf u} \, ,\\
\nonumber \partialAB{\delta{\bf u}}{t} =& -c_{s}^{2}\,\frac{\nabla \delta \rho}{\rho_{0}} - \frac{{\bf B}_{0} \times (\nabla \times \delta {\bf B})}{4\pi\,\rho_{0}} \\ 
\nonumber &+ \mu\,\left[ 
\frac{(\delta {\bf v}-\delta{\bf u})}{\langle t_{s} \rangle} 
+ \frac{(\delta {\bf v}-\delta{\bf u})\times \hat{\bf B}_{0}}{\langle t_{L} \rangle} + \frac{\driftvel\times \delta {\bf B}}{|{\bf B}_{0}|\,\langle t_{L} \rangle}
\right]\\
\nonumber &- \mu\,\frac{\driftvel}{\langle t_{s} \rangle}\,\left( \frac{\delta t_{s}}{\langle t_{s} \rangle} + \frac{\delta \rho}{\rho_{0}} - \frac{\delta \rho_{d}}{\mu\,\rho_{0}} \right) \\ 
\nonumber &- \mu\,\frac{\driftvel\times\hat{\bf B}_{0}}{\langle t_{L} \rangle}\,\left( \frac{\delta t_{L}[{\bf B}_{0}]}{\langle t_{L} \rangle} + \frac{\delta \rho}{\rho_{0}} - \frac{\delta \rho_{d}}{\mu\,\rho_{0}} \right), \\ 
\nonumber \partialAB{\delta{\bf B}}{t} =&\, \nabla \times(\delta{\bf u} \times {\bf B}_{0})  = ({\bf B}_{0}\cdot \nabla)\,\delta{\bf u} - {\bf B}_{0}\,(\nabla \cdot \delta{\bf u}) \, , \\ 
\nonumber\partialAB{\delta \rho_{d}}{t} =& -(\driftvel\cdot\nabla)\,\delta\rho_{d} -\mu\,\rho_{0}\,\nabla\cdot \delta{\bf v} \, ,\\
\nonumber \partialAB{\delta {\bf v}}{t} =& - (\driftvel\cdot\nabla)\,\delta{\bf v} -\frac{(\delta {\bf v}-\delta{\bf u})}{\langle t_{s} \rangle} + \frac{\driftvel\,\delta t_{s}}{\langle t_{s} \rangle^{2}} \\ 
\nonumber & -\frac{(\delta {\bf v}-\delta {\bf u})\times\hat{\bf B}_{0}}{\langle t_{L} \rangle} -\frac{\driftvel\times \delta {\bf B}}{|{\bf B}_{0}|\,\langle t_{L} \rangle} + \frac{(\driftvel \times \hat{\bf B}_{0})\,\delta t_{L}[{\bf B}_{0}]}{\langle t_{L} \rangle^{2}} \, ,
\end{align}
(plus the constraint $\nabla \cdot \delta{\bf B}=0$), 
where we define the usual sound speed $c_{s}^{2} \equiv \partial P/\partial \rho$ and  $\delta t_{s}$  as the linearized perturbation to $t_{s}$; i.e., $t_{s} \equiv \langle t_{s} \rangle + \delta t_{s}(\delta \rho,\,\delta {\bf v},\, ...) + \mathcal{O}(\delta^{2})$. All variables now refer to their values {\em in the free-falling frame}. The term $\delta t_{L}[{\bf B}_{0}]$ is defined for convenience here as the linearized perturbation to $t_{L}$ {\em at fixed magnetic field} ${\bf B}={\bf B}_{0}$, so it applies just to the scalar normalization of the Lorentz force (essentially, $\delta t_{L}[{\bf B}_{0}]$ captures any linear variation in the grain charge). The explicit dependence of the Lorentz force on ${\bf B}$ is  written separately, as the $\driftvel\times \delta{\bf B}$ terms.

Note that, as shown in detail in \paperone\ (\S~2 \&\ App.~B therein), the transformation between accelerating frames (moving with the homogeneous gas solution) and stationary frames has {\em no} effect on the solutions here. In particular, transforming our Fourier solutions back to the stationary frame is  mathematically equivalent to taking $\omega \rightarrow \omega + {\bf u}_{0}\cdot{\bf k} + \tilde{\bf a}\cdot{\bf k}\,t/2$ (where $\tilde{\bf a} = {\bf g} + {\bf a}\,\mu/(1+\mu)$ is the homogeneous acceleration).

\vspace{-0.05cm}
\subsection{Scalings of the Stopping Time \&\ Gyro Time}
\label{sec:stopping.time.scalings}

To define $\delta t_{s}$ we require a physical drag law. However, as shown in \papertwo, essentially all physical drag laws $t_{s} = t_{s}(\rho,\,T,\,c_{s},\,P,\,{\bf v}-{\bf u},\,...)$ can be written, assuming a barytropic equation of state under local perturbations, in the form: 
\begin{align}
\label{eqn:ts.general} \frac{\delta {t_{s}}}{\langle t_{s} \rangle} &= -\coeffTSrho\,\frac{\delta{\rho}}{\rho_{0}} - \coeffTSv\,\frac{{\driftvel}\cdot\left(\delta{\bf {v}} - \delta{\bf {u}} \right) }{|\driftvel|^{2}}	\, .
\end{align}
Likewise we expect variations of the gyro timescale (at fixed ${\bf B}$) to have the form:
\begin{align}
	\frac{\delta t_{L}[{\bf B}_{0}]}{\langle t_{L} \rangle} &= -\coeffTLrho\,\frac{\delta\rho}{\rho_{0}} \, .
\end{align}
To justify this and explore scalings in different physical regimes, we briefly describe different physical scaling laws below. 

\vspace{-0.1cm}
\subsubsection{Epstein Drag}
\label{sec:stopping.time.scalings:epstein}

For Epstein drag, valid for grain sizes smaller than the gas mean-free-path (for both super-sonic and sub-sonic drift velocities), the stopping time is $t_{s} = (\pi\gamma/8)^{1/2}\,(\bar{\rho}_{d}\,R_{d}/\rho\,c_{s})\,(1+a_{\gamma}\,|{\bf v}-{\bf u}|^{2}/c_{s}^{2})^{-1/2}$ \citep{draine.salpeter:ism.dust.dynamics}. Here $\bar{\rho}_{d}$ is the internal material density of the particle (grain), $R_{d}$ is the grain radius, $a_{\gamma} \equiv 9\pi\,\gamma/128$, and $\gamma$ is the equation-of-state parameter, needed to relate  $c_{s}$ to the  isothermal sound speed  $c_{\rm iso}$ (or temperature $T$):
\begin{align}
\gamma &\equiv \frac{c_{s}^{2}}{c_{\rm iso}^{2}} = \frac{\rho}{P}\,\partialAB{P}{\rho}.
\end{align}
Evaluating Eq.~\eqref{eqn:ts.general}, one finds
\begin{align}
\coeffTSrho^{\rm Epstein} &= \frac{\gamma+1+2\,\tilde{a}_{E}}{2\,(1+\tilde{a}_{E})}\ , \ \ \ \ \ 
 \coeffTSv^{\rm Epstein} = \frac{\tilde{a}_{E}}{1+\tilde{a}_{E}},
\end{align}
where $\tilde{a}_{E} \equiv a_{\gamma}\,(\driftvelmag/c_{s})^{2}$

As discussed in  \papertwo, because $\langle t_{s} \rangle$ (and therefore also $\tau$)  depends explicitly on $\langle {\bf v}-{\bf u} \rangle = \driftvel$, Eq.~\eqref{eqn:mean.v.offset} for the drift velocity (or Eq.~\ref{eqn:W0} for $\driftvelX$) must be solved implicitly to determine the equilibrium $\driftvel$, if one is given some ${\bf a}$ (and $\rho$, $R_{d}$, etc). If we define $t_{0} \equiv (\pi\,\gamma/8)^{1/2}\,\bar{\rho}_{d}\,R_{d}/(\rho_{0}\,c_{s})$ as the stopping time neglecting the ${\bf v}-{\bf u}$ term, then if the drift is sub-sonic ---  that is, if $|\driftvelX(t_{s}=t_{0})| = |{\bf a}|\,t_{0}/(1+\mu) \ll c_{s}$ --- then $\driftvel\approx \driftvel(t_{s}= t_{0})$; if the drift is highly super-sonic and either $\tau\ll 1$ or $\tau \gg1$, then one obtains $\driftvelmag \propto [c_{s}\,\driftvelmag(t_{s}= t_{0})]^{1/2}$.



\vspace{-0.1cm}
\subsubsection{Stokes Drag}
\label{sec:stopping.time.scalings:stokes}

The expression for drag in the Stokes limit --- which is valid for an intermediate range of grain sizes, when $R_{d} \gtrsim (9/4)\,\lambda_{\rm mfp}$ but  $\mathrm{Re}_{\mathrm{grain}}\equiv R_{d}|\driftvel|/(\lambda_{\mathrm{mfp}} c_{s})\lesssim 1$
--- is given by multiplying the Epstein expression above by $(4\,R_{d}) / (9\,\lambda_{\rm mfp})$. Here  $\lambda_{\rm mfp} \propto 1/(\rho\sigma_{\mathrm{gas}})$ is the gas mean-free-path, $\sigma_{\mathrm{gas}}$ is the gas collision cross section, and $\mathrm{Re}_{\mathrm{grain}}$ is the Reynolds number of the streaming grain. This gives
\begin{align}
\coeffTSrho^{\rm Stokes} &= \coeffTSrho^{\rm Epstein} -1	\ , \ \ \ \ \ \coeffTSv^{\rm Stokes} = \coeffTSv^{\rm Epstein}\ .
\end{align}
As discussed in \papertwo, this assumes $\sigma_{\mathrm{gas}}$ is not strongly dependent on density or temperature. Of course under some circumstances, given a specific physical model/system, it should be, in which case one can easily calculate the appropriate revised $\coeffTSrho$. Note the dependence of $t_{s}$ on ${\bf v}-{\bf u}$ is the same as Epstein, so Eq.~\eqref{eqn:mean.v.offset} must be solved non-linearly for a given ${\bf a}$.

\vspace{-0.1cm}
\subsubsection{Coulomb Drag}
\label{sec:stopping.time.scalings:coulomb}

The stopping time for Coulomb drag can be approximated as \citep{draine.salpeter:ism.dust.dynamics}, 
\begin{equation}
t_{s} = \left(\frac{\pi\gamma}{2}\right)^{1/2}\,\frac{\bar{\rho}_{d}\,R_{d}}{f_{\rm ion}\,\rho\,c_{s}\,\ln{\Lambda}}\,\left(\frac{k_{B}\,T}{z_{i}\,e\,U}\right)^{2}\,\left(1+a_{C}\,\frac{|{\bf v}-{\bf u}|^{3}}{c_{s}^{3}}\right),
\end{equation}
 where $\ln{\Lambda}$ is a Coulomb logarithm with $\Lambda \equiv [(3\,k_{B}\,T)/(2\,R_{d}\,z_{i}\,e^{2}\,U)]\,[(\mu_{i}\,m_{p}\,k_{B}\,T)/(\pi\,\,f_{\rm ion}\,\rho)]^{1/2}$,
$a_{C}\equiv \sqrt{2\gamma^{3}/9\pi}$, $e$ is the electron charge, $f_{\rm ion}$ the ionized fraction (in the gas), $z_{i}$ is the mean gas ion charge, $\mu_{i}$ is the mean molecular weight of ions, $T\propto \rho^{\gamma-1}$ is the gas temperature, and $U$ is the electrostatic potential of the grains, $U\sim Z_{\rm grain}\,e/R_{d}$ (where $Z_{\rm grain}$ is the grain charge). The behavior of $U$ is complicated and depends on a wide variety of environmental factors: in the different regimes considered in \citet{draine.salpeter:ism.dust.dynamics} they find regimes where $U\sim$\,constant and others where $U \propto Z_{\rm grain} \propto T$, we therefore parameterize the dependence by $U\propto T^{\coeffTSqCoulomb}$. This gives:
\begin{align}
\nonumber \coeffTSrho^{\rm Coulomb} &=  1 +2\,(\gamma-1)\,\coeffTSqCoulomb
- \frac{3\,(\gamma-1)}{2\,(1+\tilde{a}_{C})}
- \frac{1-(3-2\,\coeffTSqCoulomb)\,(\gamma-1)}{2\,\ln{\Lambda}}, \\ 
 \coeffTSv^{\rm Coulomb} &= -\frac{3\,\tilde{a}_{C}}{1+\tilde{a}_{C}} 
\end{align}
where $\tilde{a}_{C} \equiv \coulombcoeff\,(\driftvelmag/c_{s})^{3}$.
For relevant astrophysical conditions, $\ln{\Lambda} \sim 15-20$, so the $\ln{\Lambda}$ term in $\coeffTSrho$ is unimportant.
 As discussed in \papertwo,  when $\driftvelmag\gg c_{s}$, Coulomb drag   produces a new instability (in the absence of other drag forces), because of the large inverse dependence of $t_{s}$ on ${\bf v}-{\bf u}$ ($\coeffTSv < -1$, here); however for precisely this reason aerodynamic drag (Epstein or Stokes drag) will always dominate over Coulomb drag in this limit (the aerodynamic term becomes stronger, while the Coulomb term becomes weaker, as $\driftvelmag$ increases).

\vspace{-0.1cm}
\subsubsection{Lorentz Forces \&\ Gyro Timescales}
\label{sec:stopping.time.scalings:lorentz}

The Larmor/gyro time,  $t_{L}[{\bf B}_{0}]$, is
\begin{align}
t_{L} &\equiv -\frac{m_{\rm grain}\,c}{|q_{\rm grain}\,{\bf B}|} = \frac{4\pi\,\bar{\rho}_{d}\,R_{d}^{3}\,c}{3\,e\,|Z_{\rm grain}\,{\bf B}|} \ .
\end{align}
As noted above the grain charge distribution is complicated and determined by a range of physical processes, environmental effects, and microphysical grain properties \citep[see][]{weingartner:2001.pah.model,weingartner:2001.grain.charging.photoelectric}. Our analysis allows for an arbitrary dependence of $Z_{\rm grain}$ on these properties, but to gain some intuition we note that the dominant processes are usually ``collisional'' and/or photo-electric charging \citep{tielens:2005.book}. 
In the collision-dominated regime (grains ``sweeping up'' electrons), \citet{draine:1987.grain.charging} quote convenient results for the equilibrium charge as a function of temperature: $ \langle Z_{\rm grain} \rangle \approx -1/(1+0.037\,\tilde{T}^{-1/2}) - a_{L}\,\tilde{T} \approx -1 - a_{L}\,\tilde{T}$ where $\tilde{T} \equiv R_{d}\,k_{B}\,T/e^{2}$ and $1+a_{L} = (\mu/m_{e})^{1/2}\,\exp{(-a_{L})}$, with a minimum charge limited by field emission $Z_{\rm min} \sim -7000\,(R_{d}/0.1\,\mu{\rm m})^{2}$. Here $\mu$ is the mean molecular weight; for typical values in ionized (WIM), atomic (WNM/CNM), or molecular media, relevant values are $(\mu,\,a_{L})\approx (0.59,\,2.3)$, $(1.25,\,2.6)$, and $(2.3,\,2.8)$, respectively. 
In the photo-electric (electron-ejection dominated) regime, $\langle Z_{\rm grain} \rangle \approx -1 + \tilde{T}\,(f_{L}-1)$ where $f_{L} \equiv (F_{\rm pe}/2.6\times 10^{-4}\,{\rm erg\,cm^{-2}\,s^{-1}})\,(n_{e}/{\rm cm^{-3}})^{-1}\,(T/{\rm K})^{-1/2}$ and $F_{\rm pe}$ is the incident UV flux in the relevant (far-UV) wavelength range \citep{tielens:2005.book}, with maximum charge above which electrostatic forces prevent ejection, $Z_{\rm max} \sim 500\,(R_{d}/0.1\,\mu{\rm m})$.

For large grains ($|Z_{\rm grain}|\gg1$), between minimum/maximum charge, one can approximately interpolate between the regimes by taking $\langle Z_{\rm grain} \rangle \sim -1 + \tilde{T}\,(f_{L}-a_{L})$. If the grain-charging timescale $t_{\rm charge}$ is also short compared to the stopping time and other dynamical timescales in the system (not always guaranteed; see \S~\ref{sec:scales.validity}), then this  gives
\begin{align}
	\coeffTLrho &\approx \frac{\tilde{T}_{0}\,[a_{L}\,(\gamma-1) + f_{L}\,(3-\gamma)/2]}{1 + (a_{L}-f_{L})\,\tilde{T}_{0}} &  \\
\nonumber
&\sim 
\begin{cases}
(\gamma-1) & \hfill (f_{L} \ll 1; {\rm collision-dominated})  \\
(\gamma-3)/2 & \hfill (f_{L} \gg 1; {\rm photoelectric-dominated}) 
\end{cases}
\end{align}
using $T\propto \rho^{\gamma-1}$. If instead $t_{\rm charge}$ is long ($t_{\rm charge}\gg t_{s}$), or the grains are at their extremal charge, then $\coeffTLrho \approx 0$. 

Note that the grain charge can, in principle, also depend on velocity, which would add a term $-\zeta_{q,\,w}\,\driftvel\cdot(\delta{\bf v}-\delta{\bf u})/|\driftvel|^{2}$ to $\delta t_{L}[{\bf B}_{0}]/\langle t_{L} \rangle$ (analogous to the velocity dependence in $\zeta_{w}$). We could include this, and it does not significantly alter any of our conclusions. 
However, calculating the velocity-dependence of the charge  following  \citet{shull:grain.charge.vs.velocity} (for the collision-dominated case; see their Appendix), one obtains  $\zeta_{q,\,w}\approx -0.03\,\coeffTLrho$ for $|\driftvel|\lesssim 2\,c_{s}$, and $0 < \zeta_{q,\,w}< 0.04\,\coeffTLrho$ for all physically-relevant $|\driftvel| \gtrsim 2\,c_{s}$. 
An even weaker (or vanishing) dependence on ${\bf v}-{\bf u}$ is expected when photo-electric charging dominates and/or when grains have saturated charge. 
This is therefore a quite small correction in almost all cases, and we neglect it for simplicity.

%
%
\begin{figure*}
\begin{center}
\includegraphics[width=1.032\columnwidth]{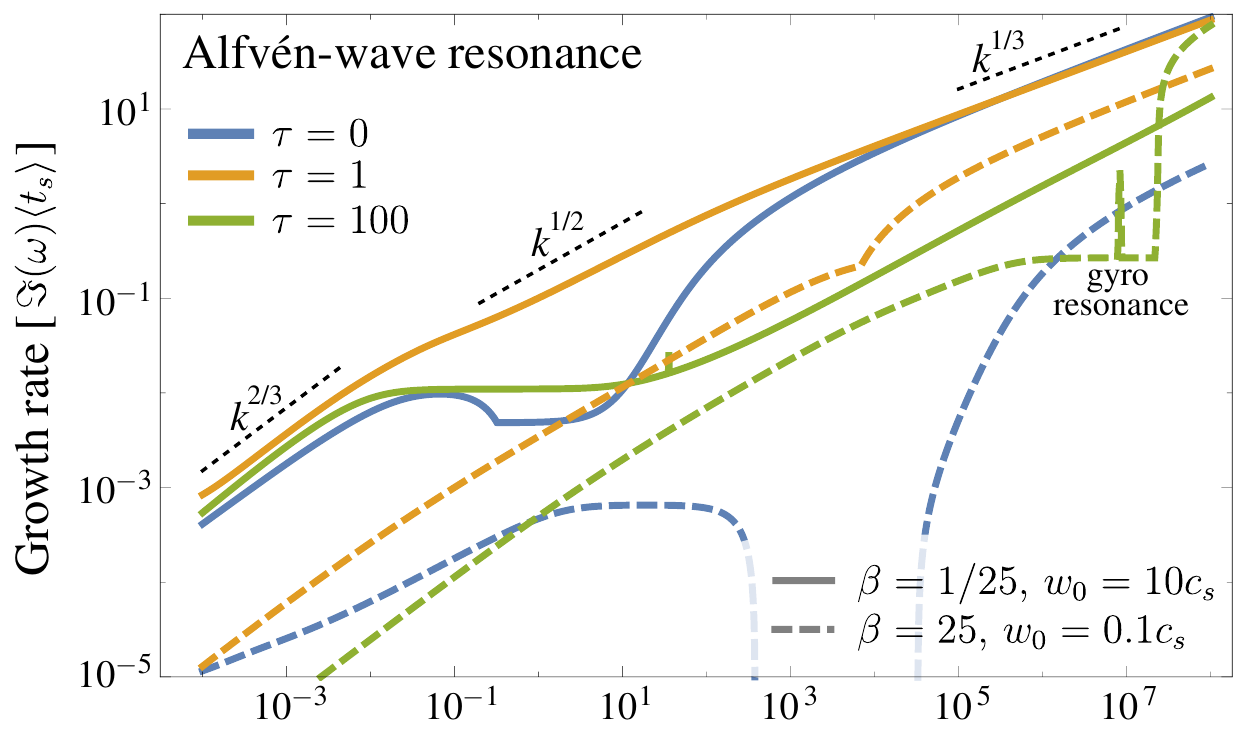}\includegraphics[width=0.97\columnwidth]{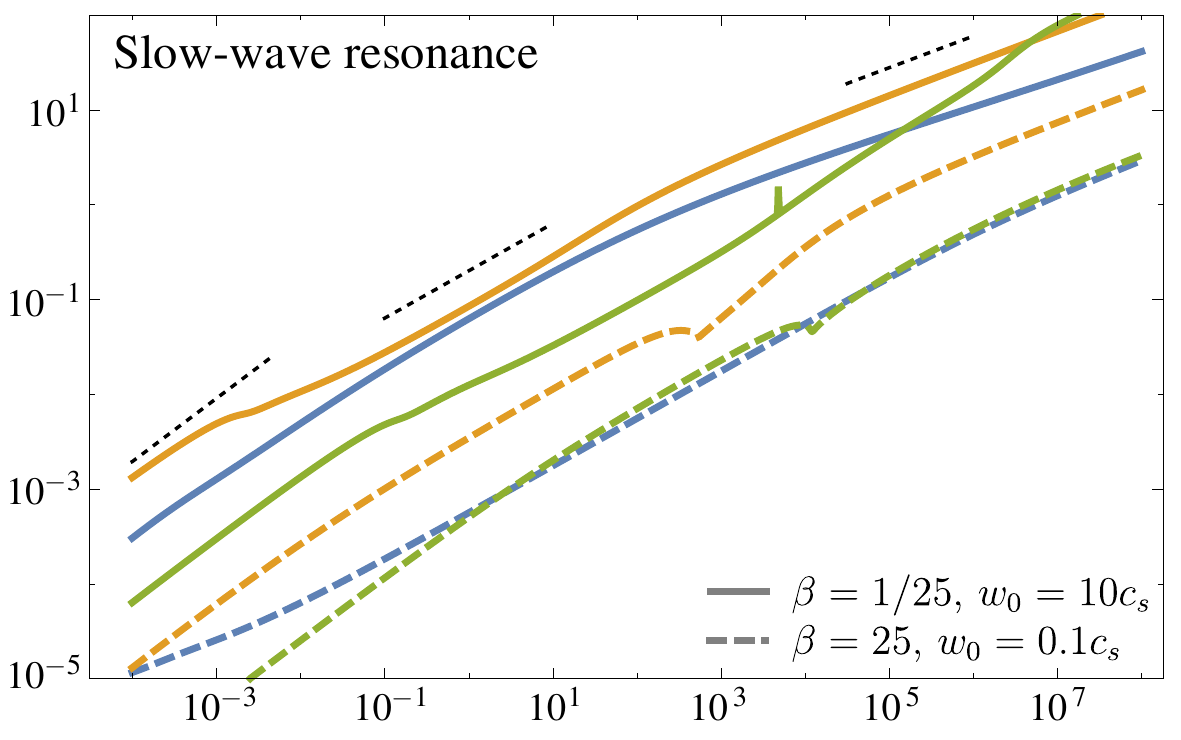}\\~\includegraphics[width=1.03\columnwidth]{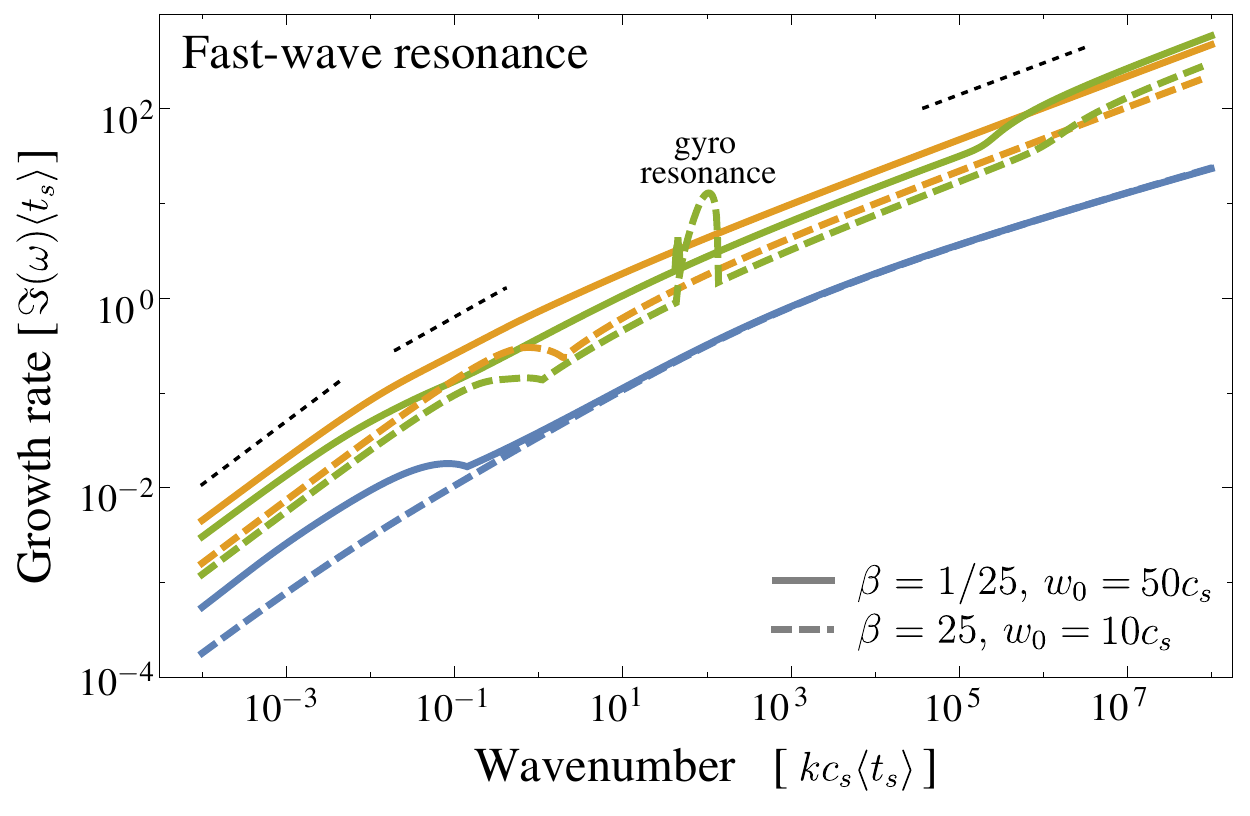} \includegraphics[width=0.97\columnwidth]{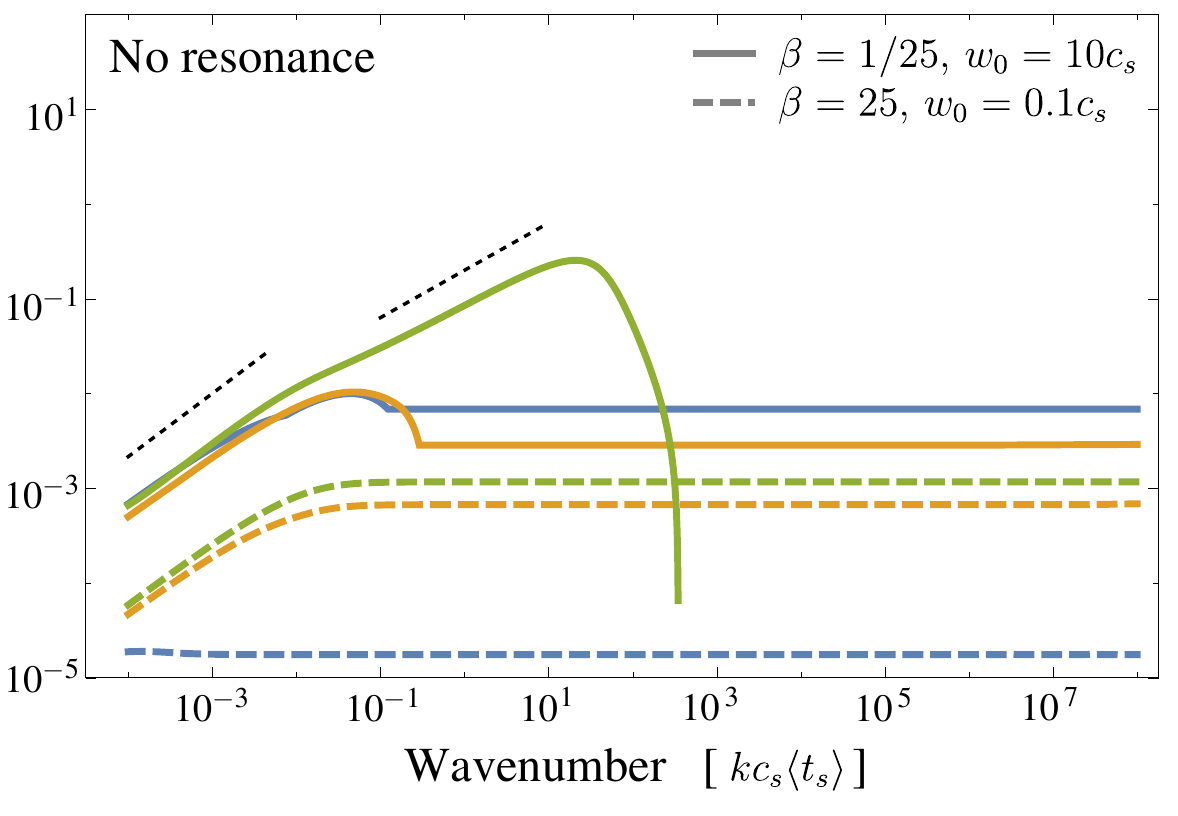}\vspace{-0.25cm}
\caption{The dispersion relation (growth rate $\Im{(\omega)}$, in units of $t_{s}$, as a function of wavenumber $k$ in units of $c_{s}\,t_{s}$) of the MHD RDIs for a variety of sample parameters. 
In the first three panels we choose the mode angle$^{\ref{foot:k.choice}}$ $\hat{{\bf k}}$ such that the mode is in resonance with the Alfv\'en wave ({\em top-left}), the slow magnetosonic wave ({\em top-right}), or the fast magnetosonic wave ({\em bottom-left}), while at {\em bottom-right} we choose $\hat{{\bf k}}$ to lie well away from any resonance, for comparison. 
In each case we plot the dispersion relation at  $\beta=1/25$ (strong magnetic fields; solid), and $\beta=25$ (weak magnetic fields; dashed), and show $\tau =0$ (unmagnetized grains; blue),  $\tau =1$ (intermediate grains; orange), and $\tau =100$ (strongly magnetized grains; green). 
In all but the fast-wave case, we show both  subsonically drifting grains  ($|\driftvelX|=0.1 c_{s}$), and supersonically  drifting grains  ($|\driftvelX|=10 c_{s}$), with $\theta_{\bf Ba}=-60^{\circ}$.
In the fast-wave case, these parameters eliminate the resonance entirely, so we set $|\driftvelX|=10 c_{s}$ and $|\driftvelX|=50 c_{s}$, with $\theta_{\bf Ba}=-80^{\circ}$. In all cases we assume Epstein drag in a gas with $\gamma=5/3$, and set the dust-to-gas ratio to $\mu=0.01$. 
We also show line segments (black dotted) illustrating the typical scalings expected for the long-wavelength (low-$k$, \S~\ref{sub: low k regime solutions}; $\Im{(\omega)}\propto k^{2/3}$), mid-wavelength (\S~\ref{sec:mhdwave.rdi:growthrates.midk}; $k^{1/2}$) and short-wavelength (high-$k$, \S~\ref{sec:mhdwave.rdi:growthrates.hik}; $k^{1/3}$) limits of the MHD-wave RDIs. 
Some gyro-resonant modes (\S~\ref{sec:gyro}) are also visible as ``spike'' in the dispersion relation when $\tau \gg 1$ (e.g., in the fast-wave dispersion relation at $k c_{s} \langle t_{s}\rangle \sim \tau \sim 100$).
\label{fig:dispersion.relation}\vspace{-0.25cm}
}
\end{center}
\end{figure*}
%
%

\vspace{-0.5cm}
\subsection{Dispersion Relation}
\label{sec:dispersion.relation}

Inserting the Fourier ansatz into Eq.~\ref{eqn:linearized}, it takes the form $\omega\,{\bf X} = \mathbb{T}\,{\bf X}$ where ${\bf X} = (\delta \rho,\,\delta\rho_{d},\, ...)$; the full expression  is given in App.~\ref{sec:appendix:dispersion.relation}. The solutions $\omega$ are given by the eigenvalues of $\mathbb{T}$, which is a $10\times 10$ matrix (the divergence constraint eliminates one degree of freedom of the perturbed variables). The general dispersion relation (the characteristic polynomial of $\mathbb{T}$) is therefore a 10th-order polynomial, where $\omega$ is a function of 14 independent variables (${\bf k}$, $\hat{\bf B}_{0}\cdot \driftvelhat$, $\coeffTSrho$, $\coeffTSv$, $\coeffTLrho$, $\driftvelmag$, $\tau$, $\beta$, $\mu$, $\rho_{0}$, $c_{s}$, $\langle t_{s} \rangle$). 

Although it is straightforward to compute this polynomial from $\mathbb{T}$ it is not in any way intuitive and must be solved numerically. We will therefore focus on exact numerical solutions and simple analytic scalings relevant in certain limits. At any given ${\bf k}$ (and fixed fluid+dust properties) there are $10$ independent modes (solution branches) for $\omega$. As discussed below, typically $\sim 3-7$ of these are unstable ($\Im{(\omega)}>0$). We focus primarily below on the fastest-growing modes (at a given $k$), since these will dominate the dynamics. 

Figs.~\ref{fig:dispersion.relation}-\ref{fig:dispersion.relation.high.tau} plot exact numerical solutions for the growth rate of various unstable modes at a given $k$, as a function of $k$, for fixed values of ($\hat{\bf B}_{0}\cdot \driftvelhat$, $\driftvelmag$, $\tau$, $\beta$, $\mu$), and ($\coeffTSrho$, $\coeffTSv$, $\coeffTLrho$) determined according to the cases in \S~\ref{sec:stopping.time.scalings}.\footnote{\label{foot:k.choice}In Fig.~\ref{fig:dispersion.relation}, for each of the resonant-mode plots, the  choice of $\hat{{\bf k}}$ is carried out as follows: first, we find the region of mode angles $\theta_{\mathrm{lower}}<\theta_{{\bf Bk}}<\theta_{\mathrm{upper}}$ where the chosen resonance is possible (considering only $0^{\circ}<\theta_{\bf Bk}<90^{\circ}$), and set $\theta_{{\bf Bk}} = (\theta_{\mathrm{lower}}+\theta_{\mathrm{upper}})/2$; then, we set the remaining component of $\hat{{\bf k}}$ to 
satisfy the resonance condition, Eq.~\ref{eqn:resonant.condition}. For example, the $\tau=1$, $\beta = 25$, $\driftvelX=0.1c_{s}$ slow-wave resonance is possible for $67^{\circ}\lesssim \theta_{{\bf Bk}}$; we set $\theta_{\bf Bk}=(67+90)/2=74.8^{\circ}$, and solve Eq.~\ref{eqn:resonant.condition} to find that $\theta_{(\driftvel\times {\bf B}){\bf k}}=-28.6^{\circ}$ is required for the slow-wave resonance. This is not, generally, the fastest-growing angle among those that satisfy the resonant condition, but is chosen to be ``typical'' (although the growth rate varies weakly within the range of angles that do satisfy the resonant condition, as we show below). In the ``No resonance'' case, we arbitrarily choose the mode to propagate at the angle $\theta_{\bf B k}=45^{\circ}$ and $\theta_{\bf (w \times B)k} = \cos^{-1}[{\bf k}\cdot (\driftvel\times {\bf B})]=-45^{\circ}$ (see \S~\ref{sub:k.hat.angle.parameterization}).} Figs.~\ref{fig:angular.structure}-\ref{fig:angle.1d} plot the growth rates of the fastest-growing mode  at a given $k$ and  similarly fixed equilibrium properties, as a function of the orientation of $\hat{\bf k}$. We see a very rich mode structure. All of the important features seen here can be understood via appropriate analytic, asymptotic expansions, which we systematically explore in the next several sections.

\subsubsection{Parameterization of $\hat{{\bf k}}$}\label{sub:k.hat.angle.parameterization}
It is worth briefly commenting on our parameterization of the mode direction $\hat{\bf k}$. While for analytic results (see \S~\ref{sec:resonant.mode.angles}) it is most convenient to use a Cartesian coordinate 
system for ${\bf k}$, a polar system is more convenient for plotting, because $k=|{\bf k}|$ is naturally kept fixed. Thus in Figs.~\ref{fig:dispersion.relation}--\ref{fig:angle.1d}, we parameterize $\hat{{\bf k}}$ through its 
angle from $\hat{\bf B}_{0}$,
\begin{equation}
\theta_{\bf Bk} = \cos^{-1}(\hat{\bf k}\cdot \hat{\bf B}_{0}),
\end{equation}
and the angle subtended from $\hat{\bf y}= \widehat{\driftvel \times {\bf B}_{0}}$,
\begin{equation}
\theta_{\bf (w \times B)k} = \cos^{-1}(\hat{\bf k}\cdot \widehat{\driftvel \times {\bf B}_{0}}),
\end{equation}
which is simply the standard azimuthal angle in spherical polar coordinates about $\hat{\bf B}_{0}$ (shifted by $90^{\circ}$).
While this parameterization is arbitrary, it has the advantage of making the resonant lines more obvious (e.g., in 
Fig.~\ref{fig:angular.structure}).

\vspace{-0.5cm}
\section{Parallel/Aligned Modes: The Pressure-Free, Cosmic-Ray Streaming, and Acoustic (Quasi-Drift \&\ Quasi-Sound) Modes}
\label{sec:parallel.modes}

We first consider the ``parallel'' modes from \S~\ref{sec:overview.modes}, where the behavior of greatest interest (e.g., fastest growth rates) occurs when $\hat{\bf k} \approx \driftvel$, as compared to the more complicated angle-dependent resonances we will discuss in subsequent sections.

\vspace{-0.5cm}
\subsection{The Long-Wavelength (low-$k$) or ``Pressure Free'' Mode}
\label{sub: low k regime solutions}

At sufficiently low-$k$, the structure of the fastest-growing unstable mode is actually rather simple (and instructive). For the acoustic RDI (neutral gas and neutral grains; \papertwo), we showed that at low-$k$, the fastest-growing mode satisfies $\omega = \omega_{\rm PF}\sim \mathcal{O}(k^{2/3})$; this is true here as well. Expanding the dispersion relation in powers of $k/(c_{s}\,\langle t_{s} \rangle) \ll \hat{\mu}$, we obtain:
\begin{align}
\label{eqn:omega.low.k} \omega_{\rm PF}^{3} =&\, \iimag\,\mathcal{F}_{\rm PF}\,\frac{\hat{\mu}\,\driftvelmag^{2}\,k^{2}}{\langle t_{s} \rangle} + \mathcal{O}(k^{4}) \, .
\end{align}
where $\mathcal{F}_{\rm PF}$ is a real scalar\footnote{The full expression for $\mathcal{F}_{\rm PF}$ in Eq.~\ref{eqn:omega.low.k} is:
\begin{align}
\nonumber \mathcal{F}_{\rm PF} =&\, \left[1 + (c_{0}^{2} + \tildeCoeffTSv^{-1}\,s_{0}^{2})\,\tau^{2} \right]^{-1}\,{\Bigl[}
\hat{k}_{z}^{2}\,(1-\coeffTSrhoV) - s_{0}\,\hat{k}_{y}\,\hat{k}_{z}\,\coeffTLrho\,\tau \\
\nonumber &- f_{2}\,\tau^{2} - f_{3}\,\tau^{3} - \tildeCoeffTSv^{-1}\,s_{0}^{2}\,(\hat{\bf k}\cdot \hat{\bf B}_{0})^{2}\,\tau^{4}  
{\Bigr]} \, , \\ 
\nonumber f_{2} =&\, s_{0}^{2}\,\left[1+ \hat{k}_{z}^{2}\,(\coeffTLrhoV-1) + \hat{k}_{y}^{2}\,(\coeffTLrho-\coeffTSrhoV) \right] - c_{0}^{2}\,\hat{k}_{z}^{2}\,(1-\coeffTSrhoV) \\
\nonumber &- s_{0}\,c_{0}\,\hat{k}_{x}\,\hat{k}_{z}\,(\coeffTLrho-\coeffTSrhoV) \, , \\ 
 f_{3} =&\, s_{0}\,\hat{k}_{y}\,\left[\hat{k}_{z}\,(s_{0}^{2}\,\coeffTLrhoV + c_{0}^{2}\,\coeffTSrhoV) + s_{0}\,c_{0}\,\hat{k}_{x}\,(\coeffTSrhoV-\coeffTLrho) \right] \, .
\end{align}
} that depends on the drift law, relative strength of Lorentz forces ($\tau$), and angles between  $\hat{\bf k}$, $\driftvelhat$, and $\hat{\bf B}_{0}$. 
The expression for $\mathcal{F}_{\rm PF}$ is complicated, but for $\coeffTSrho=\coeffTSv=\coeffTLrho=0$ (constant $t_{s}$ and $t_{L}$) it simplifies to 
\begin{align}
\label{eqn:flow.constantT}	\mathcal{F}_{\rm PF}(\coeffTSrho,\coeffTSv,\coeffTLrho=0) &\rightarrow (\hat{\bf k}\cdot\driftvelhat)^{2} - \frac{[1+(\tau\,\hat{\bf k}\cdot \hat{\bf B}_{0})^{2}]\,\sin^{2}{\theta_{\bf Bw}}}{1+\tau^{-2}} \, .
\end{align}

The dispersion relation has the form $\omega_{\rm PF}^{3} = X$; this {\em always} has an unstable ($\Im(\omega_{\rm PF}) > 0$) root, for any complex $X$ with $\| X \| \ne 0$. If we write $X = \| X \| \, \exp{(\iimag\,\phi_{X})}$ then we can define $\omega_{\rm PF} = \ispecial\,\| X \|^{1/3}$ where $\ispecial$ is the root of $\ispecial^{3} = \exp{(\iimag\,\phi_{X})}$ (so $\| \ispecial \| = 1$) with the largest imaginary part (such that $\Im(\omega_{\rm PF})>0$). For the example above, since $\mathcal{F}_{\rm PF}$ is purely real, we see that $\ispecial = \iimag$ if $\mathcal{F}_{\rm PF} < 0$ and $\ispecial = (\iimag \pm \sqrt{3})/2$ if $\mathcal{F}_{\rm PF} > 0$. 

Note that in Eq.~\ref{eqn:flow.constantT}, if we naively took $\tau\rightarrow\infty$, $\mathcal{F}_{\rm PF}$ would appear to diverge as $\tau^{2}$. However, recall from \S~\ref{sec:equilibrium} that as $\tau\rightarrow \infty$, Lorentz forces project the drift direction ($\driftvelhat$) onto the field direction ($\hat{\bf B}_{0}$), so $\sin{\theta}_{\bf Bw} \rightarrow (1/\tau)\,\tan{\theta}_{\bf Ba}$. Recall that $\theta_{\bf Ba}$  here is the angle between the field ($\hat{\bf B}_{0}$) and whatever acceleration ${\bf a}$ is  sourcing the drift (or equivalently, the direction $\driftvelXhat = \hat{\bf a}$ which the drift would have {\em without} Lorentz forces). So for a fixed external acceleration ${\bf a}$ (or $\driftvelX$), $\mathcal{F}_{\rm PF}$ remains  finite.

Thus, considering the low and high-$\tau$ limits and writing expressions  in terms of $\driftvelX$ (rather than $\driftvel$), we can express $\omega_{\rm PF}$ as,
\begin{align}
\omega_{\rm PF} &= \ispecial\, \left| \frac{f_{1}}{\tildeCoeffTSv}\,\frac{\hat{\mu}\,\driftvelXmag^{2}\,(\driftvelhat\cdot{\bf k})^{2}}{\langle t_{s} \rangle}
\right|^{1/3} \, , \\ 
\nonumber f_{1} &\equiv 
\begin{cases}
(\tildeCoeffTSv - \coeffTSrho) & \hfill (\tau \ll 1)  \\
(\tildeCoeffTSv - \coeffTSrho)\,\cos^{2}\theta_{\bf Ba} - \coeffTSrho\,\frac{\hat{k}_{y}}{\hat{k}_{z}}\,\cos\theta_{\bf Ba}\,\sin\theta_{\bf Ba} - \sin^{2}\theta_{\bf Ba} & \hfill (\tau \gg 1)
\end{cases}
\end{align}
Here $f_{1}$ has the same sign as $\mathcal{F}_{\rm PF}$ so $\ispecial = \iimag$ if $f_{1}<0$ and $\ispecial=(\iimag\pm\sqrt{3})/2$ if $f_{1}>0$.

We immediately see that the growth rates scale as $\Im(\omega)\sim (\hat{\mu}\,\driftvelmag^{2}\,k^{2}/\langle t_{s}\rangle)^{1/3}$, modulo order-unity geometric corrections, and the fastest-growing modes have ${\bf k}$ aligned with the drift $\driftvelhat$. We can see in Fig.~\ref{fig:dispersion.relation} that this provides an excellent approximation to the scalings of exact solutions for $k\,c_{s}\,\langle t_{s} \rangle \ll \hat{\mu}$. 
For $\tau \ll 1$, the scaling here is identical to the low-$k$ mode of the acoustic RDI from \papertwo. As discussed there, this is because this mode dominates at sufficiently low $k$ such that the pressure forces (which scale as $\nabla P \propto k$) become much smaller than the drag force between dust and gas ($\propto \mu$). The same is true of magnetic pressure, so the result is independent of $\beta$. For large $\tau$, the correction is essentially geometric, from the projection of the drift (by Lorentz forces) onto directions not  aligned with the ``forcing'' term ${\bf a}$. Also note  that although the growth rate for $\tau \ll 1$ appears to vanish when $\coeffTSrho = \tildeCoeffTSv$, an instability is still in fact present (but the leading-order term in our series expansion vanishes;  see  \papertwo); for $\tau \gg 1$ this does not significantly alter the growth rate because of the other terms which do not depend on $\tildeCoeffTSv-\coeffTSrho$.

Because the relevant wavelengths of this mode are larger than the ``stopping length'' $\lambda_{\rm stop} \sim \driftvelmag\,\langle t_{s} \rangle$ of the grains, and sufficiently large that perturbed pressure and MHD effects are weak, the mode structure is quite simple and essentially the same as in un-magnetized fluids (see \S~3.9 of \papertwo\ for details). The perturbation is a longitudinal ($\delta{\bf v} \propto \delta{\bf u} \propto {\bf k} \propto \driftvel$), compressible perturbation of the joint dust-and-gas fluid, which features dust and gas fluctuations nearly in-phase with one another ($\delta{\bf v} \sim \delta {\bf u}$, $\delta \rho_{d} \sim \mu\,\delta \rho$, but the velocity and density fluctuations are out-of-phase by $\sim 30\degr$), with a phase velocity $\driftvel\, (k\,\driftvelmag\,\langle t_{s} \rangle/\mu)^{1/3}$. As a result the dust-to-gas fluctuations driven by this mode are smaller than the absolute density fluctuation: however they accumulate dust in the form of enhanced ``sheets'' of dust over-density in the plane perpendicular to the drift direction (and moving with the drift).

\vspace{-0.5cm}
\subsection{The Strongly Lorentz-Dominated Limit: The Acoustic (Quasi-Sound \&\ Quasi-Drift) Modes}
\label{sec:parallel.modes:quasisound}

Under many conditions, at intermediate and high-$k$, the resonant modes described below are the fastest-growing. However, just like the non-resonant low-$k$ modes in \S~\ref{sub: low k regime solutions}, it is instructive to consider the strongly Lorentz-dominated case, where other modes may in fact be the fastest-growing. Specifically, here we refer to the case when $\tau \gg 1$ is sufficiently large that $\tau$ is much larger than any other parameter in the problem (including higher powers, e.g., $\tau \gtrsim k^{3}$ is required for the expansions below to be formally valid), and $1/\tau$ is sufficiently small that it is also an expansion parameter (e.g., $1/\tau \ll \mu$, $1/\tau \ll k$). 

In this limit, for fixed external acceleration ${\bf a}$ (or $\driftvelX$), the drift velocity $\driftvel$ becomes aligned with the magnetic field, $\driftvelhat \approx \hat{\bf B}_{0}$. The resonant modes, which mostly require $\hat{\bf k}$ to have large components perpendicular to $\hat{\bf B}$, can be suppressed. For example, we will show below that the growth rates of the mid-$k$ Alfv\'en and slow RDIs can be suppressed at large $\tau$, which, physically, is related to the Lorentz forces suppressing motion perpendicular to the field lines. While the high-$k$ Alfv\'en mode is not suppressed by large $\tau$, it may not appear until extremely large $k$ when $\tau$ is large. So in this limit the fastest-growing modes will often be the {\em parallel} modes, with $\hat{\bf k} \approx \driftvelhat \approx \hat{\bf B}_{0}$. If we assume this\footnote{More rigorously, we can use the full expression for $\driftvel$ as a function of $\driftvelX$ and $\tau$ in Eq.~\ref{eqn:W0} to obtain the dispersion relation, expand in $\tau^{-1}$, then marginalize over angles $\hat{\bf k}$, but this gives the same expressions.}
 and expand the dispersion relation in $\tau^{-1}$, we obtain two interesting branches of the dispersion relation.

One branch is {\em identical} to the dispersion relation for parallel modes ($\hat{\bf k} = \driftvelhat$) in the non-magnetized acoustic case studied in \papertwo.\footnote{Specifically, $0=c_{s} k\,\mu\,(\omega\,\tildeCoeffTSv - k\,\driftvelmag\,\coeffTSrho) + \langle t_{s} \rangle^{2} (\omega - k\,\driftvelmag)\,[(\omega - k\,\driftvelmag + i\,\tildeCoeffTSv/\langle t_{s} \rangle)\,(\omega^{2}-c_{s}^{2} k^{2}) + i\,\mu\,\langle t_{s} \rangle^{-1}(\omega^{2}\,\tildeCoeffTSv + \driftvelmag\,k\,\{ \driftvelmag\,k\,(\coeffTSrho - 1) + i\,\tildeCoeffTSv/\langle t_{s} \rangle \} - \omega\,\driftvelmag\,k\,\{\tildeCoeffTSv + \coeffTSrho - 1 \})]$, which is the dispersion relation for $\hat{\bf k} = \driftvelhat$ from \papertwo\ after factoring out the un-interesting purely damped modes.} 
Because they are longitudinal, parallel modes, the gas responds to the dust just like the pure hydrodynamic case (neither $\tau$ nor $\beta$ appears in the dispersion relation). From \papertwo, we know this has two unstable modes, the ``quasi-drift'' ($\omega_{\rm QD}$) and ``quasi-sound'' ($\omega_{\rm QS}$) modes. The details are given in \papertwo, but we summarize them here. Both are longitudinal and parallel and field-aligned in this limit ($\delta {\bf v} \propto \delta{\bf u} \propto {\bf k} \propto \driftvel \propto {\bf B}_{0}$). In both cases the mode resembles a sound wave in the gas, but the dust velocity and density fluctuations are out-of-phase both with the gas and with each other, with the dust density fluctuation lagging the gas by a phase offset $\sim 30\degr$. This in turn generates a very large dust response, with $\mu^{-1}|\delta\rho_{d} / \delta \rho| \gg 1$ at high-$k$. 

The ``quasi-drift'' mode is a modified dust-drift mode with frequency (at high $k$) given approximately by (derived in \papertwo): 
\begin{align}
 \omega_{\rm QD} &\approx k\,\driftvelmag + \iimag\,\frac{\tildeCoeffTSv}{2\,\langle t_{s} \rangle}\,\left[-1 + \left({1 + \frac{4\,{\mu}\,(\tildeCoeffTSv-\coeffTSrho)}{\tildeCoeffTSv^{2}\,(1-c_{s}^{2}/\driftvelmag^{2})}} \right)^{1/2}\right],
\end{align}
i.e., phase velocity $\sim \driftvel$ and growth rate $\Im(\omega_{\rm QD}) \sim (\mu/\langle t_{s} \rangle)\,[(1-\coeffTSrhoV)/(1-c_{s}^{2}/\driftvelmag^{2})]$. Thus, the mode is typically unstable if $\driftvelmag \ge c_{s}$ (given that Epstein drag dominates when $\driftvelmag > c_{s}$, and has $\coeffTSrhoV<1$ for essentially all physical gas equations-of-state), {\em or} if $\driftvelmag \le c_{s}$ and Coulomb drag dominates (which then gives $\coeffTSrhoV > 1$ for equations-of-state $\gamma \gtrsim 1$). 

The ``quasi-sound'' mode is a modified sound wave with: 
\begin{align}
 \omega_{\rm QS} &\approx \pm c_{s}\,k + i\,\frac{\mu}{2\,\langle t_{s} \rangle}\,\left[ \pm \frac{\driftvelmag}{c_{s}}\,(\coeffTSrho-1) -\tildeCoeffTSv  \right],
\end{align}
i.e., phase velocity $\sim c_{s}\,\hat{\bf k}$, and growth rate $\Im(\omega_{\rm QS}) \sim (\mu/2\,\langle t_{s} \rangle)\,(\driftvelmag\,|1-\coeffTSrho|/c_{s} -\tildeCoeffTSv)$.  Thus, the mode is unstable when $\driftvelmag/c_{s} \gtrsim \tildeCoeffTSv/|1-\coeffTSrho| \sim 1$ (i.e., typically when the drift is super-sonic). 

In both cases, the growth rates are essentially independent of $k$: they are the ``out of resonance'' versions of the acoustic RDI, or more precisely, the fast magnetosonic RDI (the mode structure is discussed in detail in \papertwo). However, under these high-$\tau$ conditions, they can be faster-growing than the resonant modes, because of the suppression of modes perpendicular to $\hat{\bf B}_{0}$. At high-$k$ and high-$\tau$, these are slower-growing than the ``cosmic ray''-like modes described below; however, at low-$k$, (or at some intermediate-$\tau$) these can be the fastest-growing modes (because their growth rate does not depend on $k$). 

%
%
\begin{figure}
\begin{center}
\includegraphics[width=1\columnwidth]{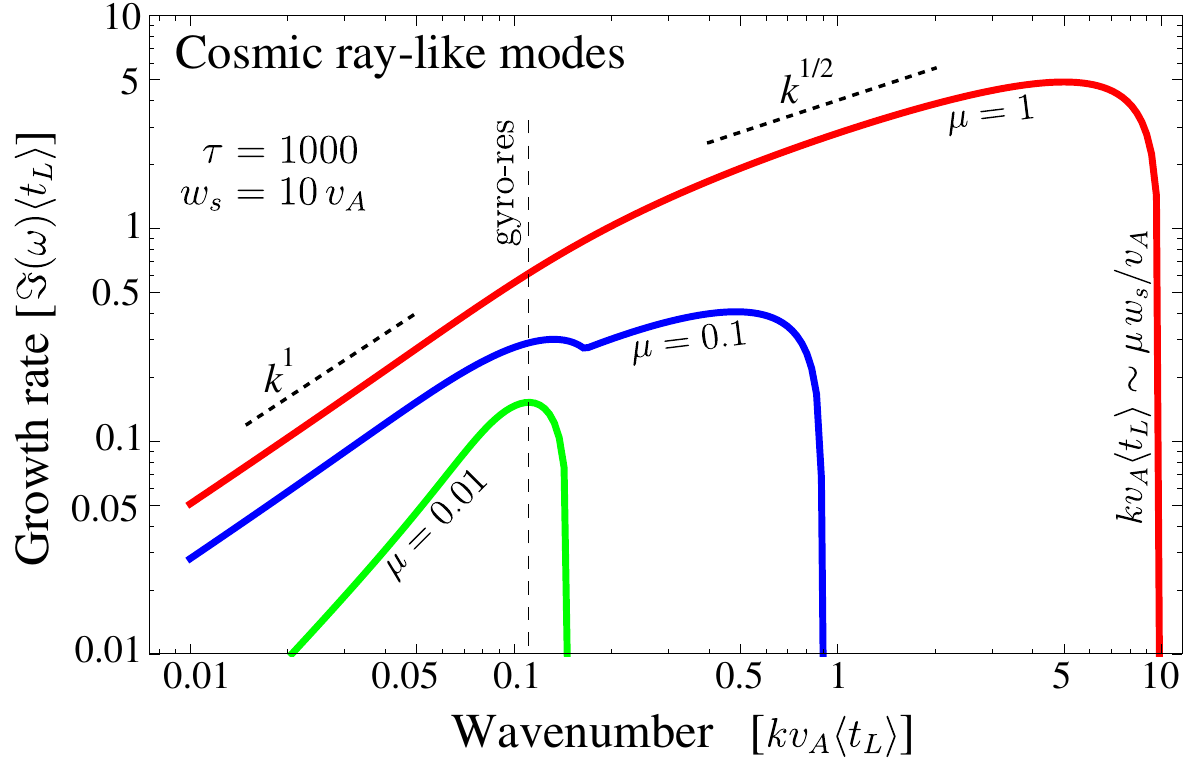}\vspace{-0.25cm}
\caption{The dispersion relation of the MHD RDI in the high-$\tau$ limit, when it becomes similar to well-known cosmic-ray instabilities (the ``cosmic-ray streaming'' modes; see \S~\ref{sub:cosmic.ray.streaming.mode}). 
Note the units of growth rate and wavelength here are the gyro time $t_{L}$ and $v_{A}\,t_{L}$.
We set the streaming aligned with the field, $\theta_{\bf Ba}=0$ (the instability is independent of this choice at sufficiently high $\tau$), with $\tau=1000$, $\driftvelmag = 10\,v_A$, and $c_s=0.1\,v_A$ ($\beta=0.01$). 
We consider modes aligned with the background drift and magnetic field, $\theta_{\bf ak}=\theta_{\bf Bk}=0$, since these are generally the fastest growing (\S~\ref{sub:cosmic.ray.streaming.mode}), and vary $\mu$ as labeled.
When $\driftvelmag < v_{A}\,\hat{\mu}^{-1/2}$ (low $\mu$ or $\driftvelmag$; green), only the resonant cosmic-ray instability \citep{kulsrud.1969:streaming.instability} is unstable. This is simply the (parallel) high-$\tau$ limit of the gyro-resonant mode (see \S~\ref{sec:gyro}), and is thus strongly peaked around the gyro-resonance (vertical dashed line, where the Larmor-frequency resonates with the Alfv\'en and fast waves: $k^{-1} \sim \driftvelmag\,\langle t_{L} \rangle$) and is unstable for $\driftvelmag>v_A$. 
Once $\driftvelmag \ge v_{A}\,\hat{\mu}^{-1/2}$ (blue \&\ red), a new branch of the dispersion relation becomes unstable, which is a manifestation of the non-resonant cosmic-ray instability of \citet{bell.2004.cosmic.rays} (c.f. their Fig.~2), and has larger growth rate at high $\mu$ and/or $\driftvelmag/v_{A}$. 
The growth rate scales as $\Im(\omega)\sim k$ (low-$k$) or $\Im(\omega) \sim k^{1/2}$  (intermediate-$k$), and it is stabilized for $k v_A \langle t_L \rangle \gtrsim \mu\,\driftvelmag/v_A$.  }
\label{fig:dispersion.relation.high.tau}\vspace{-0.5cm}
\end{center}
\end{figure}
%
%

\vspace{-0.5cm}
\subsection{The Strongly Lorentz-Dominated Limit: The ``Cosmic Ray'' Modes\label{sub:cosmic.ray.streaming.mode}}

In the strongly Lorentz-force-dominated limit (very large-$\tau$), with $\driftvelhat \approx \hat{\bf B}_{0}$, there are two other interesting branches of the dispersion relation for parallel modes ($\hat{\bf k} \approx \driftvelhat \approx \hat{\bf B}_{0}$). These modes are the manifestation of well-known cosmic-ray instabilities \citep{kulsrud.1969:streaming.instability,bell.2004.cosmic.rays}, and thus also remain unstable in the absence of drag forces (i.e., if $t_s=0$). 
In fact, because their growth rates are set by $\langle t_{L}\rangle$ (as opposed to $\langle t_{s}\rangle$),  for $\tau\gg 1$, these modes can grow much faster than the stopping time, which is the time-scale required 
for the system to reach equilibrium  (see, e.g., Fig.~\ref{fig:geometry}). Thus the results of this section 
are somewhat qualitative, and a more complete treatment would allow for arbitrary dust distribution 
functions (rather than assuming the dust to be pressure-less fluid, as done here). Many such treatments 
exist in the cosmic-ray literature (see references below).

The first of the cosmic-ray modes is simply the high-$\tau$ limit of the gyro-resonance mode. This will be discussed in detail in \S~\ref{sec:gyro}, and so here we simply note that it is related to the resonant cosmic-ray instability \citep{kulsrud.1969:streaming.instability,wentzel.1969.streaming.instability,mckenzie.1982.streaming.instability.nonlinear}, arising through the resonant interaction between an MHD wave and the dust/cosmic-ray gyro-motion.\footnote{Note that because we have assumed the dust to have zero temperature (i.e., it is 
a pressure-less fluid), our treatment captures only the $n=0$ resonance from \citet{kulsrud.1969:streaming.instability}.}  The resonance  with the Alfv\'en wave generally leads to the  fastest-growing mode, and is unstable for $\driftvelmag>v_A$.

The other cosmic-ray instability is a manifestation\footnote{Again, because we assume a zero-temperature distribution function of the dust (or cosmic rays) throughout our analysis (as well as neglecting relativistic effects) our dispersion relation is slightly different from figure 2 of \citet{bell.2004.cosmic.rays}, although it shows the same broad features.} of the non-resonant instability of \citet{bell.2004.cosmic.rays}.  These modes are fastest-growing in the  parallel direction ($\hat{\bf k} \approx \driftvelhat \approx \hat{\bf B}_{0}$), and only weakly affected by the fluid pressure (i.e., $\beta$) because they involve interactions between the dust and Alfv\'en waves. As described above (\S~\ref{sec:parallel.modes:quasisound}) for the acoustic modes, we may derive their dispersion relation  through an expansion  in $\tau\gg 1$, with $\hat{\bf k} \approx \driftvelhat \approx \hat{\bf B}_{0}$. In this limit, the full dispersion relation factors into a product of various terms and $k_z^2 [v_A^2 (\driftvelmag k_z \langle t_L\rangle+1)-\mu \driftvelmag^2]+\omega  k_z (2 \mu  \driftvelmag-v_A^2 k_z
 \langle  t_L\rangle )+\omega ^2 (-\driftvelmag k_z \langle t_L\rangle -\mu -1)+\omega ^3 \langle t_L\rangle $, which, set to zero, gives the dispersion relation of the non-resonant cosmic-ray mode.

As usual, we are particularly interested in the roots where $\omega$ has a positive imaginary component, signaling an unstable mode.  As in \citet{bell.2004.cosmic.rays}, one finds three  regimes for the solutions $\omega(k_z)$ depending on the wavelength. At longer wavelengths, $k\,v_A\langle t_L \rangle \lesssim (1+\mu) v_A/\driftvelmag $, an expansion in $k\,v_A \langle t_L\rangle$ yields the solution 
\begin{align}
\label{eqn:cosmic.ray.Bell.low.k}
\omega = \omega_{\rm Str} &\approx k\,\driftvelmag\,\hat{\mu} + i\,k\,v_{A}\,\left[ \frac{(\driftvelmag/v_{A})^{2}\,\hat{\mu} - 1}{1+\mu} \right]^{1/2},
\end{align}
while for shorter wavelengths, $(1+\mu) v_A/\driftvelmag \lesssim k\,v_A\langle t_L \rangle$, an expansion in $\driftvelmag/v_A \gg 1$ 
yields\begin{align}
\label{eqn:cosmic.ray.Bell.high.k}
\omega = \omega_{\rm Str} &\approx \frac{\mu}{\langle t_L\rangle} + i\left[ \frac{ \mu\,\driftvelmag}{\langle t_L \rangle} k_z - (k_z\,v_A )^2 -\frac{\mu^2}{\langle t_L \rangle^2} \right]^{1/2},
\end{align}
which  scales as $\Im(\omega)\sim  ( \mu\,\driftvelmag\langle t_L \rangle)^{1/2} k_z^{1/2}$ for $k\,v_A\langle t_L \rangle \lesssim \mu\,\driftvelmag/v_A$ but is 
 stabilized at the shortest wavelengths,  $k\,v_A\langle t_L \rangle \gtrsim \mu \,\driftvelmag/v_A$.  It transpires that the condition for this non-resonant mode to be unstable (across all wavelengths)  is given correctly by  Eq.~\ref{eqn:cosmic.ray.Bell.low.k} as $(\driftvelmag/v_{A})^{2}\,\hat{\mu} > 1$, i.e., $\driftvelmag > v_{A}\,\hat{\mu}^{-1/2}$.  This condition can be satisfied relatively easily in many systems \citep{bell.2004.cosmic.rays,Riquelme.2008.bell..simulations}, particularly at high $\beta$. It is also worth recalling that because $\langle t_L \rangle = \langle t_s \rangle/\tau$ and $\tau \gg 1$, this instability has very short wavelengths and fast growth rates when considered in the units of the drag time (e.g., above and in \papertwo). Thus it can grow much faster than the drag-induced quasi-sound and quasi-drift modes discussed in \S~\ref{sec:parallel.modes:quasisound}.

Considering the eigenvectors of the linear mode in detail shows that the mode resembles a mix of Alfv\'en waves, with $\driftvelmag$ playing the role of the phase velocity $v_{A}$ when  $\driftvelmag \gg  v_{A}\,\hat{\mu}^{-1/2}$. Specifically, the perturbation is very weakly compressible, with the longitudinal components of $\delta {\bf u}$, $\delta {\bf v}$ and corresponding density perturbations present but suppressed by large powers of $\mu$ and $k$. Instead it is primarily transverse, featuring dust executing gyro-motion with coupled perpendicular perturbations of the field $\delta {\bf B}_{x,\,y} \approx (\delta{\bf v}_{x,\,y} - \delta {\bf u}_{x,\,y}) / \driftvelmag$ (akin to a super-position of a forward-propagating Alfv\'en wave in the gas and backward-propagating Aflv\'en wave in dust, with $v_{A}$ replaced by $\driftvelmag$), and gas perturbations following the dust $\delta {\bf u}_{x,\,y} \approx -i\,\hat{\mu}^{1/2}\,\delta {\bf v}_{x,\,y}$ phase-shifted by $-\pi/2$, and weaker by $\hat{\mu}^{1/2}$. 


 In Fig.~\ref{fig:dispersion.relation.high.tau}, we show the numerically calculated dispersion relation in the very high-$\tau$ limit ($\tau=1000$), illustrating how the cosmic-ray modes dominate over other instabilities. We also see how at low $\mu$, the resonant instability dominates (the non-resonant instability is stable for $\hat{\mu}< (\driftvelmag/v_A)^2$), while the non-resonant instability growth rates are much larger for sufficiently large  $\mu$ and/or $\driftvelmag/v_A$ (not shown).

%
%
\begin{figure*}
\begin{center}
\includegraphics[width=0.8\columnwidth]{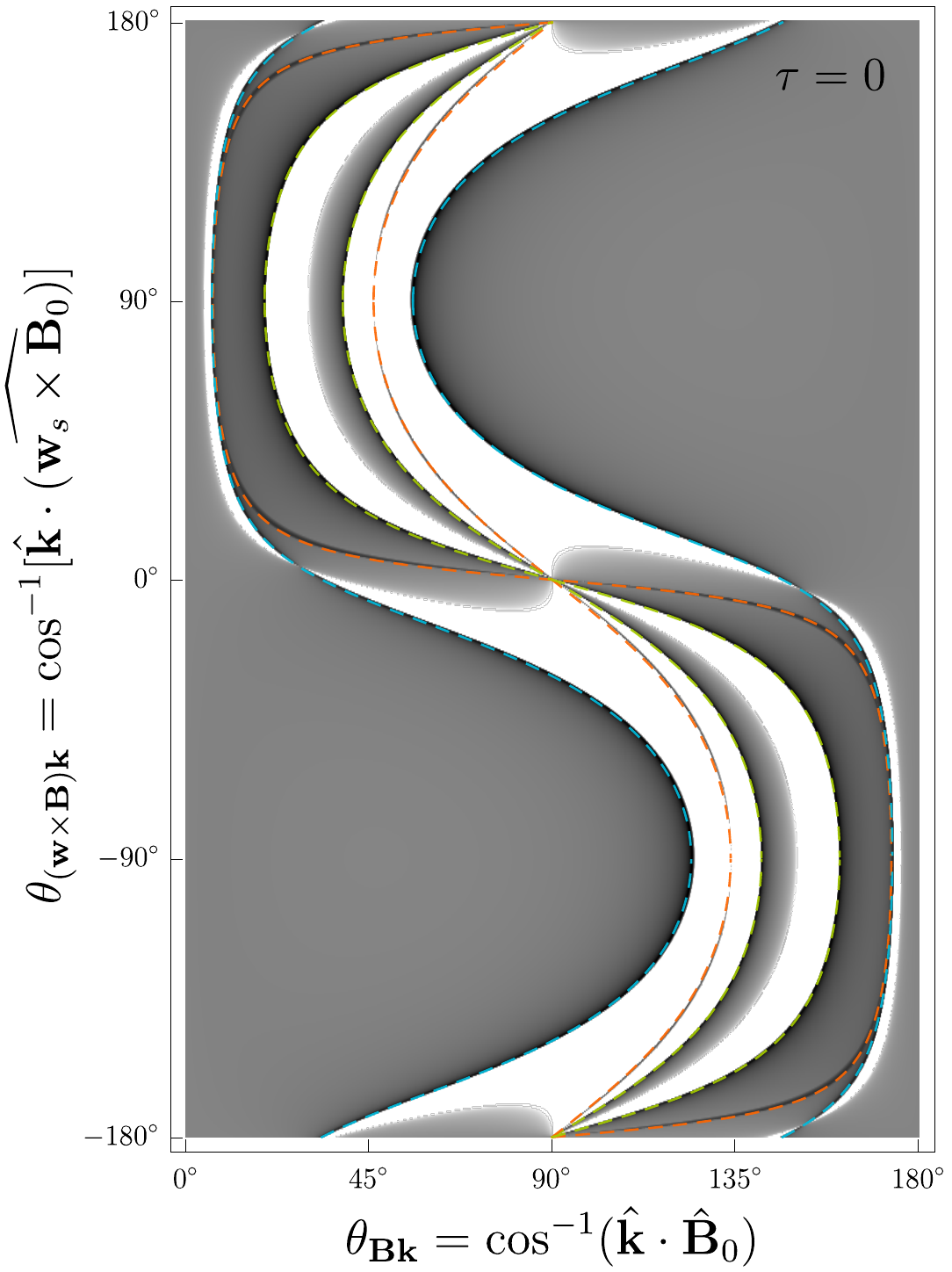}\includegraphics[width=0.655\columnwidth]{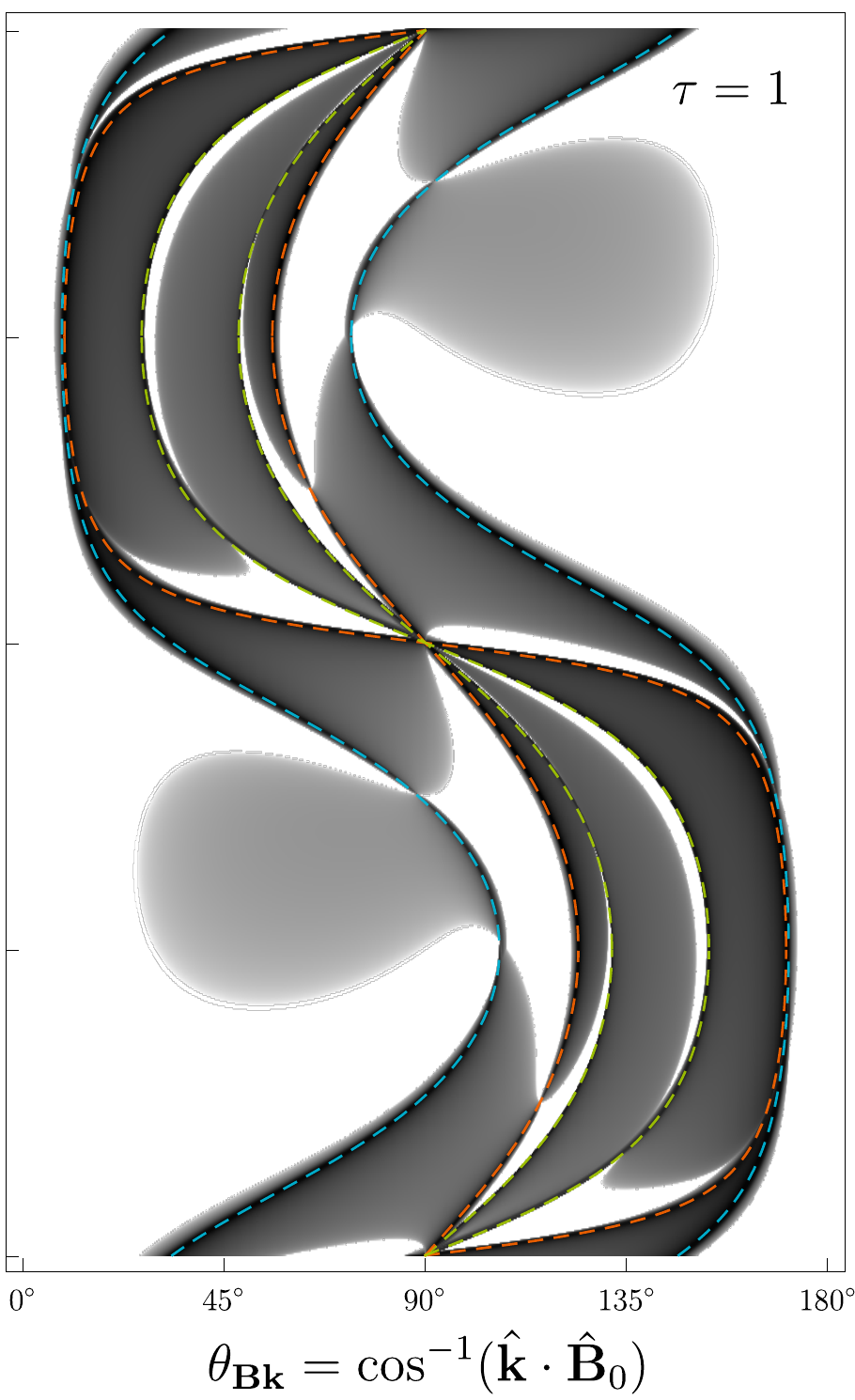}\includegraphics[width=0.655\columnwidth]{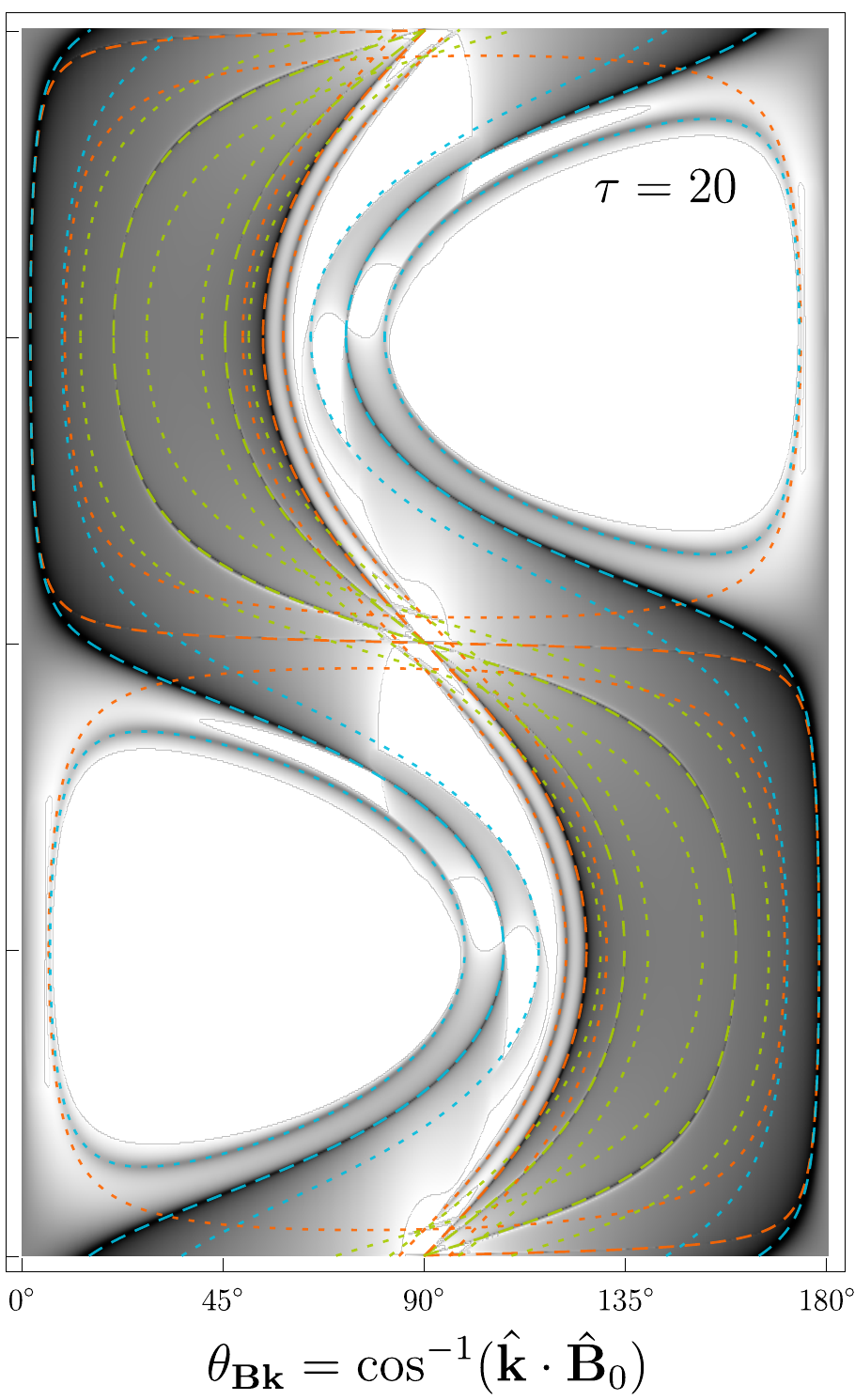}\vspace{-0.1cm}
\caption{The angular structure of the MHD RDIs, showing how the fastest-growing modes appear at the resonant angles. 
Shading in each panel shows the growth rate (increasing logarithmically from $\Im{(\omega)} \langle t_{s} \rangle < 10^{-4}$ [light] to $>1$ [dark]), evaluated at $k=|{\bf k}|=50/(c_{s}\langle t_{s}\rangle)$. The mode angle $\hat{\bf k}$ is parameterized by the angles $\theta_{\bf Bk} = \cos^{-1}(\hat{\bf k}\cdot \hat{\bf B}_{0})$ and $\theta_{\bf (w \times B)k} = \cos^{-1}(\hat{\bf k}\cdot \widehat{\driftvel \times {\bf B}_{0}})$ (see \S~\ref{sub:k.hat.angle.parameterization}). Colored lines denote the angles which satisfy the ``resonance condition'' (Eq.~\ref{eqn:resonant.condition}), for the Alfv\'en (red), slow (green), and fast (blue) wave-RDI resonances (dashed; Eq.~\ref{eqn:resonant.condition}) or gyro-RDI resonances (dotted; Eq.~\ref{eqn:gyro.condition}, only present when $\tau > 1$). 
{\em Left:} Solutions for $\tau=0$, with $|\driftvelX|=5 c_{s}$, $\theta_{\bf Ba}=-60^{\circ}$, $\beta = 1/4$, dust-to-gas ratio $\mu=0.01$, and Epstein drag with $\gamma=5/3$. 
{\em Middle:} Same parameters as {\em left}, except $\tau=1$. 
{\em Right:} Solutions for $\tau=20$, with $|\driftvelX|=60 c_{s}$, $\theta_{\bf Ba}=-88^{\circ}$, $\beta = 1/4$, $\mu=10^{-5}$, Epstein drag with $\gamma=5/3$ (we change parameters so the resonances are more visually well-separated).
The fastest-growing modes invariably occur at the predicted (resonant or gyro-resonant) mode angles, although in the $\tau=20$ case, there are certain gyro-resonances not associated with a fast-growing mode, as expected (see discussion in \S~\ref{sec:gyro:resonance}).}
\label{fig:angular.structure}\vspace{-0.5cm}
\end{center}
\end{figure*}
%
%

\vspace{-0.5cm}
\section{The MHD-Wave (Alfv\'en, Fast, and Slow) Modes}
\label{sec:mhdwave.rdi}

\subsection{Overview}
\label{sec:mhdwave.rdi:overview}

As discussed in \S~\ref{sec:intro}, the basic idea of the RDI is that, although there are often several unstable modes (at any wavelength) in the coupled dust-gas system, the fastest-growing modes at a given wavelength will (usually) be those which, to leading order, {\em simultaneously} satisfy the dispersion relation of the gas, absent dust (i.e., represent ``natural'' modes of the gas) and the dispersion relation of the dust, absent gas perturbations (i.e., ``natural'' modes of the dust). This is especially true if the modes are both undamped, so there is no natural ``competing'' damping force. A more formal discussion of these ideas is given in App.~\ref{sec:matrix resonances}.

If there were no ``back-reaction'' term (force from the dust on the gas), then the gas perturbations and dust perturbations would form two entirely separable systems. The dispersion relation for gas would simply be the usual MHD relation: 
\begin{align}
0=&\,\left(\omega_{g}^{2} - (v_{A}\,{\bf k}\cdot\hat{\bf B}_{0})^{2} \right)\,\left( \beta\,\omega_{g}^{4}-\omega_{g}^{2}\,k^{2}\,v_{A}^{2}\,(1+\beta) + v_{A}^{4}\,k^{2}\,({\bf k}\cdot\hat{\bf B}_{0})^{2} \right) 
\end{align}
which has the familiar, un-damped solutions $\omega_{g} = {\bf v}_{p} \cdot {\bf k}$ with ${\bf v}_{p} = (\pm {\bf v}_{A},\ \pm {\bf v}_{+}(\hat{\bf k}),\ \pm {\bf v}_{-}(\hat{\bf k}))$, i.e., the standard constant-phase velocity ideal MHD Alfv\'en (${\bf v}_{A}\equiv v_{A}\,\hat{\bf B}_{0}$), fast (${\bf v}_{+}$ in the direction $\hat{\bf k}$), or slow (${\bf v}_{-}$) magnetosonic waves.
Meanwhile the dust, responding to a uniform (non-perturbed) gas background, would have dispersion relation 
\begin{align}
\nonumber 0 =&\, \left( \omega_{d} - {\bf k}\cdot\driftvel \right) \times \\
\nonumber &
{\Bigl [}
\left( \omega_{d} - {\bf k}\cdot\driftvel + i\,\langle t_{s} \rangle^{-1} \right)^{2}\,
 \left( \omega_{d} - {\bf k}\cdot\driftvel + i\,\{ 1 + \coeffTSv \}\,\langle t_{s} \rangle^{-1} \right) \\ 
\label{eqn:dust.natural.modes} &\ \ \ \ \ \ \ - \frac{1}{\langle t_{L} \rangle^{2}}\,\left( \omega_{d} - {\bf k}\cdot\driftvel + i\,\{ \cos^{2}{\theta_{\bf Bw}} + \coeffTSv \}\, \langle t_{s} \rangle^{-1}
\right)
{\Bigr ]}
\end{align}
This has one un-damped mode, $\omega_{d} = {\bf k}\cdot \driftvel$ (simple advection with the drift). It also has three damped solutions, which if we take $\hat{\bf B}_{0} = \driftvelhat$ for simplicity can be easily solved as $\omega_{d} = {\bf k}\cdot\driftvel - i\,\tildeCoeffTSv\,\langle t_{s} \rangle^{-1}$ (``normal'' damped motion on the stopping time) and $\omega_{d} = {\bf k}\cdot \driftvel \pm \langle t_{L} \rangle^{-1} - i\,\langle t_{s} \rangle^{-1}$ (gyro motion damped on the stopping time). 

The MHD-wave RDI modes are those which, ``at resonance,'' simultaneously satisfy $\omega = {\bf k}\cdot\driftvel$ (the un-damped dust mode) and any of $\omega = \pm ({\bf v}_{A},\,{\bf v}_{+},\,{\bf v}_{-})\cdot {\bf k}$, to leading order. For the MHD-wave modes, this is possible when the mode propagates at an appropriate angle, so that: 
\begin{align}
\label{eqn:resonant.condition} \driftvel\cdot{\bf k} &=  \pm ({\bf v}_{A},\,{\bf v}_{+},\,{\bf v}_{-})\cdot {\bf k} \, .
\end{align}
Because the Alfv\'en and slow magnetosonic waves have phase velocities which become vanishingly small at certain angles,  solutions to Eq.~\ref{eqn:resonant.condition} {\em always} exist, for {\em any} finite $|\driftvel|$ and $\beta$.\footnote{Just as in \papertwo, we can also (much more tediously) derive the resonance condition directly from expansion of the 10th-order dispersion relation. To illustrate this, consider just the limit of arbitrarily high-$k$, where the dispersion relation to leading order becomes just: 
\begin{align}
\label{eqn:unperturbed.modes.highk} 0=&\, (\omega-{\bf k}\cdot\driftvel)^{4}\,\left(\omega^{2} - (v_{A}\,{\bf k}\cdot\hat{\bf B}_{0})^{2} \right)\\
\nonumber	&\,\left( \beta\,\omega^{4}-\omega^{2}\,k^{2}\,v_{A}^{2}\,(1+\beta) + v_{A}^{4}\,k^{2}\,({\bf k}\cdot\hat{\bf B}_{0})^{2} \right) + \mathcal{O}(\hat{\mu}\,k^{9})\, .
\end{align}
This is just the product of the (dust-free) MHD dispersion relation and $(\omega - {\bf k}\cdot \driftvel)^{4}$, so is solved either by any of the MHD modes or $\omega = {\bf k}\cdot \driftvel$. Inserting this and then expanding to next-to-leading order in $k$, we obtain the next-order correction to $\omega$. As for the acoustic RDI, these next-order terms almost always include multiple unstable modes, but for a random mode angle $\hat{\bf k}$ these have growth rates that either become independent of $k$ or are stabilized at sufficiently high $k$ (this arises from the next-to-leading term). However, if we simultaneously satisfy $\omega = {\bf k}\cdot\driftvel$ and $\omega = \omega_{g}$ (the resonant condition), we eliminate multiple additional powers in Eq.~\ref{eqn:unperturbed.modes.highk} and eliminate the next-to-leading order terms (which tend to be stabilizing). Physically, we eliminate the ``natural response'' of the gas which would otherwise suppress growth at high-$k$.}

\begin{figure}
\begin{center}
\includegraphics[width=0.98\columnwidth]{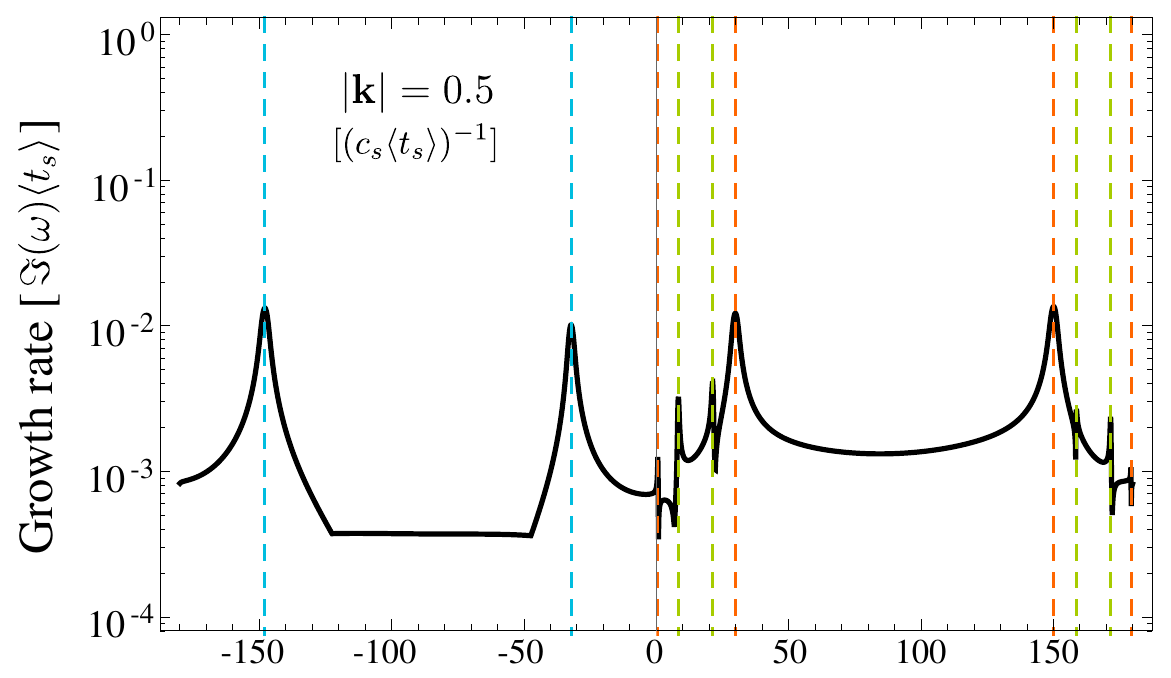}\\
\includegraphics[width=0.98\columnwidth]{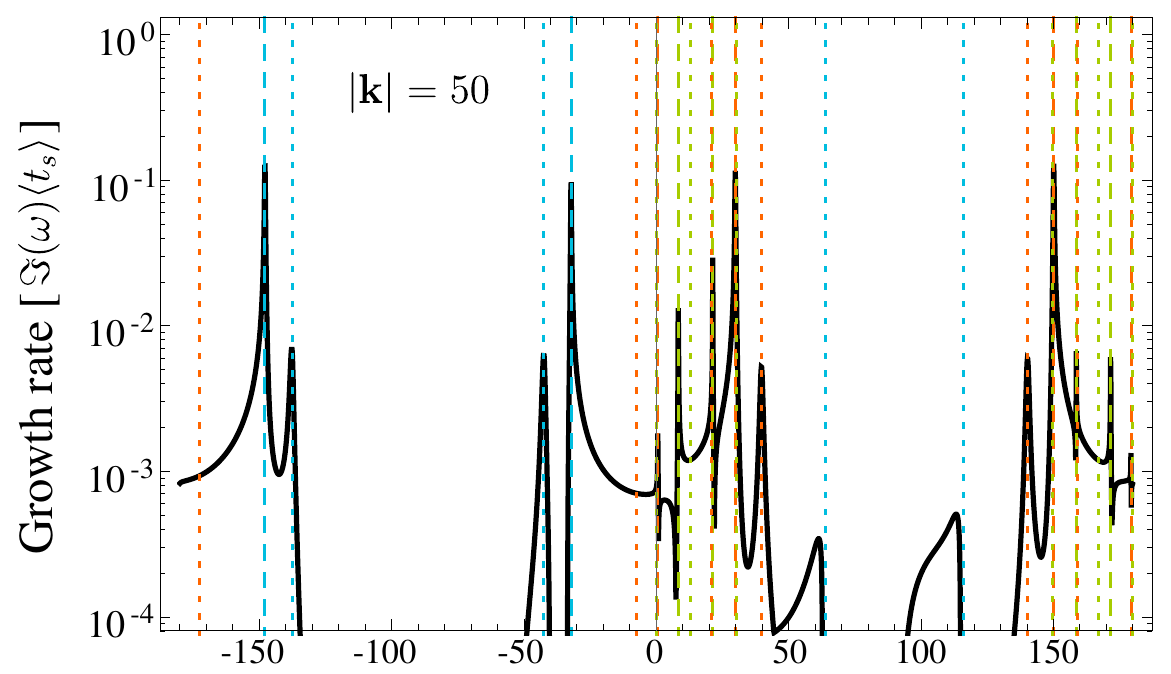}\\
\includegraphics[width=0.96\columnwidth]{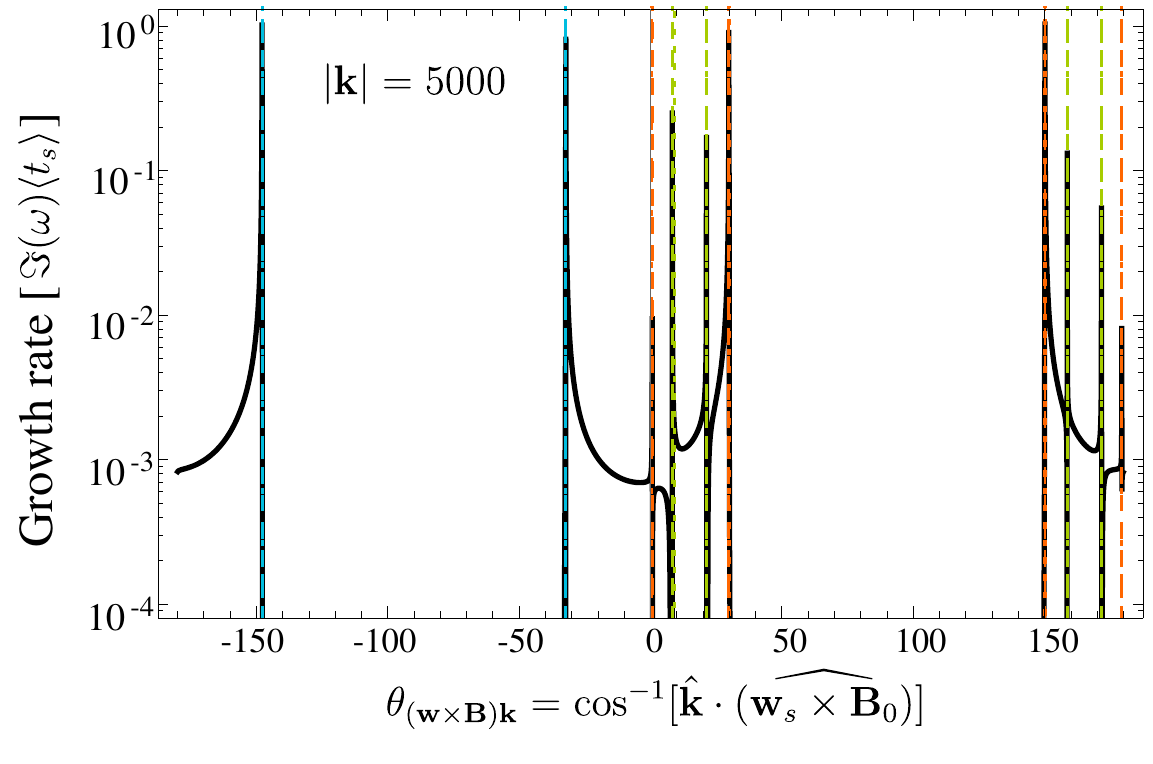}\vspace{-0.25cm}
\caption{1D angular structure of the MHD RDIs, taking a vertical ``slice'' through Fig.~\ref{fig:angular.structure}; i.e., we fix the angle between $\hat{\bf k}$ and $\hat{\bf B}_{0}$ at $\theta_{\bf Bk}=70^{\circ}$, and plot the growth rate $\Im{(\omega)}$ versus the varied free angle $\theta_{\bf (w \times B)k}$. 
Parameters are fixed to those used in the $\tau=20$ case in Fig.~\ref{fig:angular.structure} ($|\driftvelX|=60 c_{s}$, $\theta_{\bf Ba}=-88^{\circ}$, $\beta = 1/4$, $\mu=10^{-5}$), but we consider three wavenumbers: $k=|{\bf k}|=0.5/(c_{s}\,\langle t_{s} \rangle)$ ({\em top}), $k=50/(c_{s}\,\langle t_{s} \rangle)$ ({\em middle}), $k=5000/(c_{s}\,\langle t_{s} \rangle)$ ({\em bottom}). 
Colored lines show the resonant angles as in Fig.~\ref{fig:angular.structure}: Alf\'en (red), slow (green), and fast (blue) MHD-wave RDIs (dashed; Eq.~\ref{eqn:resonant.condition}) or gyro-RDIs (dotted; Eq.~\ref{eqn:gyro.condition}). 
Note that $k=0.5$ ({\em top}) is too low-$k$ for any gyro-resonant condition to be met, while the gyro-resonances become indistinguishable from the MHD-wave RDI resonances for the high-$k=5000$ case ({\em bottom}). For intermediate-$k=50$ ({\em middle}), we see that the fast-wave gyro-resonances are strongest, while the some of the Alfv\'en-wave gyro-resonances and all of the slow-wave gyro-resonances are too weak to be visible at these parameters (see \S~\ref{sec:gyro}).
\label{fig:angle.1d}}\vspace{-0.5cm}
\end{center}
\end{figure}

Figures \ref{fig:dispersion.relation}, \ref{fig:angular.structure}, and Fig.~\ref{fig:angle.1d}, show the results of direct numerical solutions of the full dispersion relation, for the MHD-wave RDI modes. As shown explicitly from Figs.~\ref{fig:angular.structure} and \ref{fig:angle.1d}, or by comparing the bottom-right panel of Fig.~\ref{fig:dispersion.relation} to the other panels,  the growth rates are  almost always maximized (at a given $k$) at the ``resonant angles'' where the condition in Eq.~\ref{eqn:resonant.condition} is met. These numerical solutions also illustrate that at any finite $k$ and $\driftvelmag$, there will generally be several resonant ``families.'' Some range of mode angles $\hat{\bf k}$ will always satisfy the resonance condition with the slow and Alfv\'en waves, so these will produce a range of angles that meet the resonance condition (with different phase velocities) for both the $\pm$ solutions of both wave families. If $\driftvelmag$ is sufficiently large, it is also possible to meet the resonance condition with the fast wave family over some range of angles. 
This rich and complex resonance structure is very different from a  pure hydrodynamical system  (the acoustic RDI in \papertwo). Because a neutral gas has only one direction-independent wavespeed ($c_{s}$), there is  only  one resonant family. This  exists only when $\driftvelmag > c_{s}$ and features just one ``resonant angle,'' $\theta_{\bf ak}=\cos^{-1}(c_{s}/\driftvelmag)$ 

\subsubsection{The Mid- and Short-Wavelength RDI Modes}\label{subsub:transition.between.mid.and.high.k}

As also occurs for the acoustic RDI (see \paperone\ and \papertwo), depending on the mode wavenumber in comparison to other scales in the problem (i.e., $c_{s}\langle t_{s}\rangle$, and various combinations of other parameters), 
there are two regimes of the MHD-wave RDIs. We term these the mid-$k$ and high-$k$ (or mid- and short-wavelength), RDIs, and explore their properties  separately in \S~\ref{sec:mhdwave.rdi:growthrates.midk:magnetosonic} and \S~\ref{sec:mhdwave.rdi:growthrates.hik:magnetosonic} respectively.\footnote{Recall the the long-wavelength, low-$k$ modes do not arise from resonances at all; see \S~\ref{sub: low k regime solutions}.} 
They are distinguished by the scaling of the growth rate with $\mu$ (and $k$): in the mid-$k$ regime $\Im(\omega)\sim \mu^{1/2}$, while in the high-$k$ regime $\Im(\omega)\sim \mu^{1/3}$.
As explained from the matrix-analysis perspective in Appendix~\ref{sec:matrix resonances} (see App.~\ref{sub:simple.matrix.resonance.algorithm} for a simple outline),
the transition between the two regimes is most simply understood by asking 
about the magnitude of the perturbation to the \emph{frequency} (i.e., effectively the magnitude of $\Im(\omega)$)
 in comparison to other parameters in the problem, as opposed to the wavenumber itself. 
 In particular,  the mid-$k$ regime generally applies when $\Im(\omega)\lesssim \langle t_{s}\rangle^{-1}$, while the high-$k$ regime applies if $\Im(\omega)\gtrsim \mathrm{MAX}( \langle t_{s}\rangle^{-1},  \langle t_{L}\rangle^{-1})$. If $\tau\gtrsim 1$, there is often a transition regime with $\langle t_{s}\rangle^{-1}\lesssim \Im(\omega)\lesssim \tau \langle t_{s}\rangle^{-1} = \langle t_{L}\rangle^{-1}$ where no clear scaling applies. 
 While these conditions can be used as a general guide, we caution that they do not
 apply near certain special points in parameter space (e.g., when certain combinations of the $\zeta_{X}$ parameters are nearly zero; see \papertwo).

This change in scaling can be clearly seen in Fig.~\ref{fig:dispersion.relation}. For the $\tau=0$ and $\tau=1$ cases in all three resonant families, there is a clear change in scaling at $\Im(\omega)\langle t_{s}\rangle\sim 1$ from mid-$k$ ($\Im(\omega)\sim k^{1/2}$) to high-$k$ ($\Im(\omega)\sim k^{1/3}$) scaling (note that the $\tau=0$ Alfv\'en-wave mid-$k$ RDI has zero growth rate, so is a special case; see \S\ref{sec:mhdwave.rdi:growthrates.midk:Alfv\'en}). For the $\tau=100$ examples, the high-$k$ scaling applies only once $\Im(\omega)\langle t_{s}\rangle\gtrsim \tau = 100$, which 
is most clearly seen on the fast-wave resonance panel (but continuing the Alfv\'en- and slow-wave resonance panels to higher wavenumbers shows that it does occur for these cases also).

%
%
%
%

\vspace{-0.5cm}
\subsection{Resonant Mode Angles}
\label{sec:resonant.mode.angles}

Because the gas dispersion relation for $\omega_{g}$ (absent dust) has 6 branches, each of which has an angle-dependent phase velocity, there are in fact always a range of angles $\hat{\bf k}$ that satisfy Eq.~\ref{eqn:resonant.condition}, and therefore produce resonance. As a general rule, the angle $\hat{\bf k}$ that produces the fastest-growing mode at a given $k$ is (usually) that which produces the largest phase velocity $|v_{p}| = |\omega_{g}|/k = |\driftvel\cdot \hat{\bf k}|$, while still satisfying the resonant condition (while somewhat difficult, it is possible to read this off of Figs.~\ref{fig:angular.structure}-\ref{fig:angle.1d}, for example). In general this must be solved numerically, but it is instructive to consider some limits.

\begin{itemize}

\item{\bf Fast Drift} ($|\driftvel | \gtrsim v_{f,\,0}$): In this case the fastest resonance is with the fast magnetosonic wave. That wave has a phase speed which is only weakly sensitive to angle (between $v_{f,\,0} = \sqrt{c_{s}^{2} + v_{A}^{2}}$ and ${\rm MAX}(v_{A},\,c_{s})$). So the resonant angle must obey, approximately:
\begin{align}
\nonumber  \cos{\theta_{\bf wk}} &= \hat{\bf k}\cdot \driftvelhat = \frac{k_{z}}{k} \approx \pm \frac{v_{f,\,0}}{ \driftvelmag }\, , \\
 \hat{\bf k}\cdot \driftvel &= \pm v_{+}(\hat{\bf k}) \approx \pm v_{f,\,0} \ .
\end{align}
The $\pm$ directions behave identically here. The weak sensitivity of $v_{+}$ to angle means that the orientation of the component of $\hat{\bf k}$ perpendicular to $\driftvelhat$ (the angle of ${\bf k}_{\bot}$ in the $x-y$ plane) usually has only a small effect on the growth rates. The phase speed of $v_{+}$ is maximized when the projection onto $\hat{\bf B}_{0}$ is minimized, so the growth rates are usually slightly higher for modes with the perpendicular component ${\bf k}_{\bot}$ oriented primarily along the direction mutually perpendicular to $\driftvelhat$ and ${\bf B}_{0}$ (i.e., the $\hat{y} \propto \driftvel \times {\bf B}_{0}$ direction), while the projection onto $k_{x}$ satisfies $k_{x}/k \approx -(\cot{\theta}_{\bf Bw})\,(k_{z}/k)$ (so $\hat{\bf k} \cdot \hat{\bf B}_{0} \approx 0$).

\item{\bf Intermediate Drift, Strongly-Magnetized} ($c_{s} \ll |\driftvel| \ll v_{A}$): In this case (which can occur at low-$\beta$), the drift is faster than the sound speed but slower than Alfv\'en, prohibiting resonance with the fast magnetosonic wave. The slow magnetosonic wave speed is maximized at ${\rm MIN}(c_{s},\,v_{A}) = c_{s}$, while the Alfv\'en wave has phase velocity $= v_{A}\,\cos{\theta}_{\bf Bk}$ (which can be much larger than the slow wave). Thus, the fastest resonance is with the Alfv\'en wave, leading to the resonance requirement $k_{x} = k_{z}\,(\driftvelmag/v_{A} -\cos{\theta}_{\bf Bw}) / \sin{\theta}_{\bf Bw}$. Now, recall we wish to maximize $| \hat{\bf k}\cdot \driftvel | \propto |k_{z}|$, but we must obey $k_{z}^{2} + k_{x}^{2} \le k^{2}$; maximizing $|k_{z}|$ subject to this constraint gives $\hat{\bf k} = (k_{x},\,k_{y},\,k_{z})$ with $k_{y}=0$ and $k_{z}=\pm [1 + (\driftvelmag/v_{A} - \cos{\theta}_{\bf Bw})^{2}/\sin^{2}\theta_{\bf Bw}]^{-1/2}$, or (to leading order): 
\begin{align}
\nonumber\hat{\bf k} &\approx \pm\left( -\cos\theta_{\bf Bw} + \frac{\driftvelmag}{v_{A}}\,\sin^{2}\theta_{\bf Bw} \, , \, 0 \, , \,  \sin\theta_{\bf Bw}\,\left[1+\frac{\driftvelmag}{v_{A}}\,\cos\theta_{\bf Bw}\right] \right) \, , \\
\hat{\bf k} & \cdot \driftvel  = {\bf v}_{A}(\hat{\bf k}) \approx \pm  \driftvelmag\,\sin{\theta}_{\bf Bw} \, ,
\end{align}
so $|\cos{\theta}_{\bf Bk}| \approx (\driftvelmag/v_{A})\,|\sin{\theta}_{\bf Bw}| \ll 1$. Note that our sign convention is such that $\sin\theta_{\bf Bw} > 0$ always (i.e., $\sin{\theta}_{\bf Bw} \equiv \sqrt{1 - \cos\theta_{\bf Bw}^{2}}$). 

In short, the fastest-growing mode is oriented almost (but not quite) perpendicular to $\hat{\bf B}_{0}$ in the $\hat{\bf B}_{0} - \driftvelhat$ plane.

\item{\bf Intermediate Drift, Weakly-Magnetized} ($v_{A} \ll |\driftvel| \ll c_{s}$): In this case, resonance with the fast wave is not possible, but resonances with either the Alfv\'en or slow modes are possible, and these modes have nearly identical phase speeds (since $\beta\gg 1$). Thus the Alfv\'en and slow resonances are essentially degenerate. We again have $\driftvel\cdot{\bf k} \approx v_{A}\,\cos{\theta}_{\bf Bk}$, or $k_{x} = k_{z}\,(\driftvelmag/v_{A} -\cos{\theta}_{\bf Bw}) / \sin{\theta}_{\bf Bw}$; maximizing $|k_{z}|$ again gives $k_{y}=0$, $k_{z}=\pm [1 + (\driftvelmag/v_{A} - \cos{\theta}_{\bf Bw})^{2}/\sin^{2}\theta_{\bf Bw}]^{-1/2}$. Then, because $\driftvelmag \gg v_{A}$, to leading order the maximum phase speed occurs at
\begin{align}
\nonumber\hat{\bf k} &\approx \pm\left( 1 - \frac{v_{A}^{2}\,\sin^{2}\theta_{\bf Bw}}{2\,\driftvelmag^{2}} \, , \, 0 \,  , \, \frac{v_{A}\,\sin\theta_{\bf Bw}}{\driftvelmag} \right) \, , \\
\hat{\bf k} & \cdot \driftvel  = {v}_{-}(\hat{\bf k}) \approx \pm  v_{A}\,\sin{\theta}_{\bf Bw} \,.
\end{align}
So, the fastest-growing mode is has $\hat{\bf k}$ primarily in the direction of ${\bf B}_{\bot}$ (the direction of ${\bf B}_{0}$ perpendicular to $\driftvel$).

\item{\bf Slow Drift} ($|\driftvel | \ll  {\rm MIN}(c_{s},\,v_{A})$): For small $|\driftvel| \lesssim {\rm MIN}(c_{s},\,v_{A})$, resonance with the fast magnetosonic wave is not possible and resonance with the slow or Alfv\'en waves requires $|\cos{\theta}_{\bf Bk}| \ll 1$ (so that the phase speed is low). Thus,  the slow-wave phase speed is again given by an expression similar to the Alfv\'en-wave phase speed, $v_{-}^{2} \approx (c_{s}\,v_{A}/v_{f,\,0})^{2}\,\cos^{2}{\theta}_{\bf Bk}$. Setting this equal to $(\hat{\bf k}\cdot \driftvel)^{2}$ we obtain the requirement $(k_{x}/k) = (k_{z}/k)\,(-\cos{\theta}_{\bf Bw} \pm \tilde{w})/|\sin{\theta}_{\bf Bw}|$ (where $\tilde{w}\equiv |\driftvel|\,v_{f,\,0} / (c_{s}\,v_{A}) \approx |\driftvel|/{\rm MIN}(c_{s},\,v_{A})$). We then obtain: 
\begin{align}
\nonumber\hat{\bf k} &\approx \pm\left( -\cos\theta_{\bf Bw} + \tilde{w}\,\sin^{2}\theta_{\bf Bw} \, , \, 0 \, , \,  \sin\theta_{\bf Bw}\,(1+\tilde{w}\,\cos\theta_{\bf Bw}) \right) \, , \\
\hat{\bf k} & \cdot \driftvel  = \pm v_{-}(\hat{\bf k}) \approx \pm  \driftvelmag\,\sin{\theta}_{\bf Bw} \, .
\end{align}
Like the intermediate-drift, strongly-magnetized case, the fastest-growing mode is oriented close to perpendicular to $\hat{\bf B}_{0}$ in the $\hat{\bf B}_{0} - \driftvelhat$ plane. Note that both the Alfv\'en and slow mode resonances have a similar resonant angle in this case, but at low $\beta$ the growth rates can be different.
	
\end{itemize}

\vspace{-0.5cm}
\subsection{Growth Rates: The Mid-Wavelength (Low-$\mu$) MHD-Wave RDI Modes}
\label{sec:mhdwave.rdi:growthrates.midk}

With \S~\ref{sec:resonant.mode.angles} in mind, if we expand the dispersion relation about $\hat{\mu} \ll 1$, {\em and} assume the resonance condition --- i.e., $\driftvel \cdot \hat{\bf k} = v_{\pm}$ or $v_{A}$ (matching the fast, slow, or Alfv\'en phase velocity) --- then we obtain a leading-order dispersion relation of the form:
\begin{align}
\label{eqn:midk.resonant.mode.general} \omega_{\pm,A} &= k\,v_{\pm,A} + \left(\iimag \pm 1\right)\, \mathcal{F}_{\rm mid} \,\left(  \frac{\hat{\mu}\,k\,v_{f,\,0}}{2\,\langle t_{s} \rangle} \right)^{1/2} + \mathcal{O}(\hat{\mu}).
\end{align}
This always has an unstable root, as with the long-wavelength mode.\footnote{In Eq.~\ref{eqn:midk.resonant.mode.general}, note that the second-from leading term (in $\iimag\pm1$) comes from solving an equation of the form $\omega_{\pm,A} = k\,v_{\pm,A} + \omegaZ$ with $\omegaZ^{2} = \pm\,\iimag\,X$ where $X = \mathcal{F}_{\rm mid}^{2}\,\hat{\mu}\,k\,v_{f,\,0}/\langle t_{s} \rangle$  is purely real. Unless $\mathcal{F}_{\rm mid} = 0$, this always has an unstable solution with roots proportional to $(\iimag \pm 1)/\sqrt{2}$ where the $\pm$ for the real part corresponds to the $\pm$ sign of $\omegaZ^{2} = \pm\,\iimag\,X$ but has no effect on the growth rate.} 
As explained in detail in Appendix~\ref{sec:matrix resonances} (see also \S\ref{subsub:transition.between.mid.and.high.k}), the expansion used to derive Eq.~\ref{eqn:midk.resonant.mode.general} is generally valid when the derived 
perturbation to $\omega_{\pm,A} $ (i.e., $\omega_{\pm,A} - k\,v_{\pm,A}$) is less than  $\sim \langle t_{s}\rangle^{-1}$, but is still less than the long-wavelength, low-$k$ growth-rate prediction (see \S~\ref{sub: low k regime solutions} and Eq.~\ref{eqn:omega.low.k}).
We will now consider the cases where the resonance is with the (fast or slow) magnetosonic, or Alfv\'en phase speeds.

\vspace{-0.1cm}
\subsubsection{ The (Fast \&\ Slow) Magnetosonic-Wave RDI}
\label{sec:mhdwave.rdi:growthrates.midk:magnetosonic}

First consider the case of modes resonant with the magnetosonic phase velocities: $\driftvel \cdot \hat{\bf k} = v_{\pm}$, where we will consider the most relevant cases of the fast-mode resonance when $\driftvelmag \gtrsim v_{f,\,0}$ (since this is the fastest-growing resonance) and slow-mode resonance when $\driftvelmag \lesssim v_{f,\,0}$. Even restricting to the magnetosonic RDI in the mid-wavelength regime, the expressions for $\mathcal{F}_{\rm mid}$ are rather un-informative, so we will further consider the limits of weak and strong Lorentz forces.

\begin{itemize}

\item{\bf Weak Lorentz Forces} ($\tau \ll 1$): If we neglect Lorentz forces, then the growth rates for this mode simplify to the general expression from \paperone:
\begin{align}
\label{eqn:weak.lorentz.mid.k.mode.scaling}
\mathcal{F}_{\rm mid} &= \left|(1-\coeffTSrhoV)\left(\frac{v_{\mp}^{2}\,\cos{\theta_{\bf Ba}}}{\cos{\theta_{\bf Bk}}\cos{\theta_{\bf ka}}} -1 \right)\,\Theta_{\pm} \right|^{1/2} \\ 
\nonumber \Theta_{\pm} &\equiv \frac{v_{\pm^{3}}\,(1-v_{\mp}^{2})}{(1+v_{\mp}^{2})^{2} 
+ v_{\pm}^{2}\,(1-2\,v_{\mp}^{2}) + \sin^{2}{\theta_{\bf Bk}}/\beta + v_{\mp}^{4}\,v_{\pm}^{2}/\cos^{2}{\theta_{\bf Bk}}}
\end{align}
(this expression is valid for any angle that satisfies the resonant condition). 
But even this is rather un-intuitive. To simplify further, consider the fastest-growing resonant angle in both the ``fast drift'' (resonance with the fast magnetosonic mode) and ``slow drift'' (slow mode resonance) limits (\S~\ref{sec:resonant.mode.angles}). Equation~\ref{eqn:weak.lorentz.mid.k.mode.scaling} then becomes,
\begin{align}
\label{eqn:msonic.rdi.midk.lotau}\mathcal{F}_{\rm mid} &\approx  
\begin{cases}
{\displaystyle |1-\coeffTSrhoV|^{1/2}} &\ \hfill (\driftvelmag \gg v_{f,\,0}) \\
\\
{\displaystyle \frac{\driftvelXmag}{v_{f,\,0}}\,{\left[ \frac{|(1-\coeffTSrhoV)\,\sin{\theta_{\bf Ba}}\,\cos{\theta_{\bf Ba}}|^{1/2}}{\beta^{1/4}} \right]}} &\ \hfill (\driftvelmag \ll v_{f,\,0}).
\end{cases}  
\end{align}

Note that for the slow-drift case, if $\sin\theta_{\bf Ba}=0$ exactly (drift and field are perfectly parallel), it becomes impossible to satisfy the slow-mode resonant condition for $\driftvelmag \lesssim {\rm MIN}(c_{s},\,v_{A})$, so the growth rate vanishes. However for $\cos\theta_{\bf Ba}=0$ (exactly perpendicular drift and field lines) the resonance does not vanish (our series expansion simply becomes inaccurate), and a more accurate derivation in the limit where $\cos{\theta}_{\bf Ba}$ is small leads to the replacement $|\cos{\theta}_{\bf Ba}| \rightarrow {\rm MAX}[|\cos{\theta}_{\bf Ba}|,\,(\driftvelmag/v_{f,\,0})]$.\footnote{Note, if $\beta$ is sufficiently large so $v_{A} \ll \driftvelmag \ll c_{s}$ (so we are not cleanly in the ``slow'' or ``fast'' regime, the scaling is modified to $\mathcal{F}_{\rm mid} \approx (1/2)\,(1-\cos^{2}{\theta}_{\bf Ba})^{1/2}\,({1-\cos^{2}\theta_{\bf Ba}})^{1/4}\,\beta^{-3/4}$.}

\item{\bf Strong Lorentz Forces} ($\tau \gg 1$): In the limit where Lorentz forces dominate drag ($\tau \gg 1$), we find:
\begin{align}
\label{eqn:msonic.rdi.midk.hitau}\mathcal{F}_{\rm mid} &\approx  
\begin{cases}
{\displaystyle 
\left| \frac{\coeffTSrho}{\tildeCoeffTSv}\left(\frac{\beta}{1+\beta} \pm \frac{\driftvelXmag}{v_{f,\,0}}\,\sin\theta_{\bf Ba} \right) + \frac{\tan^{2}\theta_{\bf Ba}}{\tildeCoeffTSv} 
 \right|^{1/2}} &\ \hfill (\driftvelmag \gg v_{f,\,0}) \\
\\
{\displaystyle \frac{\driftvelXmag}{v_{f,\,0}}\,{\left[ \frac{|(\beta+[1-\coeffTLrho])\,\sin{\theta_{\bf Ba}}\,\cos{\theta_{\bf Ba}}|^{1/2}}{\tau^{1/2}\,\beta^{1/4}} \right]}} &\ \hfill (\driftvelmag \ll v_{f,\,0}).
\end{cases}  
\end{align}
Note that if $\cos{\theta}_{\bf Ba} \rightarrow 0$, $\tan^{2}\theta_{\bf Ba}\rightarrow\infty$, but the growth rates do not actually diverge (our series expansion is simply inaccurate). A more accurate expansion gives the upper and lower limits of this term of $\tan^{2}\theta_{\bf Ba} \rightarrow {\rm MIN}\{ \tan^{2}\theta_{\bf Ba},\,  (\driftvelXmag/v_{f,\,0})^{2} / (2+\beta) \} $ (as $\theta_{\bf Ba} \rightarrow \pm \pi/2$) and $\tan^{2}\theta_{\bf Ba} \rightarrow {\rm MAX}\{ \tan^{2}\theta_{\bf Ba},\,  \tildeCoeffTSv\,(c_{s}/\driftvelXmag)^{2} \}$ (as $\theta_{\bf Ba} \rightarrow 0,\,\pi$).

\end{itemize}

\vspace{-0.1cm}
\subsubsection{The Alfv\'en-Wave RDI}
\label{sec:mhdwave.rdi:growthrates.midk:Alfv\'en}

If instead the resonance is with the Alfv\'en phase speed ($\driftvel\cdot{\bf k} = \pm v_{A}$), the character of the modes is significantly different in some regimes. 
Also, the mid-$k$ Alfv\'en RDI vanishes entirely if $\tau=0$ because it depends on the presence of Lorentz forces on the dust. As before, the general expression is rather unintuitive so we give only the  limits of weak and strong Lorentz forces.

\begin{itemize}

\item{\bf Weak Lorentz Forces} ($\tau \ll 1$): Here, the fastest-growing modes have $\hat{k}_{y}\rightarrow \pm 1$ (with a non-zero but very small projection onto the $\hat{\bf B}_{0}-\driftvel$ plane), giving:  
\begin{align}
\label{eqn:Alfv\'en.rdi.midk.lotau}\mathcal{F}_{\rm mid} &= \frac{\driftvelXmag}{v_{f,\,0}}\,\frac{(1+\beta)^{1/4}\,|\tau\,\sin\theta_{\bf Ba}\,\cos\theta_{\bf Ba}|^{1/2}}{\sqrt{2}}
\end{align}
We see this vanishes as $\tau\rightarrow 0$, unlike the magnetosonic RDI, so at very low $\tau$ this is never the fastest-growing mode. However, also note that this expression applies for {\em all} $\driftvelXmag$ (not just high or low $\driftvelXmag$). Comparing to the magnetosonic modes (Eq.~\ref{eqn:msonic.rdi.midk.lotau}), in the ``fast'' limit (where $\driftvelXmag \gg v_{f,\,0}$), this differs from the fast-RDI by a factor $\sim (\driftvelXmag/v_{f,\,0})\,(1+\beta)^{1/4}\,\tau^{1/2}$, so if $\tau$ is not {\em too} small, the Alfv\'en-wave RDI can be the fastest-growing mode in the system for sufficiently large $\driftvelmag$ or $\beta$. In the ``slow'' limit ($\driftvelXmag \ll v_{f,\,0}$) the growth rate scales similarly to the slow RDI, but with an additional factor $\sim \tau^{1/2}\,\beta^{1/4}\,(1+\beta)^{1/4}$ --- so for sufficiently large $\beta \gg 1/\tau$ this can again be the fastest-growing mode. 

\item{\bf Strong Lorentz Forces} ($\tau \gg 1$): In this regime, the fastest-growing modes have $\hat{k}_{y}=0$ (oriented in the $\hat{\bf B}_{0}-\driftvel$ plane), giving:
\begin{align}
\label{eqn:Alfv\'en.rdi.midk.hitau}\mathcal{F}_{\rm mid} &= \frac{\driftvelXmag}{v_{f,\,0}}\,\frac{(1+\beta)^{1/4}\,|\sin\theta_{\bf Ba}\,\cos\theta_{\bf Ba}|^{1/2}}{\tau^{1/2}\,\sqrt{2}}\,\left| s^{\pm} + \frac{\driftvelmag\,\tan^{2}\theta_{\bf Ba}\,}{\tildeCoeffTSv\,|\driftvelmag-v_{A}|} \right|^{1/2},
\end{align}
where $s^{\pm} = 1$ if $\driftvelmag > v_{A}$ and $s^{\pm}=-1$ if $\driftvelmag < v_{A}$. Note that this is suppressed by a power $\tau^{1/2}$. In the ``fast'' limit that suppression means this is usually slower-growing than the fast RDI (Eq.~\ref{eqn:msonic.rdi.midk.hitau}) if $\beta$ is also large; the growth rate of the Alfv\'en RDI in this limit then differs by a factor $\sim (\driftvelmag/\tau\,v_{A})^{1/2}$ so for sufficiently large $\driftvelmag \gg \tau\,v_{A}$ could still be fastest-growing (but this is usually not the case). 
In the ``intermediate'' (strongly or weakly-magnetized) or ``slow'' limits, this is very similar to the slow-RDI.

\end{itemize}

\vspace{-0.5cm}
\subsection{Growth Rates: The Short-Wavelength (High-$k$) MHD-Wave RDI Modes}
\label{sec:mhdwave.rdi:growthrates.hik}

At sufficiently short wavelengths (high $k$), we can expand the dispersion relation in powers of $k^{-1} \ll 1$. If we do this, and once again assume the resonance condition $\driftvel\cdot \hat{\bf k} = v_{\pm}$ or $v_{A}$, we obtain the leading-order dispersion relation $\omega_{\pm,A} = k\,v_{\pm,A} + \omegaZ_{\pm,A} + \mathcal{O}(k^{0})$, where $\omegaZ_{\pm,A}^{3} = \mathcal{Q}_{\pm,A}\,\mu\,k\,v_{f,\,0}/(2\,\langle t_{s} \rangle)$ so $\omegaZ \sim \mathcal{O}(k^{1/3})$. Here $\mathcal{Q}$ is a real number, so this always has unstable ($\Im(\omega)>0$) solutions unless $\| \mathcal{Q} \| = 0$ exactly. We can therefore write,
\begin{align}
\label{eqn:hik.growth.rates} \omega_{\pm,A} &= k\,v_{\pm,A} + \left(\frac{i\sqrt{3}\pm 1}{2} \right) \,\mathcal{F}_{\rm hi}\,\left| \frac{\mu\,k\,v_{f,\,0}}{2\,\langle t_{s} \rangle^{2}} \right|^{1/3} + \mathcal{O}(k^{0}) \, ,
\end{align}
where $\mathcal{F}_{\rm hi} = \| \mathcal{Q} \|^{1/3}$ and the sign of the $\pm$ is opposite the sign of $\mathcal{Q}$. 
As discussed in \S~\ref{subsub:transition.between.mid.and.high.k} and in detail in Appendix~\ref{sec:matrix resonances}, Eq.~\ref{eqn:hik.growth.rates} is generally valid for sufficiently high $k$ such that the perturbation to the growth rate (i.e., $\omega_{\pm,A} - k\,v_{\pm,A}$) is larger than $\sim \mathrm{MAX}(\langle t_{s}\rangle^{-1}, \langle t_{L}\rangle^{-1}) = \mathrm{MAX}(1,\tau) \langle t_{s}\rangle^{-1}$.
As before, we consider $\mathcal{F}_{\rm hi}$ for the fast and slow magnetosonic, or the Alfv\'en RDIs.

\vspace{-0.1cm}
\subsubsection{The (Fast \&\ Slow) Magnetosonic-Wave RDI}
\label{sec:mhdwave.rdi:growthrates.hik:magnetosonic}

As before (\S~\ref{sec:mhdwave.rdi:growthrates.midk:magnetosonic}), we first consider the magnetosonic RDI. Even with this specification, the full expression for $\mathcal{Q}$ is again rather opaque,\footnote{In full:
\begin{align}
\nonumber \mathcal{Q} &\equiv \frac{q_{0}\,(\hat{\bf k}\cdot \driftvelhat) + q_{1}\,[\tau\,\hat{\bf k} \cdot ( \driftvelhat \times \hat{\bf B}_{0} )] + q_{2}\,[\tau\,\hat{\bf k} \cdot ( \driftvelhat \times \hat{\bf B}_{0} )]^{2}}{v_{\pm}\,\left[ v_{\pm}^{2}\,(1+\beta_{i}) - \beta_{i}\, (\hat{\bf k}\cdot \hat{\bf B}_{0})^{2}\, (1+2\,v_{\pm}^{2} - \beta_{i}) \right]} \, ,\\ 
\nonumber q_{0} &\equiv v_{fs}^{2}\,\left[\hat{k}_{z}\,(\coeffTSrho-1-\hat{k}_{z}^{2}\,\coeffTSv) + \beta_{i}\,q_{0,\,a} \right] \\ 
\nonumber &+ \beta_{i}\,c_{Bk}^{2}\,\left[\hat{k}_{z}\,( 1-\coeffTSrho + \hat{k}_{z}^{2}\,\coeffTSv - c_{Bw}^{2}\,\coeffTSv\,\beta_{i} ) + c_{Bw}\,c_{Bk}\,(\coeffTSrho-1)\,\beta_{i} \right] \, , \\ 
\nonumber q_{0,\,a} &\equiv \hat{k}_{z}\,\left[\coeffTSrho-1 + \hat{k}_{z}^{2}\{c_{Bw}^{2}\,(\coeffTSrho-1)-\coeffTSv \} + \hat{k}_{x}^{2}\,(1-\coeffTSrho)\,(1-c_{Bw}^{2}) \right] \\
\nonumber &+ c_{Bw}\,c_{Bk}\,\left[1-\coeffTSrho + 2\,\hat{k}_{z}^{2}\,(\tildeCoeffTSv-\coeffTSrho) \right] \, , \\ 
\nonumber q_{1} &\equiv \beta_{i}\,c_{Bk}^{2}\,\left[\hat{k}_{z}\,(\hat{k}_{z}^{2}\,\coeffTSv -\coeffTLrho-\coeffTSrho) \right]  \\
\nonumber & + v_{fs}^{2}\left[\beta_{i}\,c_{Bw}\,c_{Bk}\{\hat{k}_{z}^{2}\,(1+\tildeCoeffTSv-2\coeffTLrho-2\coeffTSrho) - \coeffTLrho-1 \} + \hat{k}_{z}\,q_{1,\,a} \right] \, , \\ 
\nonumber q_{1,\,a} &\equiv \coeffTLrho+\coeffTSrho + \beta_{i}\left[\coeffTLrho+\coeffTSrho+(1-c_{Bw}^{2})\,\hat{k}_{x}^{2}\,(\coeffTLrho+\coeffTSrho-1) \right] \\ 
\nonumber &+ \hat{k}_{z}^{2}\,\left[1 - \tildeCoeffTSv\,(1+\beta_{i}) + \beta_{i}\,\{1 + c_{Bw}^{2}(\coeffTLrho+\coeffTSrho-1)\} \right] \, , \\ 
\nonumber q_{2} &\equiv \frac{1}{2} \left[v_{fs}^{2}\{(2+\beta_{i})\,\tildecoeffTLrho\} + \beta_{i}\{\tildecoeffTLrho + (\tildecoeffTLrho+\coeffTLrho\,v_{fs}^{2})(2\,c_{Bk}^{2}-1) \}  \right] \, , 
\end{align}
where for brevity we denoted $c_{Bk}=\cos{\theta}_{\bf Bk}$, $c_{Bw}=\cos{\theta}_{\bf Bw}$. } 
so we will consider separately the limits of weak and strong Lorentz forces.

\begin{itemize}

\item{\bf Weak Lorentz Forces} ($\tau \ll 1$): In this case\footnote{The scaling shown for the slow-drift limit in Eq.~\ref{eqn:msonic.rdi.hik.lotau} assumes $\coeffTSv \approx 0$, which is applicable for sub-sonic Epstein or Stokes or Coulomb drag. Since the slow-mode resonance limit is (by definition) sub-sonic we have expanded assuming one of these laws is true. But if the scaling of the drag law were such that $\coeffTSv$ were significantly non-zero at order larger than $\driftvelmag/v_{f,\,0}$, then there is a less-strongly-suppressed term and the leading-order term for the slow limit in Eq.~\ref{eqn:msonic.rdi.hik.lotau} is only suppressed as $\sim (\coeffTSv\,\driftvelmag/v_{f,\,0})^{1/3}$, as opposed to $\sim (\driftvelmag/v_{f,\,0})^{2/3}$.} we obtain 
\begin{align}
\label{eqn:msonic.rdi.hik.lotau}\mathcal{F}_{\rm hi} &\approx  
\begin{cases}
{\displaystyle \left| \frac{(1-\coeffTSrho)\,(\beta\pm\sin^{2}\theta_{\bf Ba})}{\beta+1} \right|^{1/3}} &\ \hfill (\driftvelmag \gg v_{f,\,0}) \\
\\
{\displaystyle \left(\frac{\driftvelXmag}{v_{f,\,0}}\right)^{2/3}\,{ \frac{|(1-\coeffTSrho)\,\sin{\theta_{\bf Ba}}\,\cos{\theta_{\bf Ba}}|^{1/3}}{\beta^{1/6}} }} &\ \hfill (\driftvelmag \ll v_{f,\,0})
\end{cases}  
\end{align}

\item{\bf Strong Lorentz Forces} ($\tau \gg 1$): And in this case we obtain: 
\begin{align}
\label{eqn:msonic.rdi.hik.hitau}\mathcal{F}_{\rm hi} &\approx  
\begin{cases}
{\displaystyle \left(\frac{\driftvelXmag\,\sin\theta_{\bf Ba}}{v_{f,\,0}} \right)^{2/3}}\,(1+\coeffTLrho)^{1/3} &\ \hfill (\driftvelmag \gg v_{f,\,0}) \\
\\
{\displaystyle \left(\frac{\driftvelXmag}{v_{f,\,0}}\right)^{2/3}\,
	{ \frac{|(\coeffTLrho - \beta)\,\sin{\theta_{\bf Ba}}\,\cos{\theta_{\bf Ba}}|^{1/3}}{\beta^{1/6}} }} &\ \hfill (\driftvelmag \ll v_{f,\,0})
\end{cases}  
\end{align}

\end{itemize}

\vspace{-0.1cm}
\subsubsection{The Alfv\'en-Wave RDI}
\label{sec:mhdwave.rdi:growthrates.hik:Alfv\'en}

As with the mid-$k$ mode, the Alfv\'en RDI exhibits significantly different character (compared to the magnetosonic RDI) in the high-$k$ regime. 
The general expression is once again not particularly informative, although we note that for $\hat{k}_{y}=0$ (not necessarily the fastest-growing case), it simplifies dramatically to $\mathcal{Q} = \hat{k}_{z}\,(\driftvelmag/v_{f,\,0})\,\tau^{2}\,[1 - 2\,(\driftvelmag/v_{A})\,\cos\theta_{\bf Bw} + (\driftvelmag/v_{A})^{2}]$.
 It is
also worth noting that the high-$k$ Alfv\'en RDI does not vanish (if $\hat{k}_{y}\neq 0$) in the limit of uncharged grains ($\tau=0$), unlike its mid-$k$ cousin.


\begin{itemize}

\item{\bf Weak Lorentz Forces} ($\tau \ll 1$): Here the growth rate vanishes if any single component of $\hat{\bf k}$ does; the maximum growth rate occurs when $\hat{k}_{y}^{2} \approx1/3\rightarrow2/3$ (depending on $\beta$). To simplify the expression, take $\hat{k}_{y}^{2}=2/3$ (the effect of changing to $\hat{k}_{y}^{2}=1/3$ is less than a factor $2$ here for all $\beta$). Then the growth rate becomes:
\begin{align}
\label{eqn:Alfv\'en.rdi.hik.lotau}\mathcal{F}_{\rm hi} &\approx \frac{(\sin\theta_{\bf Ba})^{1/2}\,| \coeffTSv|^{1/3}}{(\sqrt{3}/2)}\,{\rm min}\left\{ \frac{\driftvelmag}{v_{f,\,0}} \, , \, \frac{(1+\beta)^{-1/2}}{1-(\sin^{2}\theta_{\bf Ba})/3} \right\}^{1/3}
\end{align}
As we noted above, we see that this is independent of $\tau$, so the instability (unlike the mid-$k$ Alfv\'en wave RDI) does not vanish in the limit of uncharged 
grains ($\tau=0$). Its growth  does, however, rely on the velocity dependence of the dust drag law, since it is proportional to  $\coeffTSv=0$.

Also note that, unlike the magnetosonic RDI (where the fast and slow RDI had different scalings), the expression above applies for {\em all} values of $\driftvelXmag$. So for the ``fast'' limit ($\driftvelmag \gg v_{f,\,0}$), $\mathcal{F}_{\rm hi}$ scales very similarly to the fast magnetosonic RDI with low-$\tau$ (Eq.~\ref{eqn:msonic.rdi.hik.lotau}). Although this differs from Eq.~\ref{eqn:Alfv\'en.rdi.hik.lotau} in this limit only by order-unity constants and the scaling coefficients $\zeta$, as noted in \S~\ref{sec:scalings:zeta} below, if Epstein drag dominates (as it usually does at high-$k$) then $1-\coeffTSrho$ (which appears in Eq.~\ref{eqn:msonic.rdi.hik.lotau}) scales $\propto (c_{s}/\driftvelmag)^{2}$ for $\driftvelmag \gg c_{s}$. So the fast RDI has a somewhat suppressed growth rate in this limit, while the Alfv\'en RDI (whose pre-factor $\coeffTSv\approx 1$ for Epstein drag with $\driftvelmag \gg c_{s}$) is not suppressed by any power of $\driftvelmag$. 

In the intermediate, strongly-magnetized limit ($c_{s} \ll \driftvelmag \ll v_{A}$), where the fast RDI is not possible, however, Eq.~\ref{eqn:Alfv\'en.rdi.hik.lotau} scales as $\sim (\driftvelmag/v_{f,\,0})^{1/3}$, so can lead to a larger growth rate than the slow-RDI (Eq.~\ref{eqn:msonic.rdi.hik.lotau}) by a factor $\sim (v_{f,\,0}/\driftvelmag)^{1/3} \gg 1$. In other words, the growth rate is less strongly suppressed for the Alfv\'en RDI than for the slow RDI. This is because (as discussed in \S~\ref{sec:resonant.mode.angles}), in this particular limit the resonance with the Alfv\'en wave has much higher phase velocity  than the resonance with the slow wave because $v_{A} \gg c_{s}$. 

For the intermediate, weakly-magnetized ($v_{A} \ll \driftvelmag \ll c_{s}$) or ``slow'' limits ($\driftvelmag \ll {\rm min}\{c_{s},\,v_{A}\}$), we have $\coeffTSv \propto (\driftvelmag/c_{s})^{2}\rightarrow (\driftvelmag/c_{s})^{3} \ll 1$ (with the exponent $=2$ for Epstein/Stokes drag or $=3$ for Coulomb drag). Thus $\mathcal{F}_{\rm hi} \propto (\driftvelmag / c_{s})^{\alpha}$ with $\alpha=1$ or $4/3$, and the growth rates of the Alfv\'en RDI are suppressed relative to the slow RDI.

\item{\bf Strong Lorentz Forces} ($\tau \gg 1$): In this limit, we find:
\begin{align}
\label{eqn:Alfv\'en.rdi.hik.hitau}\mathcal{F}_{\rm hi} &= \left| \frac{\driftvelXmag}{v_{f,\,0}}\,\,\tau\,\sin\theta_{\bf Ba}\, \left(1 - \frac{\driftvelXmag\,\cos{\theta}_{\bf Ba}}{v_{A}} \right) \right|^{1/3}.
\end{align}
Again, this expression applies for {\em all} $\driftvelXmag$. This means that in the ``fast'' case ($\driftvelXmag \gg v_{A}$ or $v_{f,\,0}$), the Alfv\'en RDI has a growth rate that scales $\propto [\tau\,\driftvelXmag^{2}/(v_{f,\,0}\,v_{A})]^{1/3}$ --- dimensionally, this is larger than the fast-magnetosonic RDI (Eq.~\ref{eqn:msonic.rdi.hik.hitau}) by a factor $\sim \tau^{1/3}\,(1+\beta)^{1/6}$. For ``intermediate'' cases with $v_{A} \ll \driftvelmag \ll c_{s}\sim v_{f,\,0}$, the fast-magnetosonic RDI is not possible and the Alfv\'en RDI has a growth rate faster than the slow-magnetosonic RDI by a factor $\sim \tau^{1/3}$; for ``slow'' cases with $\driftvelmag \ll {\rm min}\{c_{s},\,v_{A}\}$, the Alfv\'en RDI has a growth rate larger than the slow-RDI by a factor $\sim \tau^{1/3}\,(v_{f,\,0}/\driftvelXmag)^{1/3}$. So at sufficiently high-$\tau$, this is usually the fastest-growing mode. Effectively, in these cases, $\langle t_{s}\rangle^{2}$ in the denominator of the growth rate (Eq.~\ref{eqn:hik.growth.rates}) is replaced by $\langle t_{s} \rangle\,\langle t_{L} \rangle$ (the geometric mean). This implies that mode growth timescales can become comparable to the gyro timescale, rather than  the (slower) stopping time. The limitation is that this large enhancement in the growth rate only occurs at very short wavelengths, $k \gtrsim \tau^{3}/(v_{f,\,0}\,\langle t_{s} \rangle)$.

\end{itemize}

\vspace{-0.5cm}
\subsection{Mode Structure}
\label{sec:mhd.wave.mode.structure}

In this section, we briefly discuss the structure of the MHD-wave RDI modes. 

\begin{itemize}

\item{\bf Fast-Wave RDI}: Around resonance, the phase velocity is approximately  that of a fast-magnetosonic wave ${\bf v}_{p}\approx {\bf v}_{+}$. Like we saw for the acoustic RDI in \papertwo, the {\em gas} perturbation $(\delta \rho,\,\delta{\bf B},\,\delta{\bf u})$ increasingly resembles a simple, pure fast-magnetosonic wave for larger $k$. Also like the acoustic RDI, the perturbed dust velocity $\delta{\bf v}$ is smaller (by $\sim \hat{\mu}^{1/2}$) and out-of-phase with $\delta{\bf u}$ (leading by $\sim 150\degr$)  while the dust density $\delta\rho_{d}$ leads $\delta\rho$ by $\sim 30\degr$ with a very large amplitude: $|\delta\rho_{d}| / |\mu\,\delta\rho| \sim \Im(\omega\,\langle t_{s} \rangle) / \mu$, which translates to $|\delta\rho_{d}| / |\mu\,\delta\rho|  \sim (k\,v_{f,\,0}\,\langle t_{s} \rangle/\mu)^{1/2}$ in the  mid-$k$ regime, and $|\delta\rho_{d}| / |\mu\,\delta\rho|\sim (k\,v_{f,\,0}\,\langle t_{s} \rangle/\mu^{2})^{1/3}$  in the high-$k$ regime. At lower $k$, the deviation from the simple structure above is more pronounced (owing to transverse components of $\driftvel$ as well as non-zero $\tau$), but these are not essential to the mode dynamics.

Qualitatively, like the acoustic RDI, the gas density peak generated by the leading fast wave decelerates the dust, generating a ``pileup.'' But this dust over-density, in turn, pushes on the gas. Because the mode is traveling in the direction $\hat{\bf k}$ with a phase velocity matched to the dust drift in that direction, these effects add coherently (if we imagine moving with a Lagrangian dust ``patch''), generating rapid growth of the instability.

\item{\bf Slow-Wave RDI}: Here the phase velocity is that of a slow wave ${\bf v}_{p}\approx {\bf v}_{-}$, and again, at short wavelengths the gas perturbation closely resembles a perturbed slow-magnetosonic wave. The dust velocity perturbations $\delta{\bf v}$ are close to in-phase with $\delta{\bf u}$ (with a somewhat smaller amplitude), and both $\delta{\bf v}$ and $\delta{\bf u}$ are primarily confined to the $\hat{\bf B}_{0}-\hat{\bf k}$ plane. The phase offset between dust and gas density perturbations is similar to the fast-magnetosonic RDI ($\sim 30\degr$) but has opposite sign, generating an even stronger proportional dust-density fluctuation, with $\delta\rho_{d} / \mu\,\delta\rho$ scaling similar to the acoustic/fast case but with an extra power $\sim (\driftvelmag/v_{f,\,0})^{-1/4}$ when $\driftvelmag \ll v_{f,\,0}$. The details therefore differ, but the qualitative scenario is quite similar to the fast mode.

\item{\bf Alfv\'en-Wave RDI}: The phase velocity is that of an Alfv\'en wave ${\bf v}_{p}\approx {\bf v}_{A}$, and at high $k$ the gas perturbation also resembles a perturbed Alfv\'en wave. For the fastest-growing mode, usually in the $\hat{\bf B}_{0}-\driftvel$ or $\hat{x}-\hat{z}$ plane, the gas perturbation is primarily an incompressible mode with $\delta{\bf u}$ and $\delta{\bf B}$ in the $\hat{y}$ direction ($180\degr$ out-of-phase). The dust introduces some weak compressibility to the gas (and $\delta{\bf u}$ and $\delta{\bf B}$ terms in the $\hat{x}-\hat{z}$ plane) but these are small if $\mu\ll 1$ (suppressed by $\sim \mu$). The dust velocity perturbation $\delta{\bf v}$ is also primarily in the same direction ($\hat{y}$) with somewhat smaller amplitude than $\delta{\bf u}$ (leading $\delta{\bf u}$ by $\sim 30\degr$); however, the compressive  (non-transverse) components of the dust velocity perturbation are not strongly suppressed (they are  smaller than $\delta {\bf v}_{y}$ by only a modest factor), so the dust still features a large density perturbation $\delta \rho_{d}$. This, like the mode growth rate in the mid-$k$ regime,
depends on the existence of non-vanishing Lorentz forces on dust, which couple the transverse ${\bf B}$-field perturbations to longitudinal dust velocity perturbations: over much of the interesting parameter space, we can approximate $|\delta\rho_{d}| \approx \mu\,\rho\,(|\delta B_{y}|/\sqrt{\rho\,c_{s}^{2}})\,[k\,c_{s}\,\langle t_{L} \rangle / \mu^{2}]^{1/3}$ (and note $\delta\rho_{d}$ is out-of-phase with $\delta {\bf B}_{y}$ by $\sim 150\degr$). So the perturbation is suppressed (like the growth rate) at low-$\tau$, and scales with $\delta{\bf B}$ via $t_{L}$, $\mu$, and $\driftvelmag$ (as opposed to $t_{s}$ or $\beta$). In the high-$k$ limit, the velocity dependence
of the drag law (parameterized through $\coeffTSv$) is able to provide the necessary coupling of the transverse gas velocity perturbations to longitudinal dust perturbations, and the mode is still able to grow even if $\tau=0$.

Essentially, this instability amounts to a similar ``pileup'' of dust pushing back on the gas, adding coherently because the phase velocity of the gas wave matches the dust drift in the same direction. However the ``pushing'' is mediated by the magnetic fields and Lorentz forces (the $\delta{\bf B}$ and dust fluctuations interacting), instead of gas pressure and aerodynamic/Coulomb drag.
	
\end{itemize}

\vspace{-0.5cm}
\section{The Gyro-Resonances}
\label{sec:gyro}

\subsection{Overview}
\label{sec:gyro:overview}

In the previous section, we considered resonances between ``advective''  dust mode(s) --- i.e., a mode with frequency  $\omega_{d} \approx {\bf k}\cdot \driftvel$ --- and the different gas modes (Alfv\'en, slow, and fast) with frequency $\omega_{g}$. The resulting RDIs are unstable over a wide range of wavelengths. However, the dust is also affected by the magnetic field, which causes it to undergo gyro-motion, and the resonance between a gas mode and a dust gyro-mode leads to a new family of RDIs --- the \emph{gyro-resonant} RDIs. 

More specifically, in addition to the undamped advection mode (with $\omega_{d} = {\bf k}\cdot \driftvel$), there are three damped dust eigenmodes in Eq.~\ref{eqn:dust.natural.modes}. In general all three depend on $\langle t_{L} \rangle$ (or $\tau$), but if we consider large $\tau \gg 1$ (the case we will show is of relevance below) or special angles (e.g., where $\cos^{2}{\theta_{\bf Bw}}=1$), then they separate into $\omega_{d} \approx {\bf k}\cdot\driftvel - i\,\langle t_{s}\rangle^{-1}\,(\cos^{2}{\theta}_{\bf Bw} + \coeffTSv)$ and $\omega_{d} \approx {\bf k}\cdot\driftvel \pm \langle t_{L} \rangle^{-1} -i\,\langle t_{s}\rangle^{-1}\,(3 - \cos^{2}{\theta}_{\bf Bw})/2$. 
As explained in Appendix~\ref{sec:matrix resonances}, the first mode, 
with $\omega_{d} \approx {\bf k}\cdot\driftvel - i\,\langle t_{s}\rangle^{-1}\,(\cos^{2}{\theta}_{\bf Bw} + \coeffTSv)$, 
leads to the high-$k$ regime of the standard RDI (see \S~\ref{subsub:transition.between.mid.and.high.k}), but
does not allow for any distinct resonances. Because it is damped, it can never exactly match the 
resonant condition with the gas, and so is only relevant once its  imaginary part (the damping) is sufficiently small compared to other terms; i.e., at high $k$, once $\Im(\omega)\gtrsim \langle t_{L}\rangle^{-1}$ (see App.~\ref{sub:dust.gas.res.in.detail}).
 This mode is present in the acoustic RDI as well (with $\cos^{2}\theta_{\bf Bw}=1$; see \paperone\ and \papertwo).

However, now let us consider the $\omega_{d} \approx {\bf k}\cdot\driftvel \pm \langle t_{L} \rangle^{-1} -i\,\langle t_{s}\rangle^{-1}\,(3 - \cos^{2}{\theta}_{\bf Bw})/2$ modes, which involve damped dust gyro-motion. For the same reason as above, these can only approximately satisfy the resonance condition when the damping term is small compared to the other terms, i.e., when $|{\bf k} \cdot \driftvel \pm \langle t_{L} \rangle^{-1} | \approx \omega_{g} \gg | \langle t_{s}\rangle^{-1}\,(3 - \cos^{2}{\theta}_{\bf Bw})/2 |$. If  also $|{\bf k} \cdot \driftvel| \gg \langle t_{L} \rangle^{-1}$, then the $\langle t_{L} \rangle^{-1}$ term is also sub-leading, the mode again reduces to an advection mode with $\omega_{d} \approx {\bf k} \cdot \driftvel$, and we simply recover the high-$k$ limit of the standard MHD-mode RDIs. However,  if $\langle t_{L} \rangle^{-1} \gg |{\bf k} \cdot \driftvel| $ and  $\langle t_{L} \rangle^{-1} \gg | \langle t_{s}\rangle^{-1}\,(3 - \cos^{2}{\theta}_{\bf Bw})/2 |$  (i.e., $\tau \gg 1$), the  resonance condition becomes 
\begin{align}
\label{eqn:gyro.condition} \omega_{d}^{\rm gyro} \approx \driftvel\cdot {\bf k} \pm \langle t_{L} \rangle^{-1}  = \pm ({\bf v}_{A},\,{\bf v}_{+},\,{\bf v}_{-})\cdot {\bf k} \, .
\end{align}
This  condition, which relies on the dust gyro-motion, has a fundamentally different character from the Alfv\'en and fast/slow RDIs above. There, if the resonant condition was satisfied at a given angle $\hat{\bf k}$, it was satisfied for all $k = |{\bf k}|$. For the gyroresonances, however, at a given angle or phase velocity $v_{p} = |{\bf v}_{p}(\hat{\bf k})|$ the condition is only satisfied around a particular $k$, namely $k^{-1} \approx \pm v_{p}(\hat{\bf k})\,\langle t_{L} \rangle$. The resonances are sharply peaked in $k$, with a specific maximum growth rate, owing to the fact that they involve a resonance with a mode of fixed physical frequency (here, the gyro frequency). This makes them much more akin to the \BV\ RDI or the epicyclic RDIs studied in \paperone\ and in detail in \citet{squire:rdi.ppd}.

\vspace{-0.5cm}
\subsection{Resonant Wavelengths and Angles}
\label{sec:gyro:resonance}

If one chooses a wave-family (Alfv\'en, slow, fast), so $v_{p}(\hat{\bf k}) = v_{A,\,+,\,-}(\hat{\bf k})$ and angle $\hat{\bf k}$, then it is trivial to solve Eq.~\ref{eqn:gyro.condition} for the wavenumber $k$ at which the gyro-resonance will occur: 
\begin{align}
	\label{eqn:k.gyro} k_{\rm gyro}^{-1} = \langle t_{L} \rangle\,\left| \driftvel\cdot\hat{\bf k} \pm v_{p}(\hat{\bf k}) \right|
\end{align}
Equivalently, we can invert this to solve for the resonant angles $\hat{\bf k}=\hat{\bf k}_{\rm gyro}$ at a given $k=k_{\rm gyro}$. Since $|\driftvel\cdot\hat{\bf k} \pm v_{p}(\hat{\bf k})| \le \driftvelmag + v_{f,\,0}$, at sufficiently low $k$ ($k^{-1} \gtrsim \langle t_{L} \rangle\,(\driftvelmag + v_{f,\,0})$), Eq.~\ref{eqn:gyro.condition} cannot be satisfied for any MHD wave family or angle, and no gyro-resonances exist. If $k_{\rm gyro}$ is sufficiently high, as noted above, Eq.~\ref{eqn:gyro.condition} becomes $\driftvel\cdot {\bf k} \approx \omega_{g}$ and the gyro-RDI modes are degenerate with the MHD-wave RDIs; this occurs when $|\driftvel\cdot{\bf k}| \sim v_{p}\,k_{\rm gyro} \gg \langle t_{L} \rangle^{-1}$, i.e., when $v_{p}/|\driftvel\cdot\hat{\bf k} \pm v_{p}| \gg 1$. Finally, if $|\driftvel\cdot{\bf k} \pm \langle t_{L} \rangle^{-1}| \sim v_{p}\,k_{\rm gyro} \lesssim (3-\cos^{2}\theta_{\bf Bw})/2\,\langle t_{s} \rangle$ (if, e.g., the $\driftvel \cdot {\bf k}$ and $\langle t_{L} \rangle^{-1}$ terms cancel, so $|\omega_{g}|$ is small), then the damping term ($-i\,\langle t_{s}\rangle^{-1}\,(3 - \cos^{2}{\theta}_{\bf Bw})/2$ above) is not small compared to the other terms in the equation, and the RDI condition is not actually valid. Thus instability requires $v_{p}/|\driftvel\cdot\hat{\bf k} \pm v_{p}| \gtrsim (3-\cos^{2}\theta_{\bf Bw})/2\,\tau$. These conditions can only be simultaneously satisfied when $\tau\gtrsim 1$. Therefore, even though Eq.~\ref{eqn:gyro.condition} is actually four differently-signed equations (each pair of $\pm$ terms being independent) for each branch (Alfv\'en, slow, fast) of $\omega_{g}$, not all of these produce interesting instabilities. Generally the ``interesting'' gyro-RDI branches occur only when $\langle t_{L} \rangle^{-1} \gtrsim |\omega_{g}| \gtrsim \langle t_{s}\rangle^{-1}$, reducing the number of interesting and unique gyro-RDI branches to 6 (two for each wave family corresponding to $v_{p}(\hat{\bf k})\,k \approx \pm \langle t_{L} \rangle^{-1}$) at intermediate $k$.


For the fast-gyro RDI, the fact that $v_{p} \approx v_{f,\,0}$ is only weakly dependent on angle means that the resonance condition is simple. For $|\driftvel\cdot\hat{\bf k}| \lesssim v_{f,\,0}$, the resonance condition can only be satisfied around a narrow range of wavenumbers: $k_{\rm gyro}^{-1} \approx \langle t_{L}\rangle v_{f,\,0}$ (nearly independently of angle). For $|\driftvel\cdot\hat{\bf k}| \gg v_{f,\,0}$, one finds $k_{\rm gyro}^{-1}\approx \langle t_{L} \rangle\,|\driftvel\cdot\hat{\bf k}|$, so the ``resonant angle'' is given by $\cos{\theta}_{\bf wk} = \driftvelhat\cdot\hat{\bf k} \approx \pm 1/(\langle t_{L} \rangle\,\driftvelmag\,k)$.\footnote{Note when $\driftvel\cdot{\bf k} \approx v_{p}(\hat{\bf k})$, the ``$+$'' branch of Eq.~\ref{eqn:k.gyro} just gives a similar solution to when $\driftvel\cdot{\bf k} \ne v_{p}(\hat{\bf k})$, while the ``$-$'' branch nearly cancels the two and produces very large $k_{\rm gyro}$. This, however, is just the limit where the gyro-RDI becomes degenerate with the MHD-wave RDI.}

The Alfv\'en-gyro RDI satisfies ${v}_{p} = v_{A}\,\cos{\theta}_{\bf Bk}$, so if $|v_{p}| \gtrsim |\driftvel\cdot\hat{\bf k}|$ then for all $k \ge 1/(v_{A}\,\langle t_{L} \rangle)$, the resonant angles are given by $\cos{\theta}_{\bf Bk} = \hat{\bf k}\cdot\hat{\bf B}_{0} \approx \pm 1/(\langle t_{L} \rangle\,v_{A}\,k)$. If $|\driftvel\cdot\hat{\bf k}| \gg v_{p}$, then the resonant angle is again just $\cos{\theta}_{\bf wk} \approx \pm 1/(\langle t_{L} \rangle\,\driftvelmag\,k)$.

The resonant angles of the slow-gyro RDI are similar to the Alfv\'en case, but with the slow phase velocity; thus, approximately, we can simply take $v_{A} \rightarrow {\rm MIN}(v_{A},\,c_{s})$ in the Alfv\'en expressions above.

Within the range of resonant angles, there is a fairly weak dependence of the growth rate on the particular angle chosen (or equivalently, on $k_{\rm gyro}$), barring the pathological cases above (where e.g., $k_{\rm gyro}\rightarrow\infty$). For this reason, we do not (as we did for the MHD-wave RDIs) attempt to estimate the fastest-growing resonant angle within the resonant branch.

\vspace{-0.5cm}
\subsection{Growth Rates \&\ (In)Stability Conditions}
\label{sec:gyro:growthrates}

Owing to the presence of the damping term discussed above, and the set of four resonant equations, exact expressions for the growth rates of the gyro-RDI at, for example,  high or low $k_{\rm gyro}$, are even more opaque than for the MHD-wave RDIs. However, since $\tau \gtrsim 1$ is required anyways for interesting behavior of these modes, if we assume $\tau \gg 1$ and expand the dispersion relation, the relevant behaviors become more clear.\footnote{Expanding at high-$\tau$, assuming $k_{\rm gyro} \propto \omega \propto \langle t_{L} \rangle^{-1} \propto \tau$ to leading order, we obtain the dispersion relation $0 = \tilde{\omega}^{2}\,\beta_{i}\,[ \{ 1-(\tilde{\omega}-a_{w})^{2}\}\,(\tilde{\omega}^{2}-a_{B}^{2})\,(\tilde{\omega}^{2}-a_{B}^{2}\,\hat{k}_{z}^{2}) + \mu\,(\tilde{\omega}-a_{w})^{2}\,\{ 2\,\tilde{\omega}^{2} + \mu\,(\tilde{\omega}-a_{w})^{2} - a_{B}^{2}\,(1+\hat{k}_{z}^{2})\}] + a_{B}^{2}\,[ 
\{ (\tilde{\omega}-a_{w})^{2} - 1\}\,(\tilde{\omega}^{2} - a_{B}^{2}\,\hat{k}_{z}^{2})^{2} - 
\mu\,(\tilde{\omega}-a_{w})^{2}\,\{ 
(1+\hat{k}_{z}^{2})\,(\tilde{\omega}^{2} - a_{B}^{2}\,\hat{k}_{z}^{2}) + \mu\,\hat{k}_{z}^{2}\,(\tilde{\omega}-a_{w})^{2}\}]$, where $\tilde{\omega}\equiv\omega\,\langle t_{L} \rangle$, $a_{w} \equiv \driftvel\cdot{\bf k}_{\rm gyro}\,\langle t_{L} \rangle$, and $a_{B}\equiv v_{A}\,k_{\rm gyro}\,\langle t_{L} \rangle$.}

A straightforward, but tedious, direct analysis of the dispersion relation at high-$\tau$ shows that when the dust-to-gas ratio $\mu$ is small, instability typically requires $\mu^{1/2}\,\tau \gtrsim 1$ and $\driftvelmag \sim v_{p}$ (or at least $\driftvelmag$ not too small compared to $v_{p}$; otherwise the damping terms in the dust eigenmode are not negligible (all $\driftvelmag$ are unstable, however, if $\mu \gtrsim 1$). If these conditions are met, then the growth rates are approximately given by 
\begin{align}
\nonumber \Im(\omega_{\rm gyro}) &\approx |\driftvelhat\cdot\hat{\bf k}|\,\frac{\mu^{1/2}\,k_{\rm gyro}\,\driftvelmag}{|k_{\rm gyro}\,\driftvelmag\,\langle t_{L} \rangle \pm 1|^{1/2}} \\ 
&= |\driftvelhat\cdot\hat{\bf k}|\,\frac{\mu^{1/2}\,\langle t_{L}\rangle^{-1}}{|1 \pm (1 \pm v_{p}(\hat{\bf k})/\driftvelmag)|^{1/2}}.\label{eqn:gyro.resonant.growth.rate}
\end{align}
So the fastest-growing gyro modes will tend to be those aligned with the dust drift ($\driftvelhat\cdot\hat{\bf k} = \pm 1$), but a wide range of angles and resonant $k_{\rm gyro}$ will have growth rates $\Im(\omega)\sim \mu^{1/2}/\langle t_{L} \rangle$. Equation~\ref{eqn:gyro.resonant.growth.rate} provides a reasonable approximation to the peak growth rates of the gyro modes in Figs.~\ref{fig:dispersion.relation}-\ref{fig:angle.1d}. 

 We stress that the instability requirements above are not a strong limit, but an approximate guide; there often exist mode angles where the gyro-RDIs are unstable despite significantly smaller $\driftvelmag$ or $\mu$ (evident in the slow-wave and Alfv\'en-wave resonance cases in Fig.~\ref{fig:dispersion.relation}). For example, if we expand in both large-$\tau$ and small-$\mu$, assuming $\omega\sim \mathcal{O}(\tau) + \mathcal{O}(\mu^{1/2}\,\tau) + ...$, we find that at the angle $\driftvelhat \cdot \hat{\bf k} = \pm  (k_{\rm gyro}\,|\driftvelmag - v_{A}|\,\langle t_{L} \rangle|)$, the term that usually stabilizes the instability  at small-$\mu$ vanishes,. This causes the gyro-RDI modes to be unstable so long as $\driftvelmag \ge v_{A}$ or $\driftvelmag \ge v_{A}/2$ (depending on which solution branch we consider), albeit with slightly different growth rates, $\Im(\omega_{\rm gyro}) \approx \mu^{1/2}\,\langle t_{L} \rangle^{-1}\,(\driftvelmag/v_{A}-1)^{1/2}/2$ or $\Im(\omega_{\rm gyro}) \approx \mu^{1/2}\,\langle t_{L} \rangle^{-1}\,(2\,\driftvelmag/v_{A}-1)^{1/2}/2$.

\vspace{-0.5cm}
\subsection{Mode Structure}
\label{sec:gyro:structure}

To leading order, the dust perturbation in ($\delta{\bf v},\, \delta\rho_{d}$) is incompressible gyro motion around $\hat{\bf B}_{0}$. The dust back-reaction produces a proportional $\delta{\bf B}$ perturbation ($|\delta{\bf B}| \sim (\rho/\beta)^{1/2}\,|\delta{\bf v}|$) which lags $\delta{\bf v}$ by a modest ($\sim 30\rightarrow 50\degr$) phase offset. This and the drag terms drive the gas into phase-lagged gyro motion and generate a compressible component (non-vanishing $\delta \rho_{d}$ and $\delta \rho$), but these are suppressed by (roughly) $\sim \mu^{1/2}$ and $\sim \mu$, respectively. 

However, the phase velocity is strongly modified from the pure gyro case (where it scales as $\driftvel\cdot{\bf k} \pm \langle t_{L} \rangle^{-1}$). It scales in a fairly complicated manner, but is of order the growth rate (smaller by another power of $\sim \mu^{1/4}$, very approximately).

\vspace{-0.5cm}
\section{Other Unstable Modes}
\label{sec:other.modes}

Recall the dispersion relation is 10th-order; at any given ${\bf k}$, there are typically $\sim 3-7$ unstable (growing) modes (i.e.,  branches of the dispersion relation). We will discuss the additional unstable modes only very briefly, because they either (a) have much smaller growth rates than those we discussed above, or (b) only appear in pathological situations. 

The additional modes include analogues of the out-of-resonance ``intermediate'' and ``slow'' modes from \papertwo; these are modes with phase velocities that (approximately) satisfy Eq.~\ref{eqn:unperturbed.modes.highk}, i.e., $v_{p} = (\driftvel\cdot \hat{\bf k},\,\pm {v}_{A},\,\pm {v}_{f},\,\pm {v}_{s})$ (matching {\em either} the dust drift, {\em or} the Alfv\'en/fast/slow-mode velocities at that ${\bf k}$). At  resonance, the mode satisfies several of these phase-velocity conditions {\em at the same time}, so a subset of the modes become degenerate and the growth rates become much larger. However, even out of resonance,  for any mode that approximately solves the  gas equation without dust, the additional corrections from the dust-gas-coupling usually lead to a positive growth rate --- i.e., the modes obey $\omega \approx v_{p}\,k + \iimag\,\mathcal{O}(\mu\,...)$,  where the growth rate is small but positive and non-zero (and usually scales as $\sim \mu$ for $\mu\ll1$, as expected from non-degenerate perturbation theory). Some of these are illustrated in Fig.~\ref{fig:dispersion.relation}, particularly in the bottom-right ``no resonance'' panel. Similarly, in Figs.~\ref{fig:angular.structure}--\ref{fig:angle.1d}, we  see that a broad range of angles away from resonance are still unstable. As discussed in \papertwo\ for the acoustic RDI, while certain combinations of the parameters $\coeffTSrho$, $\coeffTSv$, $\coeffTLrho$ can stabilize a subset of these modes out-of-resonance, there is invariably a different subset that is destabilized at the same time.

There also exist  some growing modes that are not directly related to the ``natural'' response of the un-coupled system (i.e., the uncoupled gas or dust modes), but these have very low growth rates at all $k$ (e.g., they are often suppressed by a factor of $\sim \mu\,k/(1+\tau^{2})$ at small $k$). 

Finally there are modes which can have very large growth rates but appear only for pathological parameter choices. For example, the ``decoupling'' mode from \papertwo\ is present here, if $\coeffTSv < -1$ (at low-$\tau$; at high-$\tau$ the requirement is approximately $\coeffTSv\,\cos^{2}\theta_{\bf Bw} < -1$). This large, negative $\coeffTSv$ means that as the relative dust-gas velocity increases, the total force between dust and gas rapidly becomes {\em weaker}. Thus if the dust begins to accelerate relative to the gas, the coupling becomes weaker and the two rapidly separate (formally the growth rate is $\sim |\tildeCoeffTSv|/\langle t_{s} \rangle$ at all $k$). But this physically is unlikely: as shown in \papertwo, while the scaling of $\coeffTSv$ for Coulomb drag with super-sonic $\driftvel$ formally produces this instability,  in that limit one should represent the drag via the sum of Coulomb and aerodynamic drag. The aerodynamic term (which becomes more tightly-coupled with higher drift velocity) will always dominate at large drift velocities. There are analogous instabilities that can appear with sufficiently negative $\coeffTSrho$ or $\coeffTLrho$, but we do not expect any physical dust-gas couplings to produce such values.

\vspace{-0.5cm}
\section{Scales Where Our Derivations Break Down}
\label{sec:scales.validity}

\subsection{Largest Wavelengths/Timescales} 

As discussed in \papertwo\ in detail, the scalings and derivations presented here are valid over some range of spatial and timescales. At sufficiently long wavelengths (low-$k$), the wavelengths $\lambda\sim 2\pi/k$ become comparable to some global gradient scale-length $L_{0}$ of the system, so a global solution (with appropriate boundary conditions) is obviously needed. However, as shown in \papertwo, stratification of the background does not alter the character of the modes here, so long as $k \gg L_{0}^{-1}$; i.e., so long as  we consider wavelengths short compared to the scale-length (we showed this in \papertwo\ for the acoustic RDI, but since the dimensional scaling of the growth rates and mode structure here is similar, the qualitative conclusions are identical). Likewise, if the mode growth time is comparable to (or longer than) the global evolution time of the system $t_{0}$, a global solution is needed. The relevant global evolution times can include, e.g., the timescale for the dust to drift ``through'' some global scale-length $L_{0}$, $t_{0} \sim L_{0}/\driftvelmag$. Obviously, these scales are problem-specific; we discuss them in various contexts of astrophysical applicability below (\S~\ref{sec:application}). Moreover, as discussed in \paperone\ and \citealt{squire:rdi.ppd}, if there are additional terms on large scales that need to be included (e.g., stratification, or centrifugal/coriolis forces, etc.) these almost always introduce new RDIs (even if they are unconditionally stable in a pure-gas medium), which may have faster growth rates at the largest wavelengths.

\vspace{-0.5cm}
\subsection{Smallest (Dissipation) Scales}

At sufficiently small wavelengths (high-$k$), dissipative effects in the gas (viscosity, conductivity etc.) will become important. A necessary condition for our results to be valid is that the gas wave associated with the specific RDI  in question (i.e., the Alfv\'en, slow, or fast wave) is not strongly affected by such effects. In a primarily neutral/molecular gas, the viscosity (or conductivity) is approximately isotropic, and so all waves will presumably be similarly damped, thus damping the RDIs for wavelengths  below the mean-free path scale,  $\lambda_{\rm mfp}\sim 10^{15}\,{\rm cm}\,(n_{\rm gas}/1\,{\rm cm^{-3}})^{-1}$.  However, in a magnetized   gas with a high ionization fraction,  when the mean-free path is larger than the ion gyro-radius, the viscosity and conductivity are  much larger parallel to the magnetic field than perpendicular to it \citep{Braginskii.plasma.transport}. This implies that while the slow and fast waves are damped on scales below the mean-free path,  $\lambda_{\rm mfp} \sim 10^{12}\,{\rm cm}\,(T/10^{4}\,{\rm K})^{2}\,(n_{\rm gas}/1\,{\rm cm^{-3}})^{-1}$, the Alfv\'en wave remains unmodified until its wavelength approaches ion gyro-radius scales, $\lambda_{\rm ion,gyro}\sim  10^7  {\rm cm}\,(n_{\rm gas}/1\,{\rm cm^{-3}})^{-1/2} \beta^{1/2}$. This suggests that in hot ionized plasmas, the Alfv\'en-wave RDIs (both the standard and gyro-resonant variants) can survive unmodified on  scales many orders of magnitude smaller than that where the slow and fast RDIs are damped by viscosity. This is illustrated in Fig.~\ref{fig: HII examples}, which shows the numerically calculated RDI growth rates for parameters typical of an  HII region (see \S~\ref{subsub: HII regions}), including the effect of a parallel 
(Braginskii) viscosity, $c_{s} \lambda_{\mathrm{mfp}} \nabla\cdot [\hat{\bf B}\hat{\bf B}\,(\hat{\bf B}\hat{\bf B}:\nabla{\bf u})]$, in the momentum equation. We see that the while the slow and fast RDIs  depart from their ideal scaling for $\lambda \lesssim \lambda_{\mathrm{mfp}}$, the Alfv\'en-wave RDI is unaffected because 
it  involves only perpendicular gas motions (aside from very small, $\sim \mu$, compressive corrections due to the dust; see \S~\ref{sec:mhd.wave.mode.structure}).

Another small-scale limit on our treatment arises because the fluid approximation for the dust will break down on scales comparable to the mean grain spacing, $\lambda_{\rm d,space} \sim 6000\,{\rm cm}\,(R_{d}/0.1\,\mu{\rm m})\,(n_{\rm gas}/1\,{\rm cm^{-3}})^{-1/3}\,(\mu/0.01)^{-1/3}$.  This is extremely small compared to the viscous (mean-free path) scales, and even (usually) the gyro-radius scales. 

\vspace{-0.5cm}
\subsection{Non-Ideal Effects}

Non-ideal MHD effects can  similarly cause our treatment to be invalid if ionization fractions ($f_{\rm ion}$) are sufficiently low. In the high-density, low-$f_{\rm ion}$ (strong coupling) limit, one can parameterize the dominant effects as additional terms in the induction equation: $\partial{\bf B}/\partial t = \nabla\times({\bf u}\times{\bf B}) - \nabla\times(\eta_{\rm Ohm}\,{\bf J}) - \nabla\times(\eta_{\rm Hall}\,{\bf J}\times\hat{\bf B}) - \nabla\times[\eta_{\rm AD}\,\hat{\bf B} \times ({\bf J}\times\hat{\bf B})]$ where ${\bf J}=\nabla\times{\bf B}$ is the current density and $\eta_{\rm Ohm}$, $\eta_{\rm Hall}$ and $\eta_{\rm AD}$ are the effective diffusivities for Ohmic resistivity, the Hall effect, and ambipolar diffusion. Ohmic resistivity will damp the slow and Alfv\'en modes, and modify the fast mode into a sound wave, above a wavenumber $k_{\rm Ohm}\sim v_{p}/\eta_{\rm Ohm}$. The Hall effect is not diffusive but modifies the Alfv\'en and slow wave dispersion relations for  $k\gtrsim k_{\rm Hall} \sim v_{p}/\eta_{\rm Hall}$ into a shear-Alfv\'en branch with $\omega_{g}\sim$\,constant and a Whistler branch with $\omega_{g} \propto k^{2}$. This in turn modifies the Alfv\'en and slow  RDIs into two new RDI families (see \citealt{squire:rdi.ppd}). However, for all astrophysical contexts in \S~\ref{sec:application} (except protoplanetary disks and planetary atmospheres) both $k_{\rm Ohm}$ and $k_{\rm Hall}$ correspond to wavelengths many orders of magnitude smaller than the other dissipative scales above.\footnote{Again in the low-ionization-fraction, high-density limit, we have $\eta_{\rm Ohm}\approx c^{2}\,m_{e}\,n_{t}\,\langle \sigma v\rangle_{e}/(4\pi\,e^{2}\,n_{e})$, $\eta_{\rm Hall} \approx B\,c/(2\pi^{1/2}\,e\,n_{e})$, and $\eta_{\rm AD}\approx B^{2}\,(m_{n}+m_{i})/(\langle \sigma v \rangle_{i}\,m_{i}\,m_{n}\,n_{i}\,n_{t})$ where $m_{e,i,n}$ are the electron/ion/neutral effective masses, $n_{e,i,t}$ the free electron/ion/ion+neutral number densities, and $\langle \sigma v\rangle_{e,i}$ the electron/ion-neutral collision rates. If we take typical scalings in primarily neutral (molecular/atomic) gas, this gives $k_{\rm Ohm} \sim 10^{20}\,(v_{f,\,0}\,\langle t_{s} \rangle)^{-1}\,f_{\rm ion}\,(n/{\rm cm^{-3}})^{-1}$ and $k_{\rm Hall} \sim 10^{16}\,(v_{f,\,0}\,\langle t_{s} \rangle)^{-1}\,f_{\rm ion}\,\beta^{1/2}\,(n/{\rm cm^{-3}})^{-1/2}$.} Likewise, effects of the current carried by grains themselves producing violations of ideal MHD are generally negligible.\footnote{The equations solved here (Eq.~\ref{eqn:general}) are valid if either (a) the timescale $t_{\rm charge}$ for a grain to reach local charge-exchange equilibrium with the plasma is short compared to the drag timescale $t_{s}$, and/or (b) the total charge carried by dust is negligible compared to that in ions/free electrons ($n_{\rm grain}\,|q_{\rm grain}| \ll e\,n_{e}$). Otherwise, we need to account for the current carried by grains themselves (e.g., modifying ${\bf u} \times {\bf B} \rightarrow [(n_{i}\,q_{i}\,{\bf u} + n_{\rm grain}\,q_{\rm grain}\,{\bf v}) \times {\bf B}]/(n_{e}\,e)$ in the induction equation). For the standard scalings of grain charge in \S~\ref{sec:stopping.time.scalings:lorentz}, $n_{\rm grain}\,|q_{\rm grain}|/(n_{e}\,e) \sim 10^{-10}\,(T/10^{4}\,K)\,(\mu/0.01)\,f_{\rm ion}^{-1}\,R_{0.1}^{-2}$, and (using the charging rates from \citealt{tielens:2005.book} and assuming Coulomb drag) $t_{\rm charge}/t_{s} \sim 10^{-6}\,(T/10^{4}\,K)\,R_{0.1}^{-2}$, so the grain-current corrections are indeed negligible, unless either $f_{\rm ion}$ is so low that non-ideal MHD effects would dominate anyway, or grains are so small ($\sim$\,\AA) that they cannot be described by a drag law in the first place.}

Ambipolar diffusion is more ambiguous: in primarily ionized gas it is negligible (the neutrals simply add to the effective weight of the ions). In primarily neutral gas (in the strong-coupling limit), it damps the Alfv\'en wave, but leaves the fast wave and $\hat{\bf B}$-parallel slow wave largely un-modified, while damping the slow wave (and modifying the fast wave to a sound wave) at field-perpendicular angles when $k \gtrsim k_{\rm AD} \sim v_{\rm slow} / \eta_{\rm AD}$ \citep{balsara:1996.mhd.waves.with.ambipolar.diffusion}. So the slow-RDI modes will be stabilized at large-$k$ if $\driftvelmag/v_{f,\,0}$ is sufficiently small (since the only resonant angles are nearly-perpendicular), while the fast-RDI modes are similar to those here\footnote{Alternatively (but equivalently), if we do not make a one-fluid approximation for ambipolar diffusion, but assume Epstein drag in molecular gas with $v_{p} \sim v_{f,\,0}$ (and $\sim 0.1\,\mu{\rm m}$ grains), then if $f_{\rm ion} \ll 10^{-6}\,(\driftvelmag/{\rm km\,s^{-1}})\,(\mu/0.01)$ (typical in dense GMC cores), the typical force (momentum flux) from ions on neutrals becomes much weaker than the forcing from the grains on neutrals, so we should essentially consider the RDI to be between the grains and neutral gas alone. The MHD-RDI then reduces to the un-damped acoustic RDI, eliminating the slow and Alfv\'en modes while leaving the fast mode as a sound wave.} (see also \citealt{squire:rdi.ppd}).

Finally, it is worth noting that in a well-ionized plasma, variants of  Alfv\'en RDIs could be unstable even below the ion gyro-scale, so long as there exist nearly undamped waves (e.g., kinetic-Alfv\'en-wave or Whistler-wave RDIs). Given the complexity of the dispersion relations of even the simplest kinetic plasma waves, and the very small scales on which such instabilities would be expected to operate, we shall not consider this  further here.

\vspace{-0.5cm}
\subsection{Random (Micro-physical) Grain Motions}

Our fluid approximation for the dust has assumed zero dust pressure (or temperature), \emph{viz.,}, we 
assume that, in equilibrium, all dust particles with a given size, charge, and in a given (local) spatial location move with the same local equilibrium velocity through the gas. As discussed in \paperone, 
a finite dust pressure, if it exists, causes the dust eigenmodes (in a fixed gas background) to be weakly damped because the 
dust-density eigenmode couples back to the dust bulk velocity (see App.~\ref{sec:matrix resonances}).
The damping rate is $\Im(\omega_{0,\mathrm{dust}}) \sim - c_{s,d}^{2}k^{2}\langle t_{s}\rangle$, 
where $c_{s,d}$ is the speed of the local random motions in the dust. For the same reason that the gyro-resonant mode 
is unstable only once $\Im(\omega)\gtrsim \langle t_{s}\rangle^{-1}$ (see \S~\ref{sec:gyro} and App.~\ref{sub: resonances dust eigenmodes}), an ideal RDI mode with growth rate $\Im(\omega)$ (as discussed through \S\S~\ref{sec:dispersion.relation}--\ref{sec:gyro}) will  be modified when $\Im(\omega)\langle t_{s}\rangle \lesssim \langle t_{s}\rangle|\Im(\omega_{0,\mathrm{dust}})| \sim (c_{s,d}/c_{s0})^{2} (k\,c_{s}\langle t_{s}\rangle)^{2}$. Note that the RDI modes will not necessarily be damped when this is the case, particularly at high $\tau$;
for instance, cosmic-ray instabilities (see \S~\ref{sub:cosmic.ray.streaming.mode}) are well-known to be unstable 
for more realistic and complex  cosmic-ray distributions (in fact, there are many more unstable modes than studied here; see, e.g., \citealt{kulsrud.1969:streaming.instability,wentzel.1969.streaming.instability}).

There are two general causes for random dust motions that produce an effective ``dust pressure'': (i) instabilities acting with growth times shorter than $t_{s}$, in which case they can grow before the system actually reaches the equilibrium discussed in \S~\ref{sec:equilibrium} (see, e.g., Fig.~\ref{fig:geometry} and \S~\ref{sub:cosmic.ray.streaming.mode}); or (ii) ``dust microphysics.'' The latter includes a number of effects not contained in our simple drag model (see, e.g., \citealt{draine.review.2004}). For example,  the effect of dust Brownian motion is straightforward to estimate and generally negligible: the root-mean-squared dust velocity is 
$c_{s,d}\sim 3 k_{B}T/m_{\mathrm{dust}}$, giving $(c_{s,d}/c_{s0})^{2}\sim 10^{-9}(R_{d}/0.1\,\mu\mathrm{m})^{-3}$. This low value implies that the RDIs will be affected only for very subsonic drift velocities or low $\mu$ (such that $\Im(\omega)\langle t_{s}\rangle$ is very small anyway),  very small scales (such that $k^{2}\,c_{s}^{2}\langle t_{s}\rangle^{2}$ is very large), or very small grains. Thus, other effects (e.g., gas viscosity, see above) will almost invariably be important before the Brownian motion of the grains becomes important. Other relevant dust microphysics includes photoelectric emission and photodesorption of atoms
from the dust grains, which would cause  random 
grain motions in a radiation field with a significant  isotropic component. The magnitude of these effects will depend on complicating factors such as the grain shape and molecular structure (see \citealt{weingertner.draine:photo.forces.on.dust.hard.ism.rad}), but will in general be more important for 
smaller grains, because a single electron or atom provides a proportionally larger kick to a smaller grain.

In addition, realistic grains will be drawn from a broad size distribution, and feature variations in their chemical compositions and shapes as well. This will in turn produce a range of dust drift velocities, Larmor times, and charge (even at fixed size, if composition and shape vary). While not truly random, these may have a similar effect to an effective ``dust pressure'' as above. They will likely ``smear out'' the resonances involving $\driftvelmag$ and $\tau$, potentially decreasing the maximum growth rates while increasing the range of conditions/angles where at least some sub-population of grains is near-resonance. However, it is also possible that new instabilities might appear in this regime, for example three-fluid resonances between gas, ``small'' grains, and ``large'' grains (or some other grain sub-populations). Exploring this regime will likely require simulations accounting for realistic size and charge distributions.

\vspace{-0.5cm}
\subsection{Non-Linear Effects}

If the system is strongly non-linear, or there is sufficiently sharp structure in the velocity or density fields, dust trajectories become self-intersecting and we cannot apply the fluid approximation to the dust. In this limit, numerical simulations must be used to integrate particle trajectories directly (as opposed to using a local fluid or ``terminal velocity'' approximation, sometimes seen in the literature but invalid in this limit).

%
%
\begin{figure*}
\begin{center}
\includegraphics[width=0.95\textwidth]{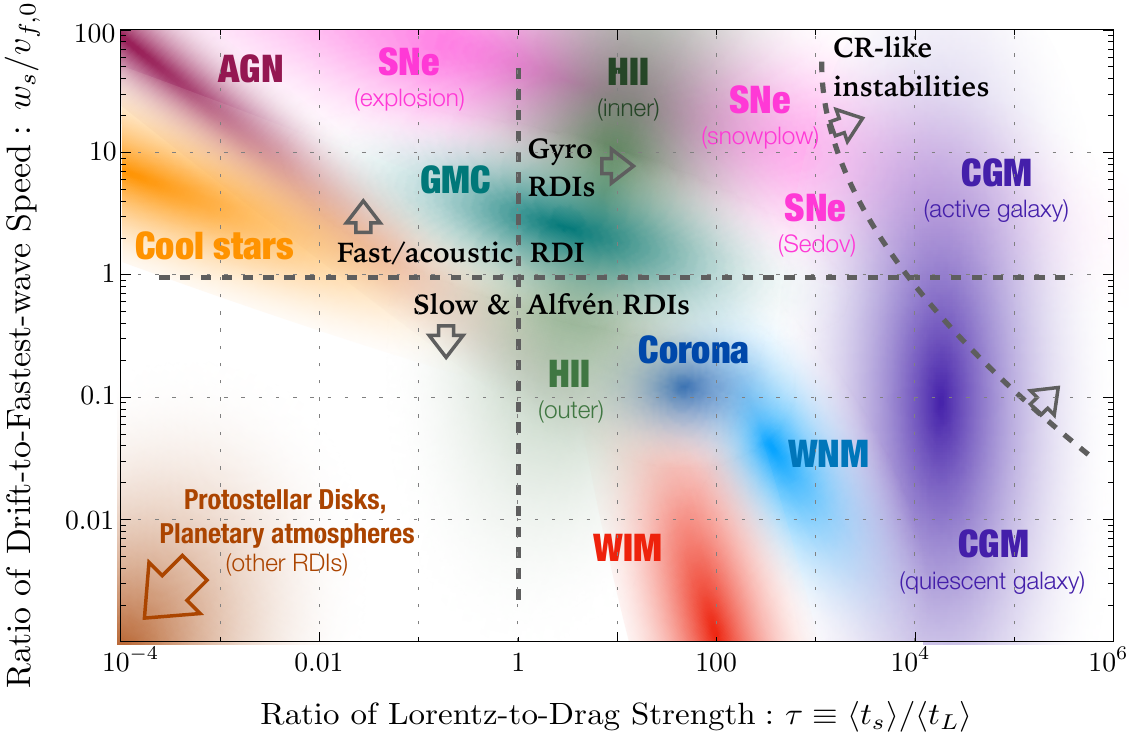}
\caption{Illustration of the two most important dimensionless parameters of the MHD RDI --- the dust drift speed normalized by the fast wave speed, $\driftvelmag/v_{f,0}$, and the ratio of stopping time to Larmor time, $\tau\equiv \langle t_s \rangle/\langle t_L \rangle$ ---   for a variety of different astrophysical environments, as discussed in the text (\S~\ref{sec:application}). As labeled on the figure, the different shaded regions illustrate: the warm-neutral medium (WNM, light blue; \S\ref{sec:application.env:wim.wnm}); the warm-ionized medium (WIM, red; \S\ref{sec:application.env:wim.wnm}); the circum-galactic and/or inter-galactic medium (CGM, purple; \S\ref{sec:application.env:cgm.igm}, around both quiescent or active/quasar-hosting galaxies); HII regions both near and far from the ionizing radiation source (HII, dark green; \S\ref{subsub: HII regions}); supernovae  in various phases of evolution (SNe, pink; \S\ref{sec:application.env:sne}); Solar coronal dust (blue; \S\ref{sec:application.env:coronae}); 
cool star photospheres and winds (orange; \S\ref{sec:application.env:coolstars}); giant molecular clouds and star-forming regions (GMC, sea green; \S\ref{sec:application.env:cold.ism}); active-galactic nucleii (AGN, maroon; see \papertwo\ \&\ \S\ref{sec:application.env:cold.ism}); and proto-stellar/planetary disks/atmospheres (brown; \S\ref{sec:application.env:ppd}), which extend off the plotted range (as labeled). We also illustrate, with the gray dotted lines and arrows, where different forms of the RDI should be unstable; the fast (acoustic) RDI is unstable for $\driftvelmag/v_{f,0}\gtrsim 1$, the gyro-resonant RDIs can be unstable for $\tau \gtrsim 1$, and the cosmic-ray like instabilities (\S\ref{sub:cosmic.ray.streaming.mode}) may dominate for very large $\tau$.}
\label{fig: regimes ws and tau}\vspace{-0.5cm}
\end{center}
\end{figure*}
%
%
%

\vspace{-0.5cm}
\section{Astrophysical Applications}
\label{sec:application}

In this section, we  consider applications of the instabilities described above to astrophysical systems. To make detailed predictions for observable or physical consequences, numerical simulations, which can explore the fully non-linear regime, will be necessary. However it is useful to consider the relevant time and spatial scales, so as to evaluate whether the instabilities can grow efficiently, and if so, which instabilities and limits apply. 

For brevity and convenience, we will introduce the following notation for this section: let the temperature $T_{x} \equiv T / 10^{x}\,{\rm K}$ (so, e.g., $T_{4} = T/10^{4}\,$K); likewise for the density $n_{x} \equiv n_{\rm gas} / 10^{x}\,{\rm cm^{-3}}$ and the grain size $R_{d} \equiv R_{0.1}\,0.1\,\mu{\rm m}$. We will also define $\tilde{\sigma} \equiv \sqrt{1+\driftvelmag^{2}/c_{s}^{2}}$ (which interpolates between $1$ for sub-sonic drift and $\driftvelmag/c_{s}$ for super-sonic drift), $f_{\rm ion} = n_{\rm ion}/n_{\rm tot}$ is the ionization fraction of the gas, and $\tilde{U} \equiv (1/2)\,(|U_{\rm grain}|/{\rm volts}) / (T/10^{4}\,{\rm K})$, which should be $\sim 1$ if $|U_{\rm grain}|\approx |Z_{\rm grain}|\,e/R_{d}$ with the scalings we have adopted in \S~\ref{sec:stopping.time.scalings:lorentz} for collisional charging (this convention follows \citealt{draine:2011.HII.region.dust.drift}, but we note the scaling of $U_{\rm grain}$ in hot gas can be complicated by a number of additional physical processes; see e.g., \citealt{draine.salpeter:ism.dust.dynamics,weingartner:2001.grain.charging.photoelectric}). Where necessary we will adopt $\theta_{\bf Ba}\sim 45^{\circ}$, since our conclusions are insensitive to the relative angle of ${\bf B}$ and ${\bf a}$ so long as it is not very close to exactly perpendicular.

\vspace{-0.5cm}
\subsection{Relevant Scalings}
\label{sec:application:scalings}

First we outline some general scalings of use below.

\vspace{-0.5cm}
\subsubsection{Stopping Time ($t_{s}$)}
\label{sec:application:scalings:ts}

The stopping time (drag coefficient) $\langle t_{s} \rangle$ is a critical parameter determining the drift velocity, relevant spatial scales for the different modes, and growth timescales. In most cases below Coulomb or Epstein drag dominates the drag force. Inserting physical values in the scalings from \S~\ref{sec:stopping.time.scalings} we find:
\begin{align}
\nonumber t_{s}^{\rm Coulomb} &\sim 10^{4}\,{\rm yr}\,\left(\frac{10^{4}\,{\rm K}}{T} \right)^{1/2}\,\left(\frac{\rm cm^{-3}}{n_{\rm gas}} \right)\,\left( \frac{ R_{d} }{0.1\,\mu m} \right)\,\frac{\tilde{\sigma}^{3}}{\tilde{U}^{2}\,f_{\rm ion}} \\
{t_{s}^{\rm Epstein}} &\sim 30\,\tilde{U}^{2}\,f_{\rm ion}\,\tilde{\sigma}^{-4}\,{t_{s}^{\rm Coulomb}}.\label{eqn:stopping.times.physical}
\end{align}
So, in largely ionized media with trans-sonic or sub-sonic drift, Coulomb drag should dominate, while in primarily neutral media ($f_{\rm ion}\ll 0.1$) and/or systems with super-sonic drift ($\driftvelmag \gtrsim 2\,c_{s}$), Epstein drag will dominate. 

\vspace{-0.5cm}
\subsubsection{Spatial Scales ($k$)}
\label{sec:application:scalings:csts}

A useful corresponding spatial scale (which allows us to estimate whether a given wavelength is in the low-$k$, mid-$k$, or high-$k$ regime) is given by some characteristic speed multiplied by $t_{s}$. It is convenient to use $c_{s}\,\langle t_{s} \rangle$ for reference, both because we used it in \papertwo\ and because it is better-known than some other possibilities, e.g., $\driftvelmag\,\langle t_{s} \rangle$. One finds:
\begin{align}
c_{s}\,t_{s}^{\rm Coulomb} &\sim 0.1\,{\rm pc}\,\left(\frac{\rm cm^{-3}}{n_{\rm gas}} \right)\,\left( \frac{ R_{d} }{0.1\,\mu m} \right)\,\frac{\tilde{\sigma}^{3}}{\tilde{U}^{2}\,f_{\rm ion}},
\end{align}
and we can also calculate $c_{s}\,t_{s}^{\rm Epstein} = c_{s}\,t_{s}^{\rm Coulomb}\,(t_{s}^{\rm Epstein}/t_{s}^{\rm Coulomb}) \sim 3\,{\rm pc}\,\tilde{\sigma}^{-1}\,n_{0}^{-1}\,R_{0.1}$, when Epstein drag dominates. 

\vspace{-0.2cm}
\subsubsection{Strength of Lorentz Forces ($\tau$)}
\label{sec:application:scalings:tau}

If one considers an ionized medium with Coulomb drag dominating (or comparable to Epstein/Stokes drag), then 
\begin{align}
\tau^{\rm Coulomb} \sim&\, 100\,\left(\frac{T}{10^{4}\,{\rm K}} \right)\,\left( \frac{0.1\,\mu m}{ R_{d} } \right)\,\left(\frac{\rm cm^{-3}}{n_{\rm gas}} \right)^{1/2}\,\left( \frac{10}{\beta} \right)^{1/2}\,\frac{\tilde{\sigma}^{3}}{\tilde{U}\,f_{\rm ion}}.
\end{align}
We can also calculate this for the Epstein case ($t_{s} = t_{s}^{\rm Epstein}$), giving: 
$\tau^{\rm Epstein} \sim 30\,T_{2}\,\tilde{U}\,(R_{0.1}\,\tilde{\sigma})^{-1}\,(n_{0}\,\beta/10)^{-1/2}$. 
Given these results, in cold, dense gas (e.g., GMCs, planetary disks, circum-nuclear or circum-AGN regions in galaxies, and cool-star winds) we expect $\tau \ll 1$, but in the diffuse ionized ISM, CGM, and some dusty atmospheres, we expect $\tau \gg 1$.

\vspace{-0.2cm}
\subsubsection{Drift Velocities ($\driftvelXmag$)}
\label{sec:application:scalings:driftvel}

The equilibrium drift velocities (ignoring the geometric corrections from Lorentz forces) are given by $\driftvelX = {\bf a}\,\langle t_{s} \rangle / (1+\mu)\sim\driftvelXmag \sim |{\bf a}|\,\langle t_{s} \rangle$ (since $\mu$ is not large). If the drift is sub-sonic, we can neglect the dependence of $\langle t_{s} \rangle$ on $\driftvel$ itself, giving $\driftvelXmag \approx \driftvelXmag^{\ast} \equiv |{\bf a}|\,\langle t_{s}(\driftvel=0) \rangle$, where $\langle t_{s}(\driftvel=0) \rangle$ is just the stopping time scaling above (Epstein or Coulomb from Eq.~\ref{eqn:stopping.times.physical}, as appropriate), with $\tilde{\sigma}=1$. If $\driftvelXmag^{\ast} \gg c_{s}$, the drift is super-sonic; in this case Epstein drag dominates, and the true drift velocity is approximately given by $\driftvelXmag \approx a_{\gamma}^{-1/4}\,(\driftvelXmag^{\ast}\,c_{s})^{1/2}$ (where $a_{\gamma} \equiv 9\pi\,\gamma/128$ is defined in \S~\ref{sec:stopping.time.scalings}, and $\driftvelXmag^{\ast}$ is evaluated for Epstein drag). 

The characteristic acceleration $|{\bf a}|$ depends on the system; recall, ${\bf a}$ is the sum of {\em any} difference in acceleration experienced by the gas and the dust (other than that from drag itself). In some regimes it is dominated by gravity (if e.g., the gas is supported hydrostatically, since the dust cannot be), and in others by radiative acceleration (of the gas, in line-driven winds, or of the dust, in continuum absorption by grains).
Other forces, some induced by external radiation (e.g., coherent photo-electric and photo-desorption effects as described in \citealt{weingertner.draine:photo.forces.on.dust.hard.ism.rad}, or Poynting-Robertson ``drag''\footnote{For our purposes, Poynting-Robertson ``drag'' is an external acceleration of the dust (and contributes to ${\bf a}$, driving instability) rather than a ``true drag force'' (like Coulomb or Epstein or Stokes drag, which appear in $t_{s}$), because momentum is exchanged between the dust and the radiation field, not between the dust and the gas. Of course, coupled dust-radiation systems with any momentum exchange terms will be subject to their own RDIs, which we will explore in future work.}), or hydrodynamic and other forces on gas (e.g., if the gas is accelerating or decelerating owing to pressure gradients or expanding explosions, or being driven by pressure from cosmic rays) are also important in some regimes. 

It is useful to consider some order-of-magnitude estimates for the drift induced by gravity and radiation pressure on dust, as these are both common sources of drift. Noting that the gravitational acceleration $a_{\rm grav} \sim G\,M/r^{2} \sim \pi\,G\,\Sigma_{\rm eff}$ where $\Sigma_{\rm eff}$ is an effective surface density (of all enclosed mass), we find,
\begin{align}
\frac{(\driftvelXmag^{\ast})^{\rm grav}}{c_{s}} &\sim 
\begin{cases}
0.006\,{R_{0.1}\,\Sigma_{500}}\,{(n_{0}\,T_{4}\,\tilde{U}^{2}\,f_{\rm ion})^{-1}} & \hfill ({\rm Coulomb})  \\
30\,{R_{0.1}\,\Sigma_{500}}\,{(n_{0}\,T_{2})^{-1}} & \hfill ({\rm Epstein}),
\end{cases}
\end{align}
where $\Sigma_{\rm 500} \equiv \Sigma_{\rm eff} / (500\,\msun\,{\rm pc^{-2}})$, and we have scaled the Epstein drag expression to the lower temperatures $T\sim 100\,$K where it typically applies (see below). 

For radiation pressure on grains with some coherent incident flux ${\bf F}$, the acceleration is $a_{\rm rad} \sim Q_{\rm abs}\,F\,\pi\,R_{d}^{2} / (c\,m_{\rm grain})$, where $Q_{\rm abs}$ is a (spectrally-averaged) absorption efficiency, $c$ the speed of light, and $m_{\rm grain} = (4\pi/3)\,\bar{\rho}_{d}\,R_{d}^{3}$. Noting that $t_{s} \propto \bar{\rho}_{d}\,R_{d}/(\rho\,c_{s})$ when $\driftvelmag=0$, for both Epstein and Coulomb drag,  we find $a_{\rm rad}\,t_{s}/c_{s} \sim Q_{\rm abs}\,(F/c)/(\rho\,c_{s}^{2}) \sim Q_{\rm abs}\,e_{\rm rad}/e_{\rm thermal}$, where $e_{\rm rad}\equiv F/c$ is the (coherent) radiation energy density, and $e_{\rm thermal} \equiv \rho\,c_{s}^{2}$ is the thermal energy density. More accurately, we can estimate:
\begin{align}
\frac{(\driftvelXmag^{\ast})^{\rm rad}}{c_{s}} &\sim Q_{\rm abs}\,\frac{e_{\rm rad}}{e_{\rm thermal}}\,
\begin{cases}
{0.03}\,{\tilde{U}^{-2}\,f_{\rm ion}^{-1}}  & \hfill ({\rm Coulomb})  \\
1 & \hfill ({\rm Epstein}).
\end{cases}
\end{align}
For reference, for a flux $F= L/(4\pi\,r^{2})$ at distance $r = r_{\rm pc}\,$pc,  with total luminosity $L = L_{6}\,10^{6}\,L_{\sun}$, we have $e_{\rm rad}/e_{\rm thermal} \sim 800\,L_{6}\,n_{0}^{-1}\,r_{\rm pc}^{-2}\,T_{4}^{-1}$. For an incident flux peaked around a wavelength $\lambda_{\rm rad}$, $Q_{\rm abs}\sim 1$ if $R_{d} \gg \lambda_{\rm rad}$ (geometric absorption), and $Q_{\rm abs} \sim (R_{d}/\lambda_{\rm rad})$ if $R_{d} \ll \lambda_{\rm rad}$.

\vspace{-0.2cm}
\subsubsection{Scaling Coefficients ($\zeta$)}
\label{sec:scalings:zeta}

The scaling coefficients $\zeta$ (given in detail in \S~\ref{sec:stopping.time.scalings}) are usually only order-unity corrections to the growth rate; however certain terms (relevant for specific modes) can become small or nearly cancel (e.g., $\coeffTSrho - \tildeCoeffTSv$) in certain regimes.

If the drift is dominated by Coulomb drag, in the sub-sonic limit (the only regime where Coulomb drag can dominate), one finds $\coeffTSrho \approx 1-1/(2\,\ln\Lambda) + (-1.4+1.9\,\zeta_{C})\,(\gamma-1) \approx 0.96 - 1.4\,(\gamma-1)$, and $\tildeCoeffTSv \approx 1-3\,a_{C}\,\driftvelmag^{3} \approx 1$. If the drift is super-sonic, Epstein drag always dominates Coulomb, and we have $\tildeCoeffTSv =  (1+2\,a_{\gamma}\,(\driftvelmag/c_{s})^{2})/(1+a_{\gamma}\,(\driftvelmag/c_{s})^{2}) \approx 2$, and $\coeffTSrho \approx 1$ (independent of the equation-of-state or $\driftvel$, to leading order). Note $1-\coeffTSrho$ (as it appears in, e.g., Eq.~\ref{eqn:msonic.rdi.hik.lotau}) scales more accurately as $1-\coeffTSrho\approx [192+(9\pi-64)\,\gamma]/[128 + 9\pi\,\gamma\,(\driftvelmag/c_{s})^{2}] \approx 1.3\,(5.4/\gamma-1)\,(\driftvelmag/c_{s})^{-2}$ in this limit (see \papertwo). Sub-sonic Epstein drag is occasionally also an important regime; in this case $\tildeCoeffTSv \rightarrow 1$ and $\coeffTSrho \rightarrow (\gamma+1)/2$ as $\driftvelmag\rightarrow 0$ (for Stokes drag, $\tildeCoeffTSv,\,\coeffTSrho \rightarrow 1,\,[\gamma-1]/2$).

For intermediate and large grains (as in most cases of interest) when collisional charging dominates, we expect the dust-charge (Lorentz-force) parameter $\coeffTLrho \approx \gamma-1$, while for sufficiently small grains $\coeffTLrho \approx 0$. The division occurs at grain size $R_{d} \sim 6$\AA$\,T_{4}^{-1}$, so the $\coeffTLrho \approx 0$ regime only occurs for interesting grain sizes when $T\lesssim100\,$K. If photo-electric charging dominates, we expect $\coeffTLrho \approx (\gamma-3)/2$ unless the grains are maximally-charged (the incident UV flux exceeds $F_{\rm pe} \gtrsim 2\times10^{-5}\,n_{0}\,T_{4}^{-1/2}\,{\rm erg\,s^{-1}\,cm^{-2}}$), in which case $\coeffTLrho \approx 0$.

\subsection{Application to Different Environments}
\label{sec:application.env}

\begin{figure}
\begin{center}
\hspace{-0.2cm}\includegraphics[width=1.0\columnwidth]{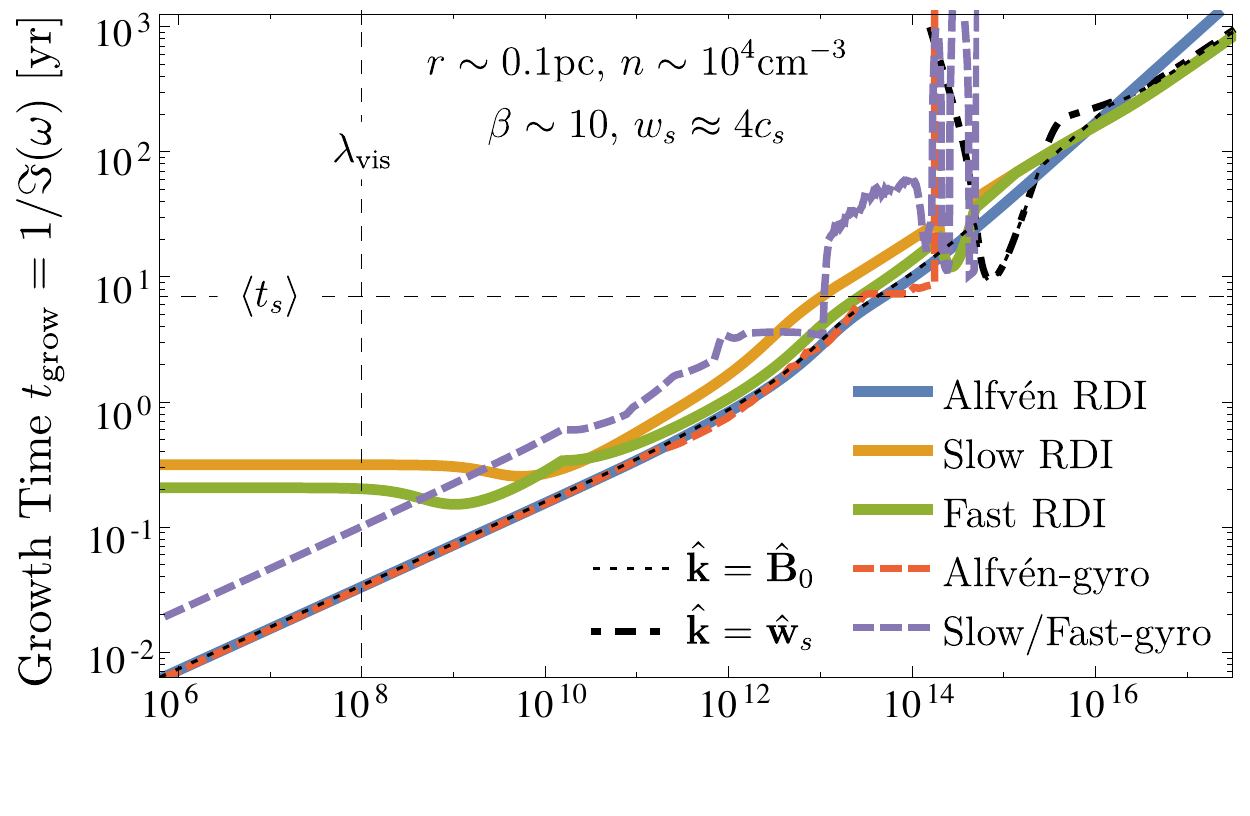}\\ \vspace{-0.45cm}\hspace{-0.2cm}\includegraphics[width=1.0\columnwidth]{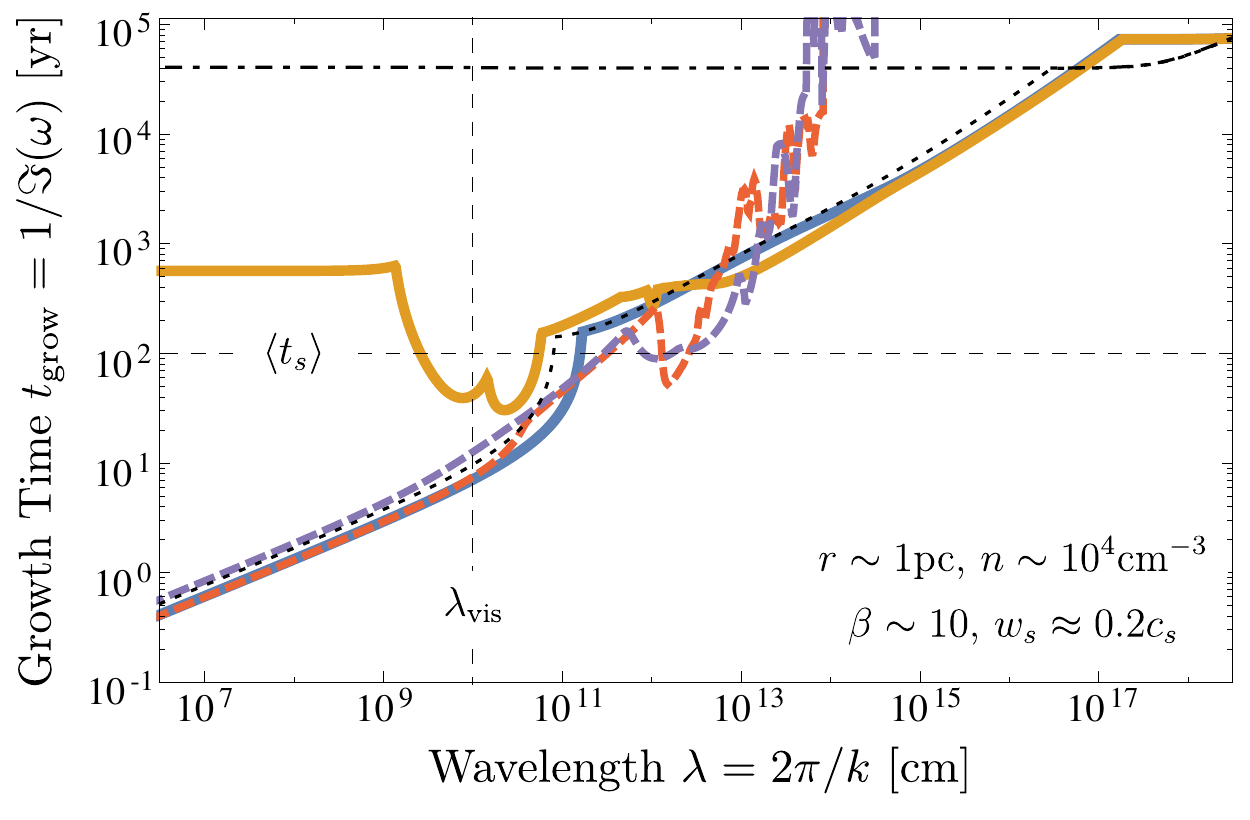}\vspace{-0.3cm}
\caption{Growth timescale $t_{\rm grow}=1/\Im{(\omega)}$ of the RDI for two illustrative examples in an HII region (see \S~\ref{subsub: HII regions}). 
{\em Top:} Parameters (labeled) appropriate near ($r\sim 0.1 {\rm pc}$) the star(s), where the drift is supersonic ($\driftvelmag \approx 4 c_{s}$). 
{\em Bottom:} Parameters appropriate further ($r\sim 1 {\rm pc}$) from the source(s) with subsonic drift ($\driftvelmag \approx 0.2 c_{s}$). 
For each line, we show the fastest-growing mode angle ($\hat{\bf k}$) which satisfies the relevant RDI condition at each $\lambda=2\pi/|{\bf k}|$. We compare the Alfv\'en-wave, slow-wave, and fast-wave RDIs (fast only unstable when $\driftvelmag > v_{f,\,0}$), Alfv\'en-gyro and slow/fast-gyro RDIs, and modes with $\hat{\bf k}=\driftvelhat$ or $\hat{\bf k}=\hat{\bf B}_{0}$. In this plot, we also include the parallel (Braginskii) viscosity in the momentum equation  ($\partial_{t}{\bf u}=\dots + c_{s} \lambda_{\mathrm{mfp}} \nabla\cdot [\hat{\bf B}\hat{\bf B}\,(\hat{\bf B}\hat{\bf B}:\nabla{\bf u})]$), which damps field-parallel motion below the viscous scale 
$\lambda_{\mathrm{vis}}\sim \lambda_{\mathrm{mfp}}$ (see \S\ref{sec:scales.validity}). This decreases the growth 
rates of the slow- and fast-wave RDIs for $\lambda\lesssim \lambda_{\mathrm{vis}}$, without affecting the Alfv\'en-wave RDI, which involves only perpendicular motions of the gas. In this example, at scales $\lambda \sim r$, the RDI growth time can be much shorter than HII region expansion times; while at short wavelengths within $r\lesssim 0.1$\,pc, growth timescales of the Alfv\'en RDI are as small as $\sim$\,days! 
}
\label{fig: HII examples}\vspace{-0.3cm}
\end{center}
\end{figure}

\subsubsection{The Warm Ionized \&\ Warm Neutral Interstellar Medium}
\label{sec:application.env:wim.wnm}

In the ``warm'' ISM, we have gas with densities $n_{\rm gas} \sim 0.1 \rightarrow 1\,{\rm cm^{-3}}$, temperatures $T\sim 10^{4}\,$K, typical grain sizes $R_{d}\sim 0.01 \rightarrow 0.1\,\mu{\rm m}$, $\beta\sim 1\rightarrow100$, and $e_{\rm rad}\sim e_{\rm thermal}$ (also with typical gravitational and radiative accelerations of grains comparable to one another). Despite modest $\beta$, this gives $\tau \sim 100 \gg 1$. For the warm ionized medium (WIM, with $f_{\rm ion}\sim 1$) Coulomb drag dominates Epstein drag, while for the warm neutral medium (WNM) the drag regime depends on the ionization fraction (Coulomb drag dominates for $f_{\rm ion} \gtrsim 0.03$, which is reasonable for the WNM). Assuming mostly-ionized gas with modest $\beta$, we have $\driftvelmag/v_{f,\,0} \sim \driftvelmag/c_{s} \sim 0.03\,f_{\rm ion}^{-1}\,e_{\rm rad}/e_{\rm thermal}$, so the typical drift velocities are expected to be modestly sub-sonic (a few tenths of ${\rm km\,s^{-1}}$; for more detailed calculations that give a very similar result, see \citealt{weingertner.draine:photo.forces.on.dust.hard.ism.rad}). 

Since $\driftvelmag \lesssim {\rm MIN}(v_{A},\,c_{s})$, we expect the dominant resonance of the RDI at intermediate and high-$k$ to be the slow-magnetosonic or Alfv\'en RDI (the fast-magnetosonic RDI will, in general, not be possible in the warm medium, except in regions near massive stars or other sources of strong local grain acceleration). Both Alfv\'en and slow modes have similar growth rates but the Alfv\'en mode may be slightly faster-growing at intermediate wavelengths. Under these conditions, all length scales are formally unstable to the RDI and related instabilities discussed here, from the viscous scale to the disk scale height (the effective ``Reynolds number'' of the instabilities will be very large). We expect $c_{s}\,\langle t_{s} \rangle \sim 0.1\,{\rm pc}\,(R_{0.1}/n_{0}\,f_{\rm ion})$, with $\mu\sim 0.01$, suggesting scales from the largest available (disk scale height) down to $\sim 1\,$pc will be in the long-wavelength (low-$k$) regime, while scales from $\sim 1\,$pc down to $\sim $\,au will be in the mid-wavelength resonant regime.

Recall, the slow-mode resonance at high $\tau$ has lower growth rates through the mid-$k$ regime, owing to the projection of the drift along $\hat{\bf B}_{0}$ (which means resonant angles with slow wavespeeds, nearly-perpendicular to $\hat{\bf B}_{0}$, are also nearly perpendicular to $\driftvelhat$). At long wavelengths ($\lambda\gtrsim$\,pc), for this particular set of parameters ($\tau$ in particular), the growth rate is only weakly dependent on scale (in part owing to coincidence of the overlap of the low-$k$ mode, gyro mode, and slow mid-$k$ mode, but primarily because the fastest-growing modes are the parallel quasi-drift and quasi-sound modes), with growth timescales 
$t_{\rm grow} \sim 70\,{\rm Myr}\,n_{0}^{-3/4}$. This is comparable to, or shorter than, dynamical times in the gas on the same scales. 
At smaller scales, ${\rm pc} \gtrsim \lambda \gtrsim {\rm au}$, the growth rates scale as 
$t_{\rm grow} \sim 0.8\,{\rm Myr}\,(\lambda/100\,{\rm au})^{1/2}\,f_{\rm ion}^{-2}\,n_{0}^{-3/4}$ (note the dependence on $\driftvelmag$, $\tau$, and $t_{s}$ is such that this is grain-size independent). Although this becomes longer than e.g., turbulent eddy turnover times in the ISM on smaller scales, at sufficiently small scales $\lambda\lesssim$\,au, the high-$k$ Alfv\'en resonance appears, whose growth rate is {\em enhanced} rather than suppressed at high-$\tau$, giving growth timescales in this regime of $t_{\rm grow} \sim 700\,{\rm yr}\,R_{0.1}\,n_{0}^{-1/2}\,(\lambda/{\rm au})^{1/3}\,(\tilde{U}/f_{\rm ion}^{2})^{1/3}$. If the instability can persist down to the thermal dust gyro radii ($\sim R_{\sun}\,(R_{0.1}/n_{0})^{1/2}$, estimated assuming Brownian motion of the dust gives $m_{\rm dust}\,\delta v_{\rm thermal}^{2} \sim k_{B}\,T_{\rm gas}$, so produces an effective ``dust pressure'' that can damp modes below this scale), then the growth time at this scale is $t_{\mathrm{grow}}\sim 100\,{\rm yr}\,R_{0.1}^{7/6}\,(\tilde{U}/f_{\rm ion}\,n_{0})^{2/3}$.\footnote{In the WIM ($f_{\rm ion}\sim1$), ideal MHD is an excellent approximation. In the WNM, if $f_{\rm ion}$ and $\driftvelmag/c_{s}$ are both sufficiently small, ambipolar diffusion may not be negligible. However the low-$k$ modes of interest act identically on both neutrals and ions, so are unaffected by this. On scales ${\rm au}\,\lesssim \lambda \lesssim 2\pi/k_{\rm AD} \sim 0.2\,{\rm pc}\,n_{0}^{-1}\,(\beta/10)^{-1/2}\,(0.01/f_{\rm ion})$, the mid-$k$ modes may be damped in the WNM if $\driftvelmag \ll c_{s}$. On scales $\lesssim {\rm au}$, the high-$k$ Alfv\'en resonance in the WNM is a resonance between ions/fields and dust, and actually has a growth time that becomes shorter than the ion-neutral coupling time, so proper treatment requires a three-fluid RDI.}


These instabilities could have a wide range of consequences. They could drive sightline-to-sightline extinction-curve variations in the WNM and WIM \citep[which have long been observed to be ubiquitous, see][for a review]{schlafly:galactic.extinction.curve.variations}, as grains are clumped on different scales in a size-dependent fashion (producing local variations in the size distribution). 
The smallest grains ($R_{0.1}\ll1$), at the smallest wavelengths ($\ll 0.1\,${\rm au}) approach growth timescales (hence likely variability timescales for the clumping in the non-linear regime) that become human-observable ($\ll 10\,$yr), and could be important for the well-known turbulent scintillation ``cascade'' apparently seen in radio \citep[for reviews, see][]{bhat:2004.pulsar.constraints.scintillation,haverkorn:2013.radio.plasma.ism.diagnostics.review}. 
Dust-to-gas ratio fluctuations on intermediate scales ($\lambda\gg 10^{3}\,$au) could  explain the well-studied order-of-magnitude (or larger) excess (relative to the ISM abundance) of large  grains ($R_{d}\gtrsim 0.1\,\mu{\rm m}$) observed  in the solar neighborhood \citep[see e.g.][and references therein]{kruger:2015.ulysses.large.grain.abundance.near.sun.review,alexashov:2016.interstellar.dust.near.heliopause.large.grains}. 
On larger scales, this may explain some tentative observational suggestions of systematic dust segregation across galactic scale height \citep{schlafly:galactic.extinction.curve.variations,gontcharov:2016.galactic.dust.segragation.toward.poles}. 
Given the large-scale and strong angle-dependence with respect to the local magnetic field angles, this will almost certainly alter predictions for grain-field alignment and therefore polarization (and ``spinning dust''), and may be necessary to explain the observational fact that polarized dust emission obeys different directional statistics (particularly in the ratio of E to B modes) to those expected of the ${\bf B}$-field in MHD turbulence, on scales of order the galactic disk height (\citealt{caldwell:2017.mhd.turb.vs.planck.dust.polarization.discrepancy.eb.modes}, but see also \citealt{kandel:2017.planck.eb.mode.dust.vs.sub.alfvenic.turb}).

\vspace{-0.5cm}
\subsubsection{The Circum-Galactic and Inter-Galactic Medium}
\label{sec:application.env:cgm.igm}

The circum-galactic medium (CGM) and inter-galactic medium (IGM) are also clearly observed to be dust-laden (see e.g., \citealt{chelouche:2007.galaxy.cluster.dust.content.vs.radius,menard:2010.sdss.dust.vs.gal.distance,menard:2012.dust.in.mgII.absorbers,peek:2015.cgm.dust.constraints,baes:2016.cgm.dust.uv.scattering}), and can exhibit a very broad range of properties.

Temperatures are in the range $T\sim 10^{5}\rightarrow10^{7}\,$K (clouds with temperatures $\sim 10^{4}\,$K also exist in the CGM closer to galaxies, but the conditions in such clouds are expected to be closer to the WIM discussed above).\footnote{In the CGM around lower-mass galaxies (halo masses $M_{\rm halo} \ll 10^{12}\,M_{\sun}$), or in the IGM beyond the virial shock, or in cold filaments (``cold flows'') accreting onto galaxies, we have $T\sim 10^{4}\,$K, but otherwise similar parameters. Under the ``fast drift'' conditions around, e.g., a QSO, the relevant scalings are quite similar to those above (the drift is even more strongly super-sonic, but this does not change the main behavior). The gravity-dominated (``slow-drift'') conditions are more interesting: under gravity alone, a simple calculation assuming the gas is hydrostatic would give a drift velocity $\driftvelmag/c_{s} \sim 8\,r_{100}^{-1}\,(M_{\rm halo}/10^{12}\,M_{\sun})^{1/2}\,(R_{0.1}/n_{-4})^{1/2}$, which (given lower densities at larger radii) could be super-sonic out to $\gg10$\,Mpc scales around $\sim M_{\ast}$ halos. With high $\tau \sim 10^{3}-10^{5}$, this would excite the fast and Alfv\'en RDIs and ``streaming-type'' mode. However, under these conditions, the gas is not hydrostatic --- instead it, and the dust, are likely to be free-falling together onto galaxies or halos. A more detailed estimate of the differential forces is therefore needed to determine the typical drift conditions.} The medium is largely ionized ($f_{\rm ion}\sim 1$), and low-density ($n_{\rm gas} \sim 10^{-6} \rightarrow 10^{-2}\,{\rm cm^{-3}}$), so $\tau \sim 10^{4}\rightarrow10^{7}\, \gg 1$, but with typical $\beta \sim 10^{2}\rightarrow10^{4}$ \citep[see][and references therein]{tumlinson:2017.cgm.review}. However, we caution that the behavior of $\tilde{U}$ in this temperature range is highly uncertain (owing to factors such as electron emission from highly-charged grains). In the classic models of \citet{draine.salpeter:ism.dust.dynamics}, for example, silicate grains range from $\tilde{U}\sim 0.01 - 0.1$, and graphite grains from $\tilde{U}\sim 0.001 - 3$, for  $T\sim 10^{5}\rightarrow10^{7}\,$K, depending on the grain size, illuminating intensity field, and background density (note  $2\,\tilde{U}$ here is approximately equal to $\phi$ defined therein). 
So Lorentz forces are always strong, but Epstein drag could dominate over Coulomb drag for $\tilde{U} \lesssim 0.1$. Typical dust-to-gas ratios are also highly uncertain, even with the observational constraints above --- in the CGM, metallicities range from $\sim 0.1\rightarrow0.3\,Z_{\sun}$ around massive galaxies \citep{leccardi:2008.cluster.metallicity.profiles,tumlinson:2017.cgm.review}, but whether the usual dust-to-metals ratio applies is largely unknown. 

On top of this, the magnitude of the grain drift is likely to vary tremendously. At one extreme, in the CGM around, say, luminous quasars with $L \sim L_{13}\,10^{13}\,L_{\sun}$, even at a distance $r \sim r_{100}\,100\,{\rm kpc}$ (assuming the QSO is not obscured) we would expect highly-supersonic drift with $\driftvelmag/c_{s} \sim 10\,r_{100}^{-1}\,(L_{13}/n_{-4}\,T_{6})^{1/2}$ (with $t_{s} \sim 30\,{\rm Myr}\,r_{100}\,R_{0.1}\,(L_{13}\,n_{-4})^{-1/2}$, and $c_{s}\,t_{s} \sim 3\,{\rm kpc}\,r_{100}\,(T_{6}/L_{13}\,n_{-4})^{1/2}$). This is an interesting case in part because it may be the mechanism by which the dust gets into the CGM in the first place \citep{murray:momentum.winds,choi:2012.bh.fb.idealized,hopkins:qso.stellar.fb.together,ishibashi:2016.dusty.outflows.cgm.agn}. At the other extreme, at such large distances from a ``faint'' galaxy such as the Milky Way (especially in somewhat denser CGM gas), the radiation pressure on grains is relatively weak and the gravitational forces dominate drift with $\driftvelmag/c_{s} \sim \driftvelXmag/c_{s} \sim 2\,(M_{\rm halo}/10^{12}\,M_{\sun})\,R_{0.1} / (n_{-4}\,T_{6}\,r_{100}^{2})$ (with $t_{s} \sim 300\,{\rm Myr}\,R_{0.1}\,n_{-4}^{-1}\,T_{6}^{-1/2}$ and $c_{s}\,t_{s} \sim 30\,{\rm kpc}\,R_{0.1}\,n_{-4}^{-1}$). 

Under the ``fast-drift'' (former) conditions, the fast RDI is unstable, but are also in the unusual ``large-$\tau$'' regime discussed in \S~\ref{sub:cosmic.ray.streaming.mode}; the drift is strongly aligned with the magnetic fields and the fastest-growing modes are also aligned. The fastest-growing modes are those analogous to the CR streaming instability, with growth rates at large wavelengths ($\gtrsim$\,kpc) $t_{\rm grow} \sim {\rm Myr}\,(\lambda/{\rm kpc})^{2/3}\,R_{0.1}\,(r_{100}/T_{6})^{1/3}\,(L_{13}\,n_{-4})^{-1/6}$ and at short wavelengths ($\lesssim$\,kpc) $t_{\rm grow} \sim 1000\,{\rm yr}\,(\lambda/{\rm pc})\,R_{0.1}\,T_{6}^{-1/2}$. This can be much faster than characteristic dynamical or free-fall or cooling times ($\sim$\,Gyr), sound-crossing times ($\sim 10\,{\rm Myr}\,(\lambda/{\rm kpc})\,T_{6}^{-1/2}$) or turbulent eddy turnover times ($\sim 50\,{\rm Myr}\,r_{100}^{1/3}\,(\lambda/{\rm kpc})^{2/3}\,T_{6}^{-1/2}$, if we assume trans-sonic Kolmogorov turbulence with driving scale $\sim r$). 

Under the ``slow-drift'' (latter) conditions the slow Alfv\'en RDIs will be unstable, but  are in the regime where the ``mid-$k$'' modes are strongly suppressed by high $\tau$. Here ``low-$k$'' corresponds to very long wavelengths ($\lambda\gtrsim 100\,$kpc), so clearly global solutions are needed in this limit. At wavelengths $30\,{\rm pc}\lesssim \lambda \lesssim 100\,$kpc, the growth rates are suppressed, with $t_{\rm grow} \gtrsim 3\,{\rm Gyr}\,R_{0.1}\,r_{100}\,n_{-4}^{-1/2}$, so the instabilities are unlikely to be important. 
If the RDI can grow at smaller scales (not obvious, owing to dissipative effects), then because $\beta$ is large the gyro resonances can appear (if $\driftvelmag \gtrsim v_{A}$) at wavelengths $\lambda\sim 2\pi\,\driftvelmag\,\langle t_{L} \rangle \sim {\rm pc}\,(\beta/1000)^{1/2}\,n_{-4}^{-1/2}\,T_{6}^{-1}\,R_{0.1}^{2}$, with growth times $t_{\mathrm{grow}}\sim 3\times10^{4}\,{\rm yr}\,(\beta/1000)^{1/2}\,(\mu/0.001)^{-1/2}\,T_{6}^{-3/2}\,n_{-4}^{-1/2}\,R_{0.1}^{2}$. 

These instabilities --- especially in the ``fast-drift'' case --- could be critical to understand how dust gets into the CGM in the first place (as needed to explain the observations above; see \citealt{ishibashi:2016.dusty.outflows.cgm.agn} and references therein), as well as how dust survives in hot CGM/IGM gas (if, for example, it clumps so as to locally self-shield, or limit/reduce thermal estimates of gas-dust collision rates), where simple sputtering estimates have long suggested it should be rapidly destroyed \citep{draine:dust.destruction}. Cooling and molecular formation processes in galactic outflows could depend critically on the presence and clumping of dust \citep[e.g.][]{richings:2018.molecular.outflows.agn.cooling.needs.dust}. It could again contribute to variations in extinction curves and polarization/scattering (as for the WIM/WNM); these may be especially important for interpretations of scattered and re-emitted light from QSOs at these large distances (e.g., observations of the transverse proximity effect; see \citealt{hennawi:2009.new.binary.qsos,martin:2010.metal.enriched.regions}). The ``streaming-type'' instabilities could allow for behavior analogous to cosmic ray streaming, e.g., transfer of energy from the radiation field (accelerating the dust) to thermal energy of the gas (if the excited Alfv\'en waves are thermalized through, e.g., turbulence).

\vspace{-0.5cm}
\subsubsection{HII Regions}\label{subsub: HII regions}

Typical HII regions have $T\sim 10^{4}\,$K, $n_{\rm gas}\sim 10^{2} \rightarrow 10^{5}\,{\rm cm^{-3}}$, modest $\beta\sim 1\rightarrow30$, $f_{\rm ion}\sim 1$, $\tilde{U}\sim1$. If we consider dust at a distance $r = r_{\rm pc}\,$pc from a source (or set of sources) of luminosity $L = L_{6}\,10^{6}\,L_{\sun}$ (e.g., an OV star), then radiation pressure dominates the acceleration of grains (over, e.g., gravity) with $Q_{\rm abs}\,e_{\rm rad}/e_{\rm thermal} \sim 8\,L_{6}\,n_{2}^{-1}\,r_{\rm pc}^{-2}$ for typical large grains. Because the acceleration falls off with radius (assuming the density falls of less-steeply than $r^{-2}$), there is a critical radius at $r_{\rm pc} \sim  0.5\,L_{6}^{1/2}\,n_{2}^{-1/2}$, outside of which the drift is sub-sonic, and inside of which it is super-sonic \citep[for more detailed calculations which give similar results, see][]{draine:2011.HII.region.dust.drift,akimkin:2015.grain.drift.in.HII.regions}. Note that the Stromgren radius scales as $r_{\rm stromgren} \sim 6\,{\rm pc}\,L_{6}^{1/3}\,n_{2}^{-2/3}$ (assuming the temperature of an OV star) so this is well within the HII region for all physical densities and luminosities. 

Outside $r_{\rm pc} \gtrsim  0.5\,L_{6}^{1/2}\,n_{2}^{-1/2}$, the drift is sub-sonic, and Coulomb drag dominates with $\driftvelmag/c_{s} \sim 0.25\,L_{6}\,n_{2}^{-1}\,r_{\rm pc}^{-2}$ and $\tau \sim 10\,R_{0.1}^{-1}\,n_{2}^{-1/2}\,\beta_{10}^{-1/2}$. We find that $t_{s}\sim 100\,{\rm yr}\,n_{2}^{-1}\,R_{0.1}$  and $c_{s}\,t_{s} \sim 0.001\,{\rm pc}\,R_{0.1}\,n_{2}^{-1}$. For wavelengths $\lambda/{\rm pc} \sim (10\,{\rm au} \rightarrow 0.6\,{\rm pc})\,R_{0.1}/n_{2}$, we are in the regime of the intermediate-wavelength or ``mid-$k$'' resonance, with the fastest-growing mode either the slow or Alfv\'en RDI (both scale similarly). This gives growth timescales $t_{\rm grow} \sim 6000\,(\lambda/0.001\,{\rm pc})^{1/2}\,r_{\rm pc}^{2}\,{\rm yr}\,n_{2}^{1/4}\,\beta_{10}^{-1/4}\,L_{6}^{-1}$; comparing to the sound-crossing time at $r$ ($t_{\rm sound} = r/c_{s} \sim 80,000\,{\rm yr}\,r_{\rm pc}$), we see that $t_{\rm grow}/t_{\rm sound} \sim 0.1\,n_{2}^{1/4}\,L_{6}^{-1}\,\beta_{10}^{-1/4}\,(\lambda/0.001\,{\rm pc})^{1/2}\,r_{\rm pc}$. Note the expansion timescale of the HII region is approximately $t_{\rm expand} = t_{\rm sound}(r = r_{\rm stromgren}) \sim 5\times10^{5}\,{\rm yr}\,L_{6}^{1/3}\,n_{2}^{-2/3}$. On the largest spatial scales $\lambda\gtrsim 0.6\,{\rm pc}\,R_{0.1}/n_{2}$, the long-wavelength mode dominates, but the growth timescales on these scales are usually  longer than the HII region expansion time. On small scales, from $\lambda\sim 10\,{\rm au}\,R_{0.1}/n_{2}$ down to the dissipation scale $\lambda\sim 0.0006\,{\rm au}/n_{2}$, we are in the high-$k$ regime. The scales are small enough, and $\tau$ modest, so the high-$k$ Alfv\'en RDI regime can be reached, with growth times around the viscous scale reaching $t_{\mathrm{grow}}\sim 10\,{\rm yr}\,\beta_{10}^{1/6}\,n_{2}^{-1/6}\,L_{6}^{-2/3}\,R_{0.1}\,r_{\rm pc}^{4/3}$

Inside the radius $r_{\rm pc} \lesssim  0.5\,L_{6}^{1/2}\,n_{2}^{-1/2}$, the drift is super-sonic, and Epstein drag dominates, with $\driftvelmag / c_{s} \sim 4\,L_{6}^{1/2}\,n_{2}^{-1/2}\,r_{\rm pc}^{-1}$ and $\tau \sim 70\,r_{\rm pc}\,R_{0.1}^{-1}\,L_{6}^{-1/2}\,\beta_{10}^{-1/2}$ \citep[for more detailed calculations which give very similar results, see][]{akimkin:2015.grain.drift.in.HII.regions}. We find $t_{s} \sim 700\,{\rm yr}\,n_{2}^{-1/2}\,L_{6}^{-1/2}\,r_{\rm pc}\,R_{0.1}$ and $c_{s}\,\langle t_{s} \rangle \sim 0.01\,{\rm pc}\,r_{\rm pc}\,R_{0.1}\,(L_{6}\,n_{2})^{-1/2}$. Under these conditions a wide range of modes are present. For example, if $r\sim 0.1\,$pc, $n\sim 10^{4}\,{\rm cm^{-3}}$, $\beta\sim 100$, $L_{6}\sim1$, then it is possible to be in all three wavelength regimes: (i) at wavelengths $\lambda\gg c_{s}\,\langle t_{s}\rangle$, the long-wavelength/pressure-free mode dominates with $t_{\rm grow}\sim 300\,{\rm yr}\,(\lambda/0.01\,{\rm pc})^{2/3}\,(r_{\rm pc}/0.1)^{1/3}\,(R_{0.1}^{2}/L_{6}\,n_{4})^{1/6}$; (ii) at scales $\lambda \sim (10^{-4}\rightarrow1)\,c_{s}\,\langle t_{s} \rangle$ the fast MHD-wave RDI is usually fastest-growing and in the mid-$k$ regime, with $t_{\rm grow} \sim 4\,{\rm yr}\,(\lambda/{\rm au})^{1/2}\,(r_{\rm pc}/0.1)^{1/2}\,(R_{0.1}^{2}/L_{6}\,n_{4})^{1/4}$; and (iii) at still smaller scales down to the dissipation scale ($\lambda_{\rm diss} \sim 10^{8}\,{\rm cm}\,n_{4}^{-1}$) we are in the high-$k$ regime with the Alfv\'en wave-RDI dominant and $t_{\rm grow} \sim 12\,{\rm days}\,(\lambda/\lambda_{\rm diss})^{1/3}\,(r_{\rm pc}/0.1)^{2/3}\,(R_{0.1}^{2}/L_{6}\,n_{4}^{2})^{1/3}$. The gyro-RDI is also present and can produce very fast growth rates at {specific} wavelengths: e.g., for these parameters, $t_{\rm grow}^{\rm gyro}\sim 5\,{\rm yr}$ around a narrow range of $\lambda\approx 60\,{\rm au}$ (a factor $\sim 6$ faster than the fastest MHD-wave RDI at the same $\lambda$). 
In Fig.~\ref{fig: HII examples}, we show the instability growth timescale for the various RDI families, calculated using parameters relevant for the inner (top panel) and outer (bottom panel) regions of an HII region.

The RDIs here could be important for a wide range of observed phenomenology. It has long been recognized that dust in HII regions is critical for their chemistry, depletion of metals (therefore accurate abundance estimates from emission-line estimators), cooling physics, and determining the emergent spectrum \citep{shields:1995.dust.HII.region.chem.rad.fx}. 
A number of studies have also noted that dust dynamics have a large impact on the expansion of HII regions via their interaction with gas and radiation pressure, and the drift and dust-gas interaction substantially alters the HII region expansion rate, densities in the central cavity, dust-to-gas ratios in the interior and swept-up-shell, and more \citep{akimkin:2017.dustgasHII.expansion.cavity.density.metallicity}. 
To date, studies of these phenomena have neglected the physics necessary to follow the RDI (3D calculations with magnetized gas, charged dust, drift, Lorentz forces, and back-reaction on gas); but the scalings above suggest these properties could be radically altered. 
Moreover, dust clumping induced by the RDI could  enhance the leakage of ionizing photons from HII regions by orders of magnitude \citep{anderson:2010.dust.HII.leakage.clumps.induced.sf.filaments}. 
The long-wavelength, pressure-free mode will directly drive dust into sharper arcs, shells, and rings concentric around the stars, as observed in many HII regions. Further, unlike explanations based purely on radiation pressure and shocks, this naturally explains observed shells inside and outside the ionized regions \citep{topchieva:2017.dust.HII.region.shape.irregular.rings.shells}, as well as operating on faster timescales. 
Meanwhile the fast-wave RDI, on shorter wavelengths, will directly drive dust into filaments or ``whiskers'' and dust lanes with extremely sharp fine-structure, as observed in essentially all sufficiently-resolved HII regions \citep{odell:2002.pne.knots.review,apai:2005.HII.region.filamentary.dust.structures}, most famously in $\eta$ Carinae \citep{morse:1998.eta.carinae.whiskers.dustlanes.filaments}. 
These may provide seeds or induce the dust condensations associated with triggered star formation around HII regions \citep[e.g.][]{anderson:2012.dust.HII.clumps.triggered.sf,deharveng:2015.HII.region.dust.condensations.filaments.triggered.SF}.
This may also naturally explain  anomalies where the dust knots or filaments do not appear coincident with gas-phase density enhancements on small scales \citep{garnett:2001.ring.nebula.dusty.knots.offset.emission.line.no.gas.overdensity}. 
The size dependence of the effects could also produce apparent observed grain-size variations across HII regions \citep{relano:2016.HII.dust.shells.depletions.grainsize.vars,hankins:2017.arches.dust.size.distrib.modified.in.HII}.

\vspace{-0.5cm}
\subsubsection{SNe Ejecta}
\label{sec:application.env:sne}

In and around SNe explosions, a wide variety of behaviors will manifest at different stages of the event.

Immediately after ($t\lesssim 1\rightarrow3\,$yr) the explosion, dust in the near vicinity of the SNe (but not within the explosion itself) suddenly sees a very bright source with $L \sim L_{9}\,10^{9}\,\msun$. Outside of a sublimation radius $r_{\rm sub} \sim 0.014\,L_{9}^{1/2}\,{\rm pc}$, if the ambient density is $n\sim 100\,n_{2}\,{\rm cm^{-3}}$, the nearby dust at radius $r$ from the SNe is radiatively accelerated to extremely high velocities $\driftvelmag \sim 4\times10^{4}\,{\rm km\,s^{-1}}\,n_{2}^{-1/2}\,T_{4}^{1/4}\,(r_{\rm sub}/r)$ \citep[for more detailed calculations including relativistic effects, see][]{hoang:2017.relativistic.dust.drag}. (Note that for $r \gg 100\,r_{\rm sub} \sim 1.4\,{\rm pc}$, the time to accelerate the grains becomes $\gg$\,yr, so the maximum velocity drops rapidly). Whether the ambient gas is typical of a cold GMC, or the diffuse WIM, we have $\driftvelmag \gg c_{s}$, with Epstein drag dominating, $\tau \ll 1$ (owing to the very large $\driftvelmag$), and $\beta \gtrsim 1$. So the rapidly-moving dust immediately triggers the fast-wave RDI, which under these conditions is essentially identical to the acoustic RDI from \paperone\ (with growth timescales $t_{\rm grow}\sim 3\,{\rm yr}\,n_{2}^{-1/4}\,T_{4}^{-1/8}\,R_{0.1}^{1/2}\,(r/r_{\rm sub})^{1/2}\,(\lambda/{\rm au})^{1/2}$, faster than the expansion time for small scales $\lambda\lesssim$\,au). 

After several years, the explosion is in the free-expansion phase and dust begins to condense in the ejecta, with what observations indicate has a high, order-unity  efficiency  \citep[e.g.][]{bianchi.2007:dust.formation.sne.ejecta,matsuura:2015.1987a.large.dust.mass,owen:2015.crab.large.dust.mass,delooze:cass.A.large.dust.mass,temim:2017.snr.rapid.massive.dust.mass}. Thus, the dust-to-gas ratio inside the ejecta could be quite high, $\mu \gtrsim 0.1$. The ejecta and dust are both free-expanding to leading order (at a radius $r_{\rm ej}\sim r_{\rm pc}\,{\rm pc}$), with internal density $n\sim 10\,(M_{\rm ej}/M_{\sun})\,r_{\rm pc}^{-3}$ and velocity $v\sim v_{4}\,10^{4}\,{\rm km\,s^{-1}}$, adiabatically cooled to $T\sim 20\rightarrow100\,$K, and likely with large magnetic fields $\beta \sim 10^{-6}\,(M_{\rm ej}/M_{\sun})\,r_{\rm pc}^{-3}\,T_{2}\,B_{0,\,{\rm mG}}^{-2}$ \citep{reynolds:2012.magnetic.fields.in.sne.reviews}, suggesting large $\tau \sim 2\times10^{4}\,r_{\rm pc}^{3}\,T_{2}^{1/2}\,R_{0.1}^{-1}\,(M_{\rm ej}/M_{\sun})^{-1}\,B_{0,\,{\rm mG}}$ (where $B_{0,\,{\rm mG}} \equiv |{\bf B}_{0}| / {\rm mG}$). 
Radiative acceleration of the dust is unimportant after several years, however, the gas ejecta may be accelerating (if being driven by relativistic winds in e.g., a pulsar wind nebula) or decelerating (from sweeping up gas in the ISM). Consider (for simplicity) the latter: if the ambient density is $n\sim 1\,{\rm cm^{-3}}$, then momentum (or energy) conservation demands that the {\em gas} has decelerated by a factor $\sim 1 - M_{\rm swept}/M_{\rm ej} \sim 1 - 0.1\,n_{0}\,r_{\rm pc}^{3}\,(M_{\rm ej}/M_{\sun})^{-1}$, which (if the gas were still cold) would imply a highly super-sonic drift velocity $\driftvelmag \sim 1000\,{\rm km\,s^{-1}}\,v_{4}$.\footnote{More accurately calculating the equilibrium drift velocity for a shell expanding into a uniform ambient medium we obtain $\driftvelmag \approx 2000\,{\rm km\,s^{-1}}\,v_{4}\,(n_{0}\,R_{0.1})^{1/2}\,r_{\rm pc}^{5/2}\,(M_{\rm ej}/M_{\sun})^{-1}$ while $\driftvelmag \ll v_{\rm ej}$.} However, the gas which is de-celerated is being processed through a reverse  shock as well, and the (Epstein) stopping time in this phase is itself comparable to the expansion time ($\sim r_{\rm ej}/v_{\rm eg}$), $t_{s}/t_{\rm expand} \sim (R_{0.1}/n_{0}\,r_{\rm pc})^{1/2}$, so it is unclear what the drift velocity in the {\em cold} portion of the ejecta will actually be without a global, time-dependent solution. 
The drift is therefore likely trans-Alfv\'enic and either trans-sonic or highly super-sonic, depending on whether the dust resides in the reverse-shocked region or remains in ``cold'' regions (or drifts ``out of'' the reverse-shocked region on a timescale $\lesssim t_{s}$). If the ``cold'' case can be realized, strong gyro-resonances are present with growth times $t_{\mathrm{grow}}\sim 40\,{\rm yr}\,R_{0.1}^{2}\,B_{0,\,{\rm mG}}^{-1}\,T_{2}^{-1}\,(\mu/0.1)^{-1/2}$ at relatively long wavelengths $\lambda\sim (0.01\rightarrow0.04)\,{\rm pc}\,R_{0.1}\,r_{\rm pc}^{3}\,T_{2}^{1/2}\,(M_{\rm ej}/M_{\sun})^{-1}\,(\driftvelmag/1000\,{\rm km\,s^{-1}})^{-1}$, while the Alfv\'en RDI dominates at shorter wavelengths with $t_{\rm grow} \sim 3\,{\rm yr}\,B_{0,\,{\rm mG}}^{-1/2}\,r_{\rm pc}^{3/2}\,R_{0.1}^{1/2}\,(\lambda/{\rm au})^{1/2}\,(M_{\rm ej}/M_{\sun})^{-1/2}\,(\driftvelmag/1000\,{\rm km\,s^{-1}})^{-1/2}$. The ``hot'' case will essentially be the next (Sedov-Taylor) phase.

Once the reverse shock has propagated through the ejecta (after it sweeps up a mass comparable to its own), the SNe remnant enters the energy-conserving Sedov phase, with shell densities $n\sim 4\,n_{\rm ISM} \sim n_{0}\,{\rm cm^{-3}}$, velocities $v_{\rm ej} \sim 840\,{\rm km\,s^{-1}}\,(E_{51}/n_{0}\,r_{10}^{3})^{1/2}$, and post-shock temperatures $T \sim 1.6 \times10^{7}\,E_{51}\,n_{0}^{-1}\,r_{10}^{-3}$ (where $E_{51}$ is the ejecta energy in $10^{51}\,{\rm erg}$ and the ejecta are at radius $r_{10}\,10\,{\rm pc}$). The gas in the ejecta is de-celerating according to energy conservation; adopting the usual Sedov solution gives a drift velocity $\driftvelmag / c_{s} \sim 0.8\,R_{0.1}\,n_{0}^{-1}\,r_{10}^{-1}\,(\tilde{U}/0.3)^{-2}$. So we expect trans- or sub-sonic drift with Coulomb drag dominating (although see the caveats about $\tilde{U}$ at these temperatures in the CGM discussion above, which could lead to Epstein drag dominating), with $c_{s}\,t_{s} \sim (1{\rm pc})\,R_{0.1}\,n_{0}^{-1}\,(\tilde{U}/0.3)^{-2}$, and $t_{s} \sim 3000\,{\rm yr}\,R_{0.1}\,n_{0}^{-1/2}\,r_{10}^{3/2}\,E_{51}^{-1/2}\,(\tilde{U}/0.3)^{-2}$. We expect modest $\beta\sim 1$, giving very large $\tau \sim 10^{6}\,E_{51}\,n_{0}^{-3/2}\,\beta^{-1/2}\,R_{0.1}^{-1}\,r_{10}^{-3}\,(\tilde{U}/0.3)^{-1}$. Importantly, under these conditions it is generally believed that a large fraction of the dust is efficiently destroyed via sputtering, so $\mu$ may be reduced to $\mu \ll 0.001$. Given the low $\mu$, subsonic or transsonic drift, and very large $\tau$ (which suppresses the mid-$k$ slow modes), the RDIs here may be heavily suppressed during these stages (with growth times $\gtrsim$\,Myr), except at very small scales ($t_{\rm grow}/t_{\rm expansion} \sim E_{51}^{1/2}\,(\mu/0.001)^{-1/2}\,r_{10}^{-5/2}\,n_{0}^{-5/4}\,\beta^{-1/4}\,(\lambda / 10^{10}\,{\rm cm})^{1/2}$ for $\lambda \ll {\rm au}$). 

After the remnant sweeps up a mass $\sim 3000\,M_{\sun}$, it enters the momentum-conserving/cooling/snowplow phase. The shell continues to decelerate (conserving momentum), with velocity $v_{\rm ej} \sim 170\,{\rm km\,s^{-1}}\,n_{0}^{-1}\,r_{30}^{-3}\,\tilde{p}$ (where $\tilde{p}$ is the ``terminal momentum'' at the cooling radius, relative to the value typically measured in simulations of $\sim 4\times10^{5}\,M_{\sun}\,{\rm km\,s^{-1}}$; \citealt{cioffi:1988.sne.remnant.evolution,thornton98,martizzi:sne.momentum.sims,walch.naab:sne.momentum,kim.ostriker:sne.momentum.injection.sims,hopkins:sne.methods}). The shell is at $r_{30}\,30\,{\rm pc}$, but it is has cooled ($T\sim 10^{2}\rightarrow10^{4}\,$K), and is more dense at the front owing to the radiative shock ($n_{\rm shell} \sim \mathcal{M}^{2}\,n_{\rm ISM}$; although since the shell is magnetized the jump may be weaker, so we will simply write $x_{\rm jump} = n_{\rm shell}/n_{\rm ISM}$). This sources a drift $\driftvelmag/c_{s} \sim 5\,\tilde{p}\,R_{0.1}^{1/2}\,n_{0}^{-3/2}\,r_{30}^{-7/2}\,T_{4}^{-1/2}\,x_{\rm jump}^{-1/2}$, with $t_{s}/t_{\rm expansion} \sim 0.3\,(R_{0.1}/n_{0}\,r_{30}\,x_{\rm jump})^{1/2}$ and $c_{s}\,t_{s} \sim {\rm pc}\,(n_{0}\,R_{0.1}\,T_{4}/x_{\rm jump})^{1/2}\,\tilde{p}^{-1}\,r_{30}^{7/2}$. Thus, over the range of temperatures and radii (or entrained masses) over which this phase is relevant, we should expect a broad range from sub-to-supersonic drift, with $\beta\sim 1$, Epstein drag dominating (since the gas is cold), and large $\tau \sim 2000\,n_{0}\,\tilde{p}^{-1}\,T_{4}^{3/2}\,\beta^{-1/2}\,R_{0.1}^{-3/2}\,r_{30}^{7/2}$. Since the gas and dust mass is primarily that which is entrained, and sputtering is no longer efficient, the dust-to-gas ratio should reflect the ISM ($\mu\sim 0.01$). Under these conditions, a wide range of modes are present: at wavelengths close to $\lambda\sim 2\pi\,\langle t_{L} \rangle\,\driftvelmag \sim 0.01\,{\rm pc}\,\tilde{p}\,x_{\rm jump}^{-1}\,n_{0}^{-2}\,R_{0.1}^{5/2}\,\beta^{1/2}\,T_{4}^{-3/2}\,r_{30}^{-7/2}$ (as well as around $\lambda\sim 2\pi\,\langle t_{L}\rangle\,v_{f,\,0}$), the gyro modes have very rapid growth timescales $t_{\mathrm{grow}}\sim \mu^{-1/2}\,\langle t_{L} \rangle \sim 300\,{\rm yr}\,R_{0.1}^{2}\,(\beta/n_{0}\,x_{\rm jump})^{1/2}\,T_{4}^{-3/2}$  (compare to the expansion time, $t_{\rm expansion} \sim 1.7\times10^{5}\,{\rm yr}\,n_{0}\,\tilde{p}^{-1}\,r_{30}^{4}$). At shorter wavelengths the mid-$k$ regime of the high-$\tau$ Alfv\'en/fast-wave RDI dominates with $t_{\rm grow}/t_{\rm expansion} \sim (\lambda/0.001\,{\rm pc})^{1/2}\,r_{30}^{3}\,n_{0}^{-5/4}\,(x_{\rm jump}/\beta)^{1/4}\,(T_{4}/\tilde{p}\,R_{0.1})$, until at very short wavelengths ($\lambda \ll\,r_{30}^{7/2}\,{\rm au}$) where the high-$k$, high-$\tau$-enhanced Alfv\'en RDI dominates with $t_{\rm grow}/t_{\rm expansion} \sim 0.001\,(\lambda/{\rm au})^{1/3}\,\tilde{p}^{1/3}\,r_{30}^{-5/3}\,R_{0.1}^{2/3}\,(n_{0}\,T_{4})^{1/2}\,(x_{\rm jump}/\beta)^{1/6}$. 

This will have a wide range of consequences. The initially-accelerated dust in the vicinity of the SNe will clump strongly, potentially shielding it against sputtering, and changing the cooling, abundance, and molecular properties of the ISM through which it moves (see the discussion of the instabilities in the warm and cold ISM). During the early free expansion phase, as many detailed theoretical studies have pointed out, clumpiness in the ejecta, and especially of the initial grain ``seeds,'' can radically alter the efficiency of grain growth, and therefore the ensuing grain size distribution, composition, and survival through the energy-conserving phase in SNe ejecta \citep{bianchi.2007:dust.formation.sne.ejecta,silvia:2010.dust.destruction.sims,hirashita:2011.dust.size.vs.growth,sarangi:2015.dust.condensation.in.sne}. This may, in part, be related to the very high condensation efficiencies observed in some SNe referenced above. Sufficiently non-linear manifestations of the highly-supersonic RDIs can lead to dust ``filaments'' which can effectively drift ``through'' the shock (Moseley et al., in prep.), potentially providing a means for some dust to avoid the reverse shock entirely. 
The strong gyro-RDI modes may also play an important role in amplifying or altering the structure of magnetic fields in SNe remnants. In both this energy-conserving stage and later (snowplow) stages, the Alfv\'en RDI will generate dust-to-gas fluctuations on small scales, which could source turbulence in the shell and the ubiquitously-observed dust ``streamers'' and micro-structure. This could, in turn, significantly alter interpretations of the far-IR and sub-mm fluxes of the remnants.

\vspace{-0.5cm}
\subsubsection{Coronal Dust}
\label{sec:application.env:coronae}

Observations have long indicated the presence of dust throughout the solar corona (the ``F-corona'') at least down to radii $r < 2\,R_{\sun}$, with the density of grains rising proportionally to $\sim r^{-1}$, similar to gas \citep[see e.g.][for reviews]{mann:2004.dust.near.the.sun.review,mann:2006.solar.system.dust.review}. This includes a range of sizes and compositions (with larger grains $R_{0.1}\gg1$ more easily observed, but smaller grains likely to appear closer to the Sun, especially if deposited by sub-grazing comets; e.g., \citealt{mukai:1974.f.corona.dust.material.models,kimura.mann:1998.f.corona.obs.review}). More recently similar coronae have been inferred around a large number of nearby stars \citep{kral:2017.exozodiacal.dust.cloud.review}. 

In the solar corona, some typical parameters are $T \sim 10^{6}\,$K, $n \sim 10^{8}\,{\rm cm^{-3}}$, $\beta\sim 0.001\,\beta_{-3}$ (scaling with distance from the sun). If dust is at a distance $r_{\sun} \equiv r / R_{\sun}$ from the sun or another star with luminosity $L_{0} = L/L_{\sun}$, then radiative acceleration produces a drift velocity\footnote{For larger ($\gg 1\,\mu{\rm m}$) grains around a solar-type star, gravity dominates over radiation pressure, so we can consider this as the dominant source of drift, but the qualitative scalings above are identical. The Poynting-Robertson effect adds another external acceleration to source drift, but this is sub-dominant by a factor $\sim \driftvelmag/c$. Even in a gravity-dominated circular orbit, gas pressure means the gas rotates at sub-Keplerian speeds, generating a relative {\em dust-gas} drift $\driftvelmag/c_{s} \sim c_{s}/V_{c}$ (the drift which matters here), which usually dominates over Poynting-Robertson.} (before projection onto the field) of $\driftvelXmag/c_{s} \sim 30\,(L_{0}\,Q_{\rm abs}/r_{\sun}\,n_{8}\,T_{6})^{1/2}$, $t_{s}\sim 260\,{\rm sec}\,R_{0.1}\,(r_{\sun}/L_{0}\,Q_{\rm abs}\,n_{8})^{1/2}$,  $c_{s}\,t_{s} \sim 0.04\,R_{\sun}\,R_{0.1}\,(T_{6}/L_{0}\,Q_{\rm abs}\,n_{8})^{1/2}$, and $\tau \sim 100\,T_{6}^{3/2}\,\tilde{U}\,R_{0.1}^{-1}\,(r_{\sun}/L_{0}\,Q_{\rm abs}\,\beta_{-3})^{1/2}$. Thus, coronal dust  is in a unique regime in several respects: $\beta \ll 1$ and the dust is drifting super-sonically (in the Epstein regime), but sub-Alfv\'enically, and the system has moderately large $\tau$. 

If we evaluate the growth rates using parameters appropriate for grains around the base of the corona ($Q_{\rm abs}\sim 0.1$, $\tilde{U}\sim 0.3$ per the CGM discussion above, at $T_{6}=\beta_{-3}=n_{8}=L_{0}=r_{\sun}=1$), then rather remarkably, the parallel, acoustic quasi-drift mode has a growth time $t_{\mathrm{grow}}\sim 1.2\,R_{0.1}\,(0.01/\mu)\,{\rm days}$ over a wide range of $\lambda \gg R_{\sun}$ (formally up to $\sim$\,au, but clearly a global solution is needed on such large scales). At shorter (intermediate) wavelengths the fastest-growing modes are the Alfv\'en-wave RDIs in the mid-$k$ regime with growth timescales $t_{\mathrm{grow}}\sim 1.7\,{\rm day}\,(\lambda/R_{\sun})^{1/2}\,(0.01/\mu)^{1/2}$ (this estimate is only weakly dependent on grain size), while at very short wavelengths $\lambda \lesssim 10\,{\rm km}$, the high-$k$ Alfv\'en RDI appears with $t_{\rm grow}\sim {\rm sec}\,(\lambda/{\rm km})^{1/3}\,(0.01/\mu)^{1/3}$. Note that we have kept the the dust-to-gas ratio $\mu$ arbitrary in these estimates, as this is highly uncertain in the near vicinity of the sun (extrapolating the models reviewed in \citealt{mann:2004.dust.near.the.sun.review} suggests $\mu\sim 10^{-4}\rightarrow10^{-3}$ at $\lesssim 4\,R_{\sun}$, but the value is expected to drop rapidly as $r\rightarrow R_{\sun}$ owing to a number of effects). 

It has long been recognized that complicated dust dynamics (formation, destruction, drift owing to radiation, gravity, and the Poynting-Robertson effect) and dust-gas interactions can produce features and enhancements in the F-corona and ``rings'' of dust observed at times \citep{mukai:1979.f.corona.model,kimura:1998.solar.dust.composition.rings,mann:2000.f.corona.enhancements.from.dynamics,kobayashi:2009.f.corona.ring.form.from.drift}. However, these studies have not included the physics necessary to follow the RDI, which could lead to much larger local enhancements (on long wavelengths, the parallel modes tend to induce ring/shell/arc structures). This could be important for dust-ring dynamics, and perhaps related to historical claims of variability \citep[e.g.][]{prasad:1995.solar.dust.ring.variability,ohgaito:2002.solar.corona.dust.1998.obs}. The Alfv\'en RDI (at shorter wavelengths) directly sources strong interactions between the dust and the magnetic fields, so may be important to understand the observed interactions of dust with the particles and magnetic fields of coronal mass ejections and the solar wind \citep{ragot:2003.f.corona.variations}. 
It is likely necessary to account for the RDI in understanding the observed fluctuations and structure functions in observed dust properties carried by the solar wind \citep{strub:2015.ulysses.fluctuations.in.dust.flow}. Additionally, more recent observations have argued for time-variable streamers and other small-scale structures in the F-corona, which could be driven by the RDI \citep{shopov:2008.f.corona.structure.interactions.streamers}.

\vspace{-0.5cm}
\subsubsection{Winds Around Cool Stars}
\label{sec:application.env:coolstars}

Around cool, giant stars, dust forms in the photosphere and may be critical for launching winds. These conditions are high-density ($n\gtrsim 10^{12}\,{\rm cm^{-3}}$), low temperature ($T\lesssim 1000\,$K) and modestly magnetized (intermediate $\beta$), with $R_{d}\sim 0.001\rightarrow1\,\mu{\rm m}$, suggesting $\tau \ll 1$ with Epstein drag dominating over Coulomb drag. Both simple analytic arguments\footnote{To summarize from \papertwo: in a wind with $\rho=\dot{M}/(4\pi\,r^{2}\,v_{\rm wind})$, with $v_{\rm wind} = v_{10}\,10\,{\rm km\,s^{-1}}$, $r=r_{100}\,100\,R_{\sun}$, $\dot{M}\sim \dot{M}_{-3}\,10^{-3}\,\msun\,{\rm yr^{-1}}$, and $T\sim 1000\,$K, around a giant with luminosity $L\sim L_{5}\,10^{5}\,L_{\sun}$, assuming geometric absorption gives $\driftvelmag/c_{s} \sim 2\,(L_{5}\,v_{10}/\dot{M}_{-3}\,T_{3})^{1/2}$ (with corresponding $t_{s} \sim 1\,{\rm sec}\,R_{0.1}\,r_{100}^{2}\,(v_{10}/L_{5}\,\dot{M}_{-3})^{1/2}$ and $c_{s}\,t_{s} \sim 3\times10^{5}\,{\rm cm}\,T_{3}^{1/2}\,(t_{s}/{\rm sec})$). The long-wavelength ($\lambda \gtrsim 10^{8}\,{\rm cm}$), mid-$k$, and short-wavelength (from $10^{4}\,{\rm cm}$ to the viscous scale $\sim 10\,{\rm cm}$) regimes at  are all present and relevant, with growth rates at long wavelengths of $t_{\rm grow}/t_{\rm wind} \sim 0.02\,v_{10}^{4/3}\,(R_{0.1}\,r_{100}/\dot{M}_{-3})^{1/3}\,T_{3}^{-1/2}\,(\lambda / r)^{2/3}$ (where $t_{\rm wind}=r/v_{\rm wind}$ is the wind expansion time) and at short wavelengths (approaching the viscous scale) of $t_{\rm grow} \sim 0.1\,{\rm sec}\,R_{0.1}^{2/3}\,T_{3}^{-1/2}$} and detailed simulations \citep{macgregor:grains.cool.star.eventually.decouple,hartquist:bfield.dust.coupling.cool.star.winds,woitke:2d.rad.pressure.dust.agb.wind.models} have shown that the drift velocities are transsonic, with $\driftvelmag \sim 0.1\rightarrow10\,c_{s}$, depending on local conditions (e.g., different stars or different regions within the same wind).

Under these conditions, the regions with super-sonic drift will be dominated by the fast-magnetosonic mode resonance, which for low-$\tau$ and $\driftvelmag \gg v_{f,\,0}$ behaves effectively identically to the acoustic RDI studied in detail in \papertwo. Even if $\beta \ll 1$, the only difference in this regime from our \papertwo\ scalings is the replacement $c_{s} \rightarrow v_{f,\,0}$. Under these conditions we expect the acoustic or fast-magnetosonic RDI to be important, with growth timescales much faster than wind expansion times even for the largest scale modes (wavelengths of order the distance from the star), reaching growth timescales as fast as $t_{\mathrm{grow}}\ll 1\,$second on the smallest (viscous) scales ($\lambda\ll 1\,$m). Essentially all scales are unstable, and the interesting modes span the long-wavelength, mid-$k$ and high-$k$ resonant regimes. For more detailed discussion of these cases, we refer to \papertwo.

However, the addition of the slow-mode resonance, when we consider MHD, is of particular interest for the sub-sonic drift regime. Recall, for modest $\beta$ and low $\tau$, the slow-mode resonant RDI has growth rates that scale similarly to the fast-mode RDI, except for being suppressed by a factor $\sim (\driftvelmag/v_{f,\,0})^{\alpha}$ with $\alpha$ in the range $\sim 2/3\rightarrow 1$. But since the drifts expected in cool-star winds are only modestly sub-sonic, this is not a large suppression factor (this also likely means they will not be strongly-suppressed by ambipolar diffusion, if present). This situation is quite different than  the neutral hydrodynamic system (acoustic instability; \papertwo), where there is a very sharp change in the character of the instabilities for $\driftvelmag < c_{s}$ (at which point the resonances cease to exist entirely in the absence of magnetic fields). So we expect a rather smooth transition between the two cases: the behavior for sub-sonic or super-sonic drift will be similar, but with faster growth rates in regions of faster drift. In these systems the RDI can source both large-scale features commonly seen in the outflows (arcs/shells/rings, and global dust asymmetries; see e.g., \citealt{morris:1993.cool.wind.dust.drag.instability.slow.saturated.mode,1994A&A...288..255W,deguchi:1997.dust.envelope.pne.spherical.drag.instability.quasi.resonant,balick:2002.pne.shape.structure.review}), as well as the ubiquitous observed dust knots/filaments/fliers/streamers \citep{odell:1996.pne.cometary.knots.from.instabilities,odell:2002.pne.knots.review,balick:1998.knots.fliers.cometary.tails.pne,matsuura:2009.pne.knots}. In addition, this provides a natural mechanism to generate observed gas clumpiness and turbulence in outflows \citep{2003ApJ...582L..39F,young:2003.clumpy.wind.models,2007Natur.447.1094Z,2010ApJ...724L.133A,2012A&A...537A..35C}.

\vspace{-0.5cm}
\subsubsection{The Neutral and Cold Interstellar Medium (GMCs, Star-Forming Regions, and AGN Torii)}
\label{sec:application.env:cold.ism}

In the cold ($T\ll 1000\,$K) ISM (e.g., the cold-neutral and molecular media, including molecular clouds, star forming regions, galactic nuclei, the dusty obscuring ``torii'' and narrow-line regions around AGN), we have $T\sim 10\rightarrow1000\,$K, $R_{d}\sim 0.01\rightarrow10\,\mu{\rm m}$, $n_{\rm gas}\sim 10\rightarrow10^{6}\,{\rm cm^{-3}}$, $\beta\sim 1$, and $f_{\rm ion}\sim 10^{-8} \rightarrow 10^{-5}$. Under these conditions, Epstein drag dominates over Coulomb drag, Lorentz forces are weak ($\tau \ll 1$), and the drift velocities tend to vary from trans-sonic for the smallest grains, through to highly super-sonic for large grains (especially around bright sources like massive stars or star clusters or AGN, where radiation pressure can be sufficient to launch strong outflows of gas and dust). For example, one finds $\driftvelmag \gtrsim 100\,c_{s} \gg v_{f,\,0}$ in quasar-driven outflows in the torus or narrow-line region. 
%

Given the high drift velocities, we generally expect the fast-magnetosonic RDI with $\tau \ll 1$ to dominate; as noted above this is essentially identical to the acoustic RDI from \papertwo. As shown there, the growth timescale can be many orders of magnitude faster than competing dynamical timescales in starburst and AGN nuclei, and the most interesting modes are usually the long-wavelength and mid-$k$ resonant modes. Since the scalings are essentially unchanged, we refer interested readers to \papertwo\ for detailed discussion of these cases.\footnote{To summarize the scalings in \papertwo: for an AGN with luminosity $L\sim L_{13}\,10^{13}\,L_{\sun}$, and dusty torus with inner radius around the dust sublimation radius $\sim 0.6\,{\rm pc}\,L_{13}^{1/2}$, midplane column density $N_{26}\,10^{26}\,{\rm cm^{-2}}$, temperature $T\sim 1000\,$K, the (Epstein) stopping time is $t_{s} \sim 12\,{\rm hr}\,R_{0.1}\,L_{13}^{1/4}\,N_{26}^{-1/2}$, $c_{s}\,t_{s}\sim 10^{10}\,{\rm cm}\,R_{0.1}\,L_{13}^{1/4}\,(T_{3}/N_{26})^{1/2}$, $\driftvelmag/c_{s} \sim 100\,L_{13}^{1/4}\,N_{26}^{-1/2}$, and $t_{\rm grow} \sim 10\,{\rm yr}\,R_{0.1}^{1/3}\,L_{13}^{-1/12}\,N_{26}^{1/6}\,(Z/Z_{\sun})^{-1/3}\,(\lambda/0.1\,{\rm pc})^{2/3}$ down to $\lambda \lesssim {\rm au}$, with $t_{\rm grow} \sim t_{s} \sim 10\,$hr for modes approaching the viscous scale. For a GMC ($T\sim 10-100\,$K, $n\sim 1-10^{3}\,{\rm cm^{-3}}$) which has converted a fraction $\sim 0.1\,\epsilon_{0.1}$ of its mass into stars (assuming a standard stellar initial mass function and that none have exploded, to obtain $e_{\rm rad}$), and total size $r\sim r_{10}\,10\,{\rm pc}$, we estimate $\driftvelmag \sim 10\,r_{10}^{1/2}\,\epsilon_{0.1}^{1/2}$, $t_{s} \sim 4\times10^{4}\,{\rm yr}\,R_{0.1}\,n_{2}^{-1}\,(r_{10}\,\epsilon_{0.1})^{-1/2}$, $c_{s}\,t_{s} \sim 0.04\,{\rm pc}\,R_{0.1}\,n_{2}^{-1}\,(T_{2}/r_{10}\,\epsilon_{0.1})^{1/2}$, giving $t_{\rm grow} \sim 0.3\,{\rm Myr}\,(\lambda/0.1\,{\rm pc})^{1/2}\,(R_{0.1}/n_{2})^{1/2}\,(r_{10}\,T_{2}\,\epsilon_{0.1})^{1/4}$.}

As discussed in \papertwo, the RDI  could be an important part of observed phenomena ranging from the well-studied clumpiness, sub-structure and turbulence in the dusty AGN ``torus'' \citep[see e.g.][and references therein]{krolik:clumpy.torii,nenkova:clumpy.torus.model.1,mor:2009.torus.structure.from.fitting.obs,hoenig:clumpy.torus.modeling}, time-variability in AGN dust obscuration \citep{mckernan:1998.agn.occultation.by.clumpy.outflow,risaliti:nh.column.variability}, AGN winds driven by radiation pressure on dust \citep{murray:momentum.winds,elitzur:torus.wind,miller:2008.clumpy.agn.disk.wind,roth:2012.rad.transfer.agn,wada:torus.mol.gas.hydro.sims}, observed dust-gas segregation in GMCs \citep{padoan:dust.fluct.taurus.vs.sims} and abundance anomalies sourced by these \citep{hopkins:totally.metal.stars,hopkins.conroy.2015:metal.poor.star.abundances.dust}, dust growth and coagulation (believed to occur primarily  in the cold, dense ISM; \citealt{draine:2003.dust.review} and references therein), dust chemistry/cooling physics critical for star formation and formation of complex organic compounds and molecules \citep{goldsmith:molecular.dust.cooling.gmcs,dopke.2013:fragmentation.all.dust.levels.but.enhanced.with.crit.dust,ji:2014.si.dust.cooling.threshold.for.early.stars,chiaki:2014.critical.dust.abundance.for.cooling}, radiation-pressure driven outflows from massive stars \citep{murray:momentum.winds,thompson:rad.pressure,krumholz:2007.rhd.protostar.modes,hopkins:rad.pressure.sf.fb,grudic:sfe.cluster.form.surface.density}, and thermal regulation of proto-star formation via heating dust in coupled dust-gas cores \citep{guszejnov.2015:feedback.imf.invariance}. 

Of course, the drift velocities will be sub-sonic for sufficiently small grains, and even for larger grains in certain regions (given the highly inhomogenous nature of these systems). In this case, as noted in the discussion of cool stars  above, the slow-magnetosonic RDI does become interesting, and (since the drift is only modestly sub-sonic) produces instabilities with similar behavior and growth rates to the fast-mode RDI. However as noted in \S~\ref{sec:scales.validity}, ambipolar diffusion will suppress the slow resonance when $\driftvelmag \ll c_{s}$ and $f_{\rm ion} \ll 10^{-6}$ (it has little or no effect on the fast resonances discussed above).

\vspace{-0.5cm}
\subsubsection{Proto-Planetary Disks, Proto-Stellar Disks, \&\ Planetary Atmospheres}
\label{sec:application.env:ppd}

In proto-planetary or proto-stellar disks, the high densities and very low ionized fractions mean that Epstein (or Stokes, for large $\gtrsim\,$cm-sized pebbles) drag strongly dominates over Coulomb drag. Similarly Lorentz forces are extremely weak ($\tau \ll 10^{-10}$), and magnetic fields (even in the active regions of the disk) are likely relatively weak also ($\beta \gg 1$). Drift velocities are highly sub-sonic $\driftvelmag \ll 10^{-2}\,c_{s}$ \citep{chiang:2010.planetesimal.formation.review}, so the acoustic RDI from \papertwo\ has very low maximum growth rates  and is unlikely to be important. In principle, the slow-mode RDI, even with $\beta \gg 1$,  has growth   rates that become arbitrarily large at high-$k$. However, as discussed in detail in \citet{squire:rdi.ppd}, the ionization fractions in proto-planetary disks  are sufficiently low that  non-ideal magnetic dissipation terms (e.g., Ohmic resistivity and ambipolar diffusion) cannot be neglected. As discussed above (\S~\ref{sec:scales.validity}), these terms suppress the  short-wavelength slow-mode growth. Instead, \citet{squire:rdi.ppd} focuses on several other classes of RDI that could be  important for planetesimal growth, including the the well-studied ``streaming instability'' \citep{youdin.goodman:2005.streaming.instability.derivation}, which is an RDI with disk epicyclic oscillations, the related vertical-epicyclic RDI or ``settling instability'' (which has growth times comparable to the disk dynamical time at all grain sizes), and the \BV\ RDI. 

The situation is similar in most planetary atmospheres. There, the  comparatively weak centrifugal/coriolis forces suggest that the \BV\ RDI is likely to be more important than RDIs associated with epicyclic oscillations.

\vspace{-0.5cm}
\section{Discussion}
\label{sec:discussion}

We have studied the linear stability of dust-gas systems where the gas obeys the ideal MHD equations and dust experiences some combination of an arbitrary drag law and Lorentz forces. We show that such systems   exhibit a broad spectrum of ``resonant drag instabilities'' (RDIs), as generically predicted in \paperone, with several unstable modes at all wavenumbers. We identify a large number of different instability families, each of which has distinct behavior. These are summarized in \S~\ref{sec:overview.modes}, but include MHD-wave RDI families (the ``Alfv\'en RDI,'' ``Fast-Magnetosonic RDI,'' and ``Slow-Magnetosonic RDI''), the ``gyro-resonant RDI'' families (one familiy for each MHD wave family: Alfv\'en-gyro RDI, fast-gyro RDI, and slow-gyro RDI), the ``pressure free'' mode (at sufficiently long wavelengths), the aligned acoustic modes (out-of-resonance modes), and the ``cosmic-ray streaming'' modes. Each of these families has distinct behaviors, structure, and possible resonances, and can be the fastest-growing instability over some range of wavelengths under different conditions. 

Although some of these instabilities are, in a  general sense, related to certain well-known instabilities (for example, of cosmic rays), none of them has (to our knowledge)  been previously recognized before \paperone. Although we first identified the existence of the MHD-wave RDI families in \paperone, several 
features of this work are novel compared to previous works, in particular the inclusion of grain charge (grain Lorentz forces). This significantly modifies the phenomenology of the instabilities, increasing the importance of the Alfv\'en-wave related modes, and destabilizing the gyro-resonant families and the ``cosmic ray streaming'' mode (which depend on Lorentz forces on dust).

In a very broad sense, the qualitative behavior of the pressure-free mode and MHD-wave RDI families is similar to that of the ``acoustic RDI'' studied in \papertwo. At sufficiently long wavelengths, the pressure-free mode has the fastest growth rates (and is nearly identical in non-magnetized fluids). At intermediate and short wavelengths, the fastest-growing MHD-RDI modes are usually those with wavevector ${\bf k}$ oriented at a ``resonant angle,'' such that $\driftvel \cdot {\bf k} = v_{p}(\hat{\bf k})\,k$, where $v_{p}(\hat{\bf k})$ is the phase speed of the corresponding MHD wave at that angle. The growth rates increase without limit towards short wavelengths, and are only weakly dependent on the dust-to-gas mass ratio $\mu$ (as $\sim \mu^{1/3}$ or $\sim\mu^{1/2}$), the strength ($\beta$) and orientation ($\hat{\bf B}_{0}$) of magnetic fields,  the details of the dust drag law (e.g., Epstein, Stokes, Coulomb, or other drag laws, which we parameterize through $\coeffTSrho$, $\coeffTSv$, etc.), and the dust charge (including any dependence on temperature, parameterized through $\coeffTLrho$). We show that Lorentz forces on charged grains do not generally suppress these instabilities; in fact Lorentz forces can enhance the growth rates in some regimes. The most important difference between the MHD-wave RDI cases studied here and the acoustic RDI (\papertwo) is the existence of the slow and Alfv\'en waves. Because these have phase velocities that can be arbitrarily small, there is  {\em always} a range of angles  that satisfy the resonance condition  and produce rapid growth rates,  at {\em any} non-zero drift velocity. In contrast, in the un-magnetized acoustic case, the drift must be super-sonic ($\driftvelmag \ge c_{s}$) in order to satisfy the resonance condition.

All of the MHD-wave RDI families generate compressive modes in the dust which directly source exponentially growing dust-to-gas fluctuations. For the magnetosonic (fast and slow) RDIs, this occurs via the interaction of dust drag with the acoustic gas density fluctuations. For the Alfv\'en RDI the perturbed gas mode is, to leading order, incompressible, but the presence of Lorentz forces on dust means the transverse magnetic field fluctuations interact with the longitudinal dust velocities, generating a strong compressible dust response.

The gyro-resonant RDI families have a fundamentally different character. These have their fastest-growth when the combination of drift velocity and gas wave phase velocity matches the gyro velocity along the same direction, i.e., $|\driftvel\cdot{\bf k} \pm v_{p}(\hat{\bf k})\,k| = \langle t_{L} \rangle^{-1}$. Unlike the MHD-wave RDIs, for a mode propagating in a given direction $\hat{\bf k}$, this resonance occurs at a single specific wavenumber $k_{\rm gyro}$. The growth rates are sharply-peaked around $k_{\rm gyro}$ with growth rates $\Im(\omega)\sim \mu^{1/2}/\langle t_{L} \rangle$, which is again sub-linear in the dust-to-gas ratio, but independent of wavenumber (dependent only on the Larmor time). Thus they are more akin to the \BV\ modes discussed in \paperone\ and \citet{squire:rdi.ppd}, or resonant cosmic-ray instabilities \citep{kulsrud.1969:streaming.instability}. Because the gyro RDIs must overcome resistance by drag, these modes are only unstable when the Larmor time is shorter than the drag time ($\tau\gtrsim 1$). The gyro-RDI modes resemble perturbed gyro motion and are only weakly compressible in both dust and gas, primarily involving unstable growth of the dust gyro motion and the perturbed transverse field lines. 

In the limit where the Lorentz forces become extremely strongly-dominant over drag, the drift is tightly aligned along field lines and various ``parallel'' instabilities dominate. For example if the dust is streaming sufficiently super-Alfv\'enically, instabilities resembling the cosmic-ray streaming instabilities appear, with growth rates $\sim \mu^{1/2}\,\driftvelmag\,k$ increasing rapidly at short wavelengths.

We consider a broad range of astrophysical applications (\S~\ref{sec:application}). Based on simple order-of-magnitude estimates, we expect these instabilities to be important in a very broad range of astrophysical systems with dust, including: cool-star winds, the solar and other stellar coronae, SNe explosions and remnants (with qualitatively different behaviors for dust in or around the ejecta at each stage of the initial explosion and subsequent blastwave/remnant evolution), HII regions, star-forming GMCs and galactic nuclei, AGN obscuring torii and narrow-line regions, the diffuse warm ($T\sim 10^{4}\,$K) neutral and ionized medium, and the circum-galactic and inter-galactic medium (in regions laden with dust). Over the very broad dynamic range spanned by these systems, growth timescales of the interesting instabilities can be as short as $t_{\mathrm{grow}}\lesssim 0.1\,$seconds or as long as $t_{\mathrm{grow}}\sim 10^{9}\,$yr. Most importantly, in many cases, the growth timescales are much shorter than other relevant dynamical times of the systems. The non-linear outcomes are likely to be quite different depending on both the system properties and which of the various different RDIs is the fastest-growing (dominant) mode. For example, both the ``pressure-free'' mode and the magnetosonic-wave RDIs directly source large dust-to-gas fluctuations but in a very different manner: in the pressure-free mode the dust is collected in large scale arcs/shells/planes perpendicular to the drift direction, whereas the magnetosonic-wave RDIs concentrate dust into filaments and ``streamers'' aligned with propagation (see \papertwo). The ``cosmic ray streaming''-type instability, on the other hand, may excite high-frequency Alfv\'en waves which scatter grains and realign magnetic fields.

Any of these outcomes could have important implications. Grain-grain collisions, coagulation, shattering, grain-ISM chemistry, molecule formation, grain polarization (alignment of spinning dust with magnetic fields), ISM cooling in dusty gas, launching of radiation-pressure driven winds (via radiation impinging on dust, in particular), retention of dust grains in regions with strong radiation fields, visual and extinction morphologies of dusty systems, local extinction curve variations, magnetic field structure in around bright sources and in the distant CGM/IGM --- each of  of these could, in principle, be substantially altered by the instabilities here. Related physics may also play a role in  resolving several long-standing observational puzzles in these areas (for examples, see \S~\ref{sec:application.env}). However, to make detailed predictions and understand the observational consequences of the RDI for these systems we require numerical simulations, which are the only way to realistically explore the non-linear saturation of the instabilities.

Obvious analytic extensions of the work here include considering non-ideal MHD (as discussed in very limited fashion in \citealt{squire:rdi.ppd}), relevant when ionization fractions are low (but not vanishingly small, where we would simply have the acoustic case in \papertwo). As noted in \S~\ref{sec:scales.validity}, Ohmic resistivity is purely diffusive, but Hall MHD is non-diffusive and produces new wave families which generate new RDIs (e.g., the Whistler RDIs), while a proper treatment of ambipolar diffusion (where important)  requires a three-fluid  theoretical treatment (with ions, neutrals, and grains). Another possible extension is to include kinetic plasma effects: Braginskii viscosity and conduction are dissipative but anisotropic, so introduce additional terms that depend on the mode angles relative to the fields. As seen in a limited fashion above (see Fig.~\ref{fig: HII examples}) some modes may be damped, but others not, and new instabilities may appear that relate the RDI to other known instabilities of anisotropically conducting systems. A rich phenomenology of RDIs remains largely unexplored.

\vspace{-0.7cm}
\acknowledgments 
We would like to thank E.~S.~Phinney and E.~Quataert for helpful discussions, as well as our anonymous referee. Support for PFH \&\ JS was provided by an Alfred P. Sloan Research Fellowship, NASA ATP Grant NNX14AH35G, and NSF Collaborative Research Grant \#1411920 and CAREER grant \#1455342. JS was funded in part by the Gordon and Betty Moore Foundation
through Grant GBMF5076 to Lars Bildsten, Eliot Quataert and E. Sterl
Phinney.\\
\vspace{-0.2cm}
\bibliography{/Users/phopkins/Dropbox/Public/ms}

\appendix

\clearpage

\onecolumn

\vspace{-0.5cm}
\section{Dispersion Relation}
\label{sec:appendix:dispersion.relation}

To obtain the dispersion relation, begin with the linearized equations (Eq.~\ref{eqn:linearized}),\footnote{For stratified media we can explicitly include a pressure equation $D\delta P/Dt = c_{s}^{2}\,D\delta \rho/Dt$ (where $D/Dt$ is the comoving derivative) but for our homogeneous background this trivially evaluates to $\delta P = c_{s}^{2}\,\delta \rho$ so we simply insert this directly in Eq.~\ref{eqn:linearized}.} then make the Fourier {\em ansatz} as in the text, $\delta X = \delta X_{0}\, \exp{[\iimag\,({\bf k}\cdot {\bf x} - \omega\,t)]})$. Use the divergence constraint $0 = \nabla\cdot \delta{\bf B} = \iimag\,{\bf k} \cdot \delta {\bf B}$ to eliminate one component of $\delta{\bf B}$ ($\delta {\bf B}_{z} = -(k_{x}\,\delta {\bf B}_{x} + k_{y}\,\delta {\bf B}_{y})/k_{z}$), where we define the axes $\hat{x}$, $\hat{y}$, $\hat{z}$ as the unit vectors parallel to $(\driftvel\times {\bf B}_{0})\times\driftvel$, $\driftvel\times{\bf B}_{0}$, and $\driftvel$, respectively. For convenience we will make the problem dimensionless, by defining the units of density, velocity, and time equal to the homogeneous values of $\rho_{0}$, $c_{s}$, and $\langle t_{s} \rangle$; we also conveniently write $\delta{\bf B}$ in units of $|{\bf B}_{0}|$. 

Inserting a single mode, Eq.~\ref{eqn:linearized} can be written: $\omega\,{\bf X} = \mathbb{T}\, {\bf X}$, where ${\bf X} = (\delta \rho_{d},\ \delta {\bf v}_{x},\ \delta {\bf v}_{y},\ \delta {\bf v}_{z},\ \delta\rho,\ \delta {\bf u}_{x},\ \delta {\bf u}_{y},\ \delta {\bf u}_{z},\ \delta {\bf B}_{x},\ \delta {\bf B}_{y})$ (recall we eliminated $\delta {\bf B}_{z}$ already), and the matrix $\mathbb{T}$ is given by:
\begin{scriptsize}
\begin{align}
	\begin{bmatrix}
	k_{z} \driftvelmag & k_x & k_y & k_z & 0 & 0 & 0 & 0 & 0 & 0 \\
 0 & k_{z} \driftvelmag -i & -\iimag c_0 \tau  & 0 & 0 & \iimag & \iimag c_0 \tau  & 0 & 0 & \iimag \tau  \driftvelmag \\
 0 & \iimag c_0 \tau  & k_{z} \driftvelmag -\iimag & -\iimag s_0 \tau  & -\iimag \coeffTLrho s_0 \tau  \driftvelmag & -\iimag c_0 \tau  & \iimag & \iimag
   s_0 \tau  & -\iimag \tau  \driftvelmag & 0 \\
 0 & 0 & \iimag s_0 \tau  & k_{z} \driftvelmag -\iimag (\coeffTSv+1) & -\iimag \coeffTSrho \driftvelmag & 0 & -\iimag s_0 \tau  & \iimag
   (\coeffTSv+1) & 0 & 0 \\
 0 & 0 & 0 & 0 & 0 & k_x & k_y & k_z & 0 & 0 \\
 0 & \iimag \mu  & \iimag c_0 \mu  \tau  & 0 & k_x & -\iimag \mu  & -\iimag c_0 \mu  \tau  & 0 & -\frac{c_0
   \left(k_x^2+k_z^2\right)}{\beta  k_z} & -\frac{c_0 k_x k_y}{\beta  k_z}-\iimag \mu  \tau  \driftvelmag \\
 \iimag \mu  s_0 \tau  \driftvelmag & -\iimag c_0 \mu  \tau  & \iimag \mu  & \iimag \mu  s_0 \tau  & k_y+\iimag (\coeffTLrho-1) \mu  s_0
   \tau  \driftvelmag & \iimag c_0 \mu  \tau  & -\iimag \mu  & -\iimag \mu  s_0 \tau  & \frac{k_y \left(s_0 k_z-c_0 k_x\right)}{\beta 
   k_z}+\iimag \mu  \tau  \driftvelmag & -\frac{c_0 \left(k_y^2+k_z^2\right)+s_0 k_x k_z}{\beta  k_z} \\
 \iimag \mu  \driftvelmag & 0 & -\iimag \mu  s_0 \tau  & \iimag (\coeffTSv+1) \mu  & k_z+\iimag (\coeffTSrho-1) \mu  \driftvelmag &
   0 & \iimag \mu  s_0 \tau  & -\iimag (\coeffTSv+1) \mu  & \frac{s_0 \left(k_x^2+k_z^2\right)}{\beta  k_z} & \frac{s_0
   k_x k_y}{\beta  k_z} \\
 0 & 0 & 0 & 0 & 0 & -c_0 k_z & s_0 k_y & s_0 k_z & 0 & 0 \\
 0 & 0 & 0 & 0 & 0 & 0 & -c_0 k_z-s_0 k_x & 0 & 0 & 0 \\
	\end{bmatrix} \label{eq: full matrix form}
\end{align}
\end{scriptsize}
where $c_{0} \equiv \cos{\theta}_{\bf Bw} = \hat{\bf B}_{0}\cdot \driftvelhat$ and $s_{0} \equiv (1-c_{0}^{2})^{1/2}$. 

The solutions $\omega$ are the eigenvalues of $\mathbb{T}$. We can write the full dispersion relation as the characteristic polynomial of this matrix, but the resulting 10th-order polynomial is not instructive. We instead focus in the text on numerical solutions to the general equation, and intuitive analytic expressions which apply under appropriate limits.

 Since this is a 10x10 sparse matrix and we consider limits where some parameters are much larger than others, in numerically evaluating $\omega$, care is needed. Numerical solutions in the text use the Python {\small mpmath} package which allows for arbitrary floating-point precision, retaining 30 significant figures, and were subsequently verified directly.

 \vspace{-0.5cm}
\section{Dust and Gas Resonances}
\label{sec:matrix resonances}

In this appendix, we discuss the origin of the mid-$k$, high-$k$, and gyroresonant modes, and 
how these relate to the matrix theory introduced in \paperone.
Our starting point is the full dust-gas matrix operator, Eq.~\eqref{eq: full matrix form}, and we use the dimensionless units of \S~\ref{sec:appendix:dispersion.relation}. Because we have 
organized the variables as ${\bf X} = ({\bf X}_{d},\ {\bf X}_{g})$, where ${\bf X}_{d}$ and ${\bf X}_{g}$ represent the dust and gas variables respectively, $\mathbb{T}$ has the form,
\begin{equation}
\mathbb{T} =\mathbb{T}_{0}+\mu\mathbb{T}^{(1)}=\left(\begin{matrix}
\mathcal{A} & \mathcal{C} \\
0 & \mathcal{F}
\end{matrix}\right) + \mu \left(\begin{matrix}
\mathcal{T}^{(1)}_{AA}  & \mathcal{T}^{(1)}_{AF}  \\
\mathcal{T}^{(1)}_{FA} & \mathcal{T}^{(1)}_{FF} 
\end{matrix}\right)\label{eq: basic matrix blocks}
\end{equation}
where $\mathcal{A}$ (the top-left $4\times 4$ block in Eq.~\eqref{eq: full matrix form}) represents the effect of the dust on the dust itself, $\mathcal{F}$  the effect of the gas on the gas, 
$\mathcal{C}$  the effect of the gas on the dust, and $\mathcal{T}^{(1)}_{FA}$ the backreaction from
the dust back onto the gas. We will calculate the eigenvalues of Eq.~\eqref{eq: basic matrix blocks} by 
considering $\mu\mathbb{T}^{(1)}$ (with $\mu\ll1 $) to be a perturbation to $\mathbb{T}_{0}$, 
analyzing the resulting perturbation ($\omega^{(1)}$) to the eigenvalues  of $\mathbb{T}_{0}$ (termed $\omega_{0}$) using perturbation theory. In other 
words, we shall find $\omega$, the eigenvalue of $\mathbb{T}$, through the expansion
\begin{equation}
\omega = \omega_{0} + \omega^{(1)}+\dots, \end{equation}
where $ \omega^{(1)}$ scales with some power of $\mu$.
 The basic result of \paperone\ was that if $\mathcal{A}$ and $\mathcal{F}$ share
an eigenmode (i.e., there is a resonance), then the system is likely unstable, $\Im(\omega^{(1)})\neq 0$, and that $\omega^{(1)}$ (which determines the growth rate of the resulting instability) scales
as $\omega^{(1)}\sim\mu^{1/2}$ (or $\omega^{(1)}\sim\mu^{1/3}$), rather than the usual perturbation theory expectation, 
$\omega^{(1)}\sim \mu$.
In this work, we have extended the dust model used by \paperone\ to account for charge on the grains. This 
changes the matrix $\mathcal{A}$, allowing a richer mode structure and the appearance of the gyro-resonance 
modes in the dust. We thus straightforwardly extend the theory of \paperone\ to account for this more complex dust physics.

Before continuing, we caution that these ideas only apply at sufficiently small $\mu$ such that $\mu\mathcal{T}^{(1)}$ can be considered a perturbation. With so many combinations of parameters, even values 
one might usually consider ``small'' (e.g., $\mu\sim 0.01$) may not be sufficiently small, and non-resonant modes can dominate in certain regimes (for instance the low-$k$, ``pressure-free'' modes discussed 
in \S~\ref{sub: low k regime solutions} fall into this category). This is particularly true at high $\tau$, where the full mode structure
(at arbitrary $\mu$, $k$, $\driftvelmag/c_{s}$, $v_{A}/c_{s}$ etc.) becomes very complex and difficult to classify; see, e.g., \S\S~\ref{sec:parallel.modes:quasisound}--\ref{sub:cosmic.ray.streaming.mode} for some examples.

\vspace{-0.2cm}
\subsection{Dust Eigenmode Structure}\label{sub: resonances dust eigenmodes}
A key aspect of understanding the structure of the RDI solutions to the full matrix Eq.~\eqref{eq: full matrix form} is understanding the eigenmodes of the dust matrix $\mathcal{A}$. This can be written in the form 
\begin{equation}
\mathcal{A} = \left(\begin{matrix}
k_{z} \driftvelmag & {\bf k}^{T} \\ 
\bf{0} & k_{z} \driftvelmag \mathbb{I} + \mathcal{D}_{\mathrm{drag}} + \tau \mathcal{D}_{\mathrm{gyro}} 
\end{matrix}\right).
\end{equation}
The submatrices $\mathcal{D}_{\mathrm{drag}}$ and $\tau \mathcal{D}_{\mathrm{gyro}}$ describe, respectively, the effect of drag forces on the dust (i.e., the $-{(\delta {\bf v}-\delta{\bf u})}/{\langle t_{s} \rangle}$ term in Eq.~\eqref{eqn:linearized}) and the effect of the magnetic field on the dust (the $-{(\delta {\bf v}-\delta {\bf u})\times\hat{\bf B}_{0}}/{\langle t_{L} \rangle} $  term in Eq.~\eqref{eqn:linearized}), while $\mathbb{I}$ is the 
identity matrix.
With $\coeffTSv=0$, large-$\tau$, or for various special angles (e.g., $\cos^2\theta_{\bf Bw} =1$), the eigenvalues of $\mathcal{A}$ are particularly simple\footnote{In the general case (e.g., arbitrary $\tau$ or $\coeffTSv$) the fundamental character of the   eigenvalue solutions is unchanged, 
but  their form becomes  complicated and unintuitive.}; see \S~\ref{sec:gyro}. For example, with $\coeffTSv=0$ (i.e., when $t_s$ is independent of $|{\bf v}- {\bf u}|$) they are: (i) a dust density perturbation, with no associated velocity perturbation and frequency $\omega_{\mathcal{A}} = k_{z} \driftvelmag$ (due to the Doppler shift from $\driftvel$); (ii) a damped velocity 
perturbation along the magnetic field with $\omega_{\mathcal{A}} = k_{z} \driftvelmag - \iimag$ (recall that, because we work in units where $\langle t_{s}\rangle=1$, this is damping at the rate $\Im(\omega)=-\langle t_{s}\rangle^{-1}$); and
(iii) two damped gyration modes, which involve magnetic-field-influenced motion perpendicular to the field and have 
$\omega_{\mathcal{A}} = k_{z} \driftvelmag \pm \tau - \iimag$.

\vspace{-0.2cm}
\subsection{Resonance between the Dust and the Fluid}\label{sub:dust.gas.res.in.detail}

The matrix-resonance theory of \paperone\ then says that we should  attempt to match eigenmodes of the gas, denoted here by $\omega_{\mathcal{F}}$, to 
those of $\mathcal{A}$, so as to find the regions of parameter space where $\omega^{(1)}$ is largest (i.e., where  any instability grows the fastest).
As discussed 
in \S~\ref{sec:mhdwave.rdi:overview}, the fluid part of Eq.~\eqref{eq: full matrix form} (the bottom-right $6\times 6$ block)
supports 6 real oscillation modes (eigenmodes): shear-Alfv\'en waves with  $\omega_{\mathcal{F}} = \pm {\bf v}_{A}\cdot {\bf k}$, slow waves with $\omega_{\mathcal{F}} = \pm {\bf v}_{-}\cdot {\bf k}$, and fast waves with $\omega_{\mathcal{F}} = \pm {\bf v}_{+}\cdot {\bf k}$. To make progress, we must choose one of these eigenmodes and match it to a dust eigenmode $\omega_{\mathcal{A}}\approx \omega_{\mathcal{F}}$.
For the remainder of this section, we shall assume that we have done this, and denote the mode's eigenvalue by $\omega_{\mathcal{F}}$ and the corresponding left and right eigenmodes by $\xi_{\mathcal{F}}^{L}$ and  $\xi_{\mathcal{F}}^{R}$ respectively (these  satisfy $\xi^{L}_{\mathcal{F}} (\mathcal{F}-\omega_{\mathcal{F}} \mathbb{I})$ and $ (\mathcal{F}-\omega_{\mathcal{F}} \mathbb{I})\xi^{R}_{\mathcal{F}}$, as well as $\xi^{L}_{\mathcal{F}}\mathcal{F}\xi^{R}_{\mathcal{F}} = \omega_{\mathcal{F}}$).

 Let us consider the dust modes ($\omega_{\mathcal{A}}$), so as to understand 
 the different regimes of the MHD RDI.
The dust density mode, which satisfies  $\omega_{\mathcal{A}} = k_{z} \driftvelmag$, is the simplest: the resonance 
condition $\omega_{\mathcal{F}} = \omega_{\mathcal{A}}={\bf k}\cdot \driftvel$ can generally be satisfied 
exactly for some choice of ${\bf \hat{k}}$, leading
to the standard ``mid-$k$'' RDI, described in \S~\ref{sec:mhdwave.rdi:growthrates.midk}.
The other dust modes, which involve a velocity perturbation, are damped by
 the drag on the gas (with $\Im(\omega_{\mathcal{A}})\sim -\langle t_{s}\rangle^{-1}$), and thus cannot resonate exactly with the (undamped) MHD modes.
However, it transpires that we may consider $\mathcal{D}_{\mathrm{drag}}$ to be {a part of the perturbation} (i.e., effectively part of $\mu \mathbb{T}^{(1)}$), \emph{so long as the perturbation to the eigenmode ($\omega^{(1)}$) is larger than the change to $\omega_{\mathcal{A}}$ that arises from including $\mathcal{D}_{\mathrm{drag}}$}. 
In other words, if part of the $\mathcal{A}$ matrix ($\mathcal{D}_{\mathrm{drag}}$ or $\mathcal{D}_{\mathrm{gyro}}$) is smaller than $\mu \mathbb{T}^{(1)}$, we should  consider this part as belonging to the perturbation, rather 
than to the $\mathcal{A}$ matrix itself.\footnote{\label{foot:app.B.determinant}
The easiest way to understand that this should be the case is to put the matrix \eqref{eq: basic matrix blocks}
into the fluid eigenmode basis, by making the transformation,
\begin{equation}
\mathbb{T}_{\mathcal{F}} = 
\left(\begin{array}{cc}
\mathbb{I} & 0 \\
0 & \xi^{L}_{\mathcal{F}} 
\end{array}\right) \mathbb{T}\left(\begin{array}{cc}\mathbb{I} & 0 \\
0 & \xi^{R}_{\mathcal{F}} 
\end{array}\right) = \left(\begin{array}{cc}
\mathcal{A} & \mathcal{C}\xi^{R}_{\mathcal{F}} \\
\mu\, \xi^{L}_{\mathcal{F}} \mathcal{T}^{(1)}_{FA} & \omega_{\mathcal{F}}
\end{array}\right).\label{eq: footnote matrix in e-mode basis}
\end{equation}
Note that we have included only the $\mathcal{T}^{(1)}_{FA}$ part of $\mathbb{T}^{(1)}$ in  Eq.~\eqref{eq: footnote matrix in e-mode basis}, as appropriate for computation of the lowest-order $\mu$ perturbation (see \paperone). A direct computation of the determinant of $\mathbb{T}_{\mathcal{F}}-(\omega_{\mathcal{F}}+\omega^{(1)})\mathbb{I}$ captures both the mid-$k$ and high-$k$ RDI growth 
rates (and the gyroresonant mode). Specifically, if $\omega^{(1)}\ll 1$, the $\mathcal{D}_{\mathrm{drag}}$ contribution is important and we obtain only the mid-$k$ mode. If $\omega^{(1)}\gg 1$, $\mathcal{D}_{\mathrm{drag}}$ does not contribute and we can  obtain the gyro-resonant mode (as a different root of the resulting polynomial in $\omega^{(1)}$). Finally, if $\omega^{(1)}\gg \mathrm{MAX}(1,\tau)$, neither $\mathcal{D}_{\mathrm{drag}}$ nor $\mathcal{D}_{\mathrm{gyro}}$ contributes, and we obtain the high-$k$ mode $\omega^{(1)}\sim \mu^{1/3}$.
}
This implies that if $\omega^{(1)} \gtrsim 1$ --- i.e., if the perturbation to the eigenvalues from $\mathbb{T}^{(1)}$ is larger than that  due to $\mathcal{D}_{\mathrm{drag}}$ --- then we should not include   $\mathcal{D}_{\mathrm{drag}}$ in the calculation 
of the eigenmodes of $\mathcal{A}$ (physically, this condition, $\omega^{(1)} \gtrsim 1 = \langle t_{s}\rangle^{-1}$, is simply that the instability growth time  is faster than the stopping time).  
We then see that the gyroresonant modes  occur when $\pm \tau +{\bf k}\cdot \driftvel \approx \omega_{F}$ (see Eq.~\ref{eqn:gyro.condition} ), and
when the resulting perturbation, $\omega^{(1)}$, satisfies $\omega^{(1)}\gtrsim 1$ (see \S~\ref{sec:gyro} for further discussion). This
is only possible for $\tau>1$, because otherwise the damping of the dust gyration modes is larger than the
effect of the magnetic field (i.e., the gyration).
The high-$k$ RDI (\S~\ref{sec:mhdwave.rdi:growthrates.hik}) is slightly different, 
arising from a \emph{triple resonance} between the fluid mode, the dust density perturbation mode ($\omega_{\mathcal{A}}={\bf k}\cdot \driftvel $), and the (damped) dust velocity perturbation mode ($\omega_{\mathcal{A}}={\bf k}\cdot \driftvel -\iimag$). The matrix is then triply defective, leading to the different high-$k$ RDI scaling, $\omega^{(1)}\sim \mu^{1/3}$ (rather than $\omega^{(1)}\sim \mu^{1/2}$; see \paperone). It transpires\textsuperscript{\ref{foot:app.B.determinant}} that this triple 
resonance requires that the $\omega^{(1)}$ perturbation is larger than both $\mathcal{D}_{\mathrm{drag}}$ and $\mathcal{D}_{\mathrm{gyro}}$, so the high-$k$ scaling applies once $\omega^{(1)}\gtrsim \mathrm{MAX}(1,\tau)$. When $\tau>1$ there is a transition range $1\lesssim \omega^{(1)}\lesssim \tau$ where there is no clear universal scaling of the mode; see, for example, the $\tau=100$ cases in Fig.~\ref{fig:dispersion.relation} (in particular the fast-mode panel, which reaches sufficiently high growth rates to show the $\omega^{(1)}\gtrsim \tau$ transition).

Once one has decided which of the modes one wishes to study, the growth rate 
can be computed using the techniques introduced in \paperone. 
For the mid-$k$ RDI and the gyro-resonance mode, which arise from the double resonance (between one dust eigenmode and the chosen gas eigenmode), the eigenvalue perturbation is given by
\begin{equation}
\omega^{(1)} \approx  \mu^{1/2}\left[\left(\xi_{\mathcal{F}}^{L}\, \mathcal{T}^{(1)}_{FA}\, \xi_{\mathcal{A}}^{R}\right) \left(\xi_{\mathcal{A}}^{L}\, \mathcal{C}\, \xi_{\mathcal{F}}^{R}\right)\right]^{1/2},\label{eqn: double resonant general}
\end{equation}
where $\xi_{\mathcal{A}}^L$ and $\xi_{\mathcal{A}}^R$ are the left and right eigenvectors for the chosen dust mode (mid-$k$ RDI or gyro-resonance mode). For the high-$k$ RDI, which arises from the triple resonance (between two dust eigenmodes and the chosen gas eigenmodes), the perturbation is 
\begin{equation}
\omega^{(1)}\approx s_3\,\mu^{1/3 }\left[ (\xi_{\mathcal{F}}^{L} \mathcal{T}^{(1)}_{\rho_{d}})\,(\bf{k}^{T}\mathcal{C}_{\bf{v}}\xi_{\mathcal{F}}^{R})\right]^{1/3},\label{eqn: triple resonant dust}
\end{equation}
where $\mathcal{T}_{\rho_d}^{(1)}$ is the left column of  $\mathcal{T}_{FA}^{(1)}$ (this arises from the $\omega_{\mathcal{A}} = k_{z} \driftvelmag$ dust eigenmode), $\mathcal{C}_{\bf v}$ denotes the lower three columns of $\mathcal{C}$, and we have also assumed $\omega^{(1)}\gg \tau$ for simplicity (so as to easily compute the dust eigenmodes). 
Such methods  --- i.e., using Eqs.~\eqref{eqn: double resonant general} and \eqref{eqn: triple resonant dust} --- are commensurate
with the dispersion-relation expansions used throughout the main text: in some cases 
these methods provide a simpler way of obtaining the RDI growth rates, in other cases 
the dispersion relation expansions are simpler. While Eq.~\eqref{eqn: double resonant general} (for both the mid-$k$ and gyro-resonant mode) and Eq.~\eqref{eqn: triple resonant dust} can be straightforwardly, if tediously, computed from the matrix Eq.~\eqref{eq: full matrix form}, the resulting expressions are complex enough so 
as to require subsidiary expansions to reach forms similar to those given in the main text. 
Given this, we do not provide these expressions in full here.

\subsection{An algorithm to find magnetic RDIs} \label{sub:simple.matrix.resonance.algorithm}

Let us summarize the rather technical discussion of the previous paragraphs with a simple algorithm
for ``choosing'' the relevant dust-gas resonance, using the matrix resonance theory of \paperone.
\begin{enumerate}
\item Choose a gas wave and calculate its eigenfrequencies and eigenmodes.
\item Calculate the perturbed eigenvalues $\omega^{(1)}$ of the mid-$k$, high-$k$, and gyroresonant modes from Eqs.~\eqref{eqn: double resonant general} and \eqref{eqn: triple resonant dust}. (Note that the low-$k$ modes do not arise from a resonance at all; see \S~\ref{sub: low k regime solutions} and \papertwo).
\item If the gyroresonant mode satisfies $\omega^{(1)}\gtrsim 1$ then it can grow, and the 
expression is valid. If not, it is damped by the gas drag (i.e., the neglected $\mathcal{D}_{\mathrm{drag}}$ term in $\mathcal{A}$ is important).
\item If the  mid-$k$ RDI (Eq.~\ref{eqn: double resonant general}) satisfies $\omega^{(1)}\lesssim 1$, then this is 
the correct expression and the mode is in the mid-$k$ regime. Otherwise, if the high-$k$ result (Eq.~\ref{eqn: triple resonant dust}) satisfies $\omega^{(1)}\gtrsim \tau$, the expression is valid and the mode is in the high-$k$ regime. If  $1 \lesssim \omega^{(1)}\lesssim  \tau$, it is likely that neither the mid-$k$ or the high-$k$ results are correct.
\end{enumerate}
This series of steps can usually be used to understand the transitions between regimes discussed in \S~\ref{sec:mhdwave.rdi:growthrates.midk}-\ref{sec:mhdwave.rdi:growthrates.hik} and \S~\ref{sec:gyro}, although inaccuracies can arise near certain special points in parameter space (e.g., for certain combinations of the $\zeta_{X}$ parameters). Most importantly, it shows that the transition between the mid-$k$ ($\Im(\omega)\sim k^{1/2}$) and high-$k$ ($\Im(\omega)\sim k^{1/3}$) regimes occurs when $\Im(\omega)\sim \langle t_s \rangle^{-1}$ if $\tau\lesssim 1$, and that the gyro-resonant mode always requires $\tau\gtrsim 1$ to grow.
The method does not, however, find non-resonant modes (e.g., the Bell instability in \S~\ref{sub:cosmic.ray.streaming.mode}, or the low-$k$ modes of \S~\ref{sub: low k regime solutions}), which may be the fastest-growing modes in some regimes (e.g., $\tau\gg 1$ and/or larger $\mu$).

\end{document}